\documentclass[12pt]{iopart} 
\usepackage{catmac,amssymb,bbm,amsthm,amscd}
\newtheorem{Theorem}{Theorem}[section]

\theoremstyle{remark}

\newcommand{\CC}{{\mathbb{C}}}

\newcommand{\II}{{\mathbbm{1}}}

\newcommand{\RR}{{\mathbb{R}}}

\newcommand{\ZZ}{{\mathbb{Z}}}
\renewcommand{\gcd}{g}

\newcommand{\hMD}{h^\rmM_D}
\newcommand{\hSD}{h^\rmS_D} 
\newcommand{\hUD}{h^\rmU_D}

\newcommand{\MC}[1]{\rmM_{#1}(\CC)}
\newcommand{\MJ}{\MC{J}}
\newcommand{\MJp}{\MC{J'}}
\newcommand{\prM}[2]{\pr^{\rmM}_{#1,#2}}
\newcommand{\prU}[2]{\pr^{\rmU}_{#1,#2}}

\newcommand{\lSJ}{\lambda^\rmS_J}
\newcommand{\lSJp}{\lambda^\rmS_{J'}}
\newcommand{\lUJ}{\lambda^\rmU_J}

\newcommand{\ab}[2]{#1 \,\!^{[\,#2\,]}}

\newcommand{\Ad}{{\rm Ad}}

\newcommand{\bfb}{{\bf b}}

\newcommand{\bfh}{{\bf h}}

\newcommand{\bfk}{{\bf k}}

\newcommand{\bfm}{{\bf m}}

\newcommand{\bfv}{{\bf v}}

\newcommand{\BSUJ}{\rmB\SUJ}
\newcommand{\BUJ}{\rmB\UJ}
\newcommand{\Bun}[2]{{\mbox{\rm Bun}(#1,#2)}}

\newcommand{\calY}{{\cal Y}}

\newcommand{\con}{{\cal A}}
\newcommand{\const}{{\mbox{\rm const}}}

\newcommand{\CP}{\CC{\rm P}}
\newcommand{\CW}{{\it CW}}

\newcommand{\diag}{{\mbox{\rm diag}}}
\newcommand{\Diff}{\mbox{\rm Diff}}
\newcommand{\diff}{{\rm d}}

\newcommand{\gau}{{\cal G}}
\newcommand{\gref}[1]{{\rm (\ref{#1})}}

\newcommand{\Hom}{{\mbox{\rm Hom}}}

\newcommand{\id}{{\rm id}}
\newcommand{\im}{{\rm im}\,}

\newcommand{\ka}[1]{f_{#1}}

\newcommand{\lens}[2]{{{\rm L}_{#1}^{#2}}}

\newcommand{\LRA}{\mbox{~~$\Leftrightarrow$~~}}

\newcommand{\mod}{{\mbox{\rm ~mod~}}}
\newcommand{\mf}{\mathfrak}
\newcommand{\onepoint}{\ast}
\newcommand{\orb}{{\cal M}}

\newcommand{\pr}{{\rm pr}}
\newcommand{\RA}{\mbox{$\Rightarrow$~~}}

\newcommand{\rmB}{{\rm B}} 

\newcommand{\rmC}{{\rm C}}

\newcommand{\rmE}{{\rm E}}

\newcommand{\rmG}{{\rm G}}

\newcommand{\rmK}{{\rm K}}

\newcommand{\rmL}{{\rm L}}

\newcommand{\rmM}{{\rm M}}

\newcommand{\rmN}{{\rm N}}

\newcommand{\rmP}{{\rm P}}
\newcommand{\rmp}{{\rm p}}

\newcommand{\rmS}{{\rm S}}

\newcommand{\rmT}{{\rm T}}

\newcommand{\rmU}{{\rm U}}
\newcommand{\rmu}{{\rm u}}

\newcommand{\rmZ}{{\rm Z}}

\newcommand{\rmeven}{{\rm even}}

\newcommand{\rmgl}{{\rm gl}}

\newcommand{\rmSO}{{\rm SO}}

\newcommand{\rmSU}{{\rm SU}}
\newcommand{\rmsu}{{\rm su}}
\newcommand{\rref}[1]{{\rm \ref{#1}}}

\renewcommand{\smile}{}

\newcommand{\sphere}[1]{{{\rm S}^{#1}}}

\newcommand{\SUJ}{{\rmSU(J)}}
\newcommand{\SUJp}{{\rmSU(J')}}
\newcommand{\torus}[1]{{{\rm T}^{#1}}}
\newcommand{\tsl}{{\cal T}}
\newcommand{\tw}{\tilde}
\newcommand{\twA}{{\tw A}}
\newcommand{\twF}{{\tw F}}
\newcommand{\twH}{{\tw H}}
\newcommand{\twm}{{\tw m}}
\newcommand{\twP}{{\tw P}}
\newcommand{\twQ}{{\tw Q}}

\newcommand{\twgamma}{{\tw{\gamma}}}

\newcommand{\twphi}{{\tw{\phi}}}

\newcommand{\type}{{\rm Type}}
\newcommand{\UJ}{{\rmU(J)}}
\newcommand{\UJp}{{\rmU(J')}} 
\newcommand{\ve}{\varepsilon}
\newcommand{\vp}{\varphi}
\newcommand{\vr}{\varrho}

\newcommand{\wt}{\widetilde}
\newcommand{\ezmatrix}[2]{\left(\begin{array}{cc}   
#1&#2\end{array}\right)}
\newcommand{\zematrix}[2]{\left(\begin{array}{c}    
#1\\#2\end{array}\right)}
\newcommand{\zzmatrix}[4]{\left(\begin{array}{cc}   
#1&#2\\#3&#4\end{array}\right)}
\newcommand{\vveckmatrix}[9]{\left(\begin{array}{cccc}  
#1&#2&\cdots&#3\\#4&#5&\cdots&#6\\          
\vdots&\vdots&\ddots&\vdots\\#7&#8&\cdots&#9
\end{array}\right)}
\newcommand{\congeq}[1]{\con^{\geq#1}}
\newcommand{\conleq}[1]{\con^{\leq#1}}
\newcommand{\hori}{{\mf H}}
\newcommand{\Inv}{{\rm Inv}}
\newcommand{\Met}{\mbox{\rm Met}}
\newcommand{\orbgeq}[1]{\orb^{\geq#1}}   
\newcommand{\orbleq}[1]{\orb^{\leq#1}}
\newcommand{\slice}{{\cal S}}  
\newcommand{\tube}{{\cal U}}
\renewcommand{\con}{{\cal C}}
\renewcommand{\gau}{{\cal G}}
\renewcommand{\orb}{{\cal M}}
\renewcommand{\tsl}{{\cal T}}

\newcommand{\twbfm}{\wt\bfm}
\renewcommand{\vert}{{\mf V}}
\newcommand{\punkt}[1]{\put(#1){\circle*{0.06}}}
\newcommand{\marke}[3]{\put(#1){\put(0.05,0.1){\makebox(-0.1,-0.2)[#2]{\tiny
   $#3$}}}}
\newcommand{\vpunkte}[1]{\put(#1){\multiput(0,-0.1)(0,0.1){3}{\circle*{0.01}}}}
\newcommand{\linie}[3]{\put(#1){\line(#2){#3}}}

\newcommand{\whole}[3]{\put(#1){\punkt{0,0}\put(0.05,0.1){\makebox(-0.1,
   -0.2)[#2]{\tiny $#3$}}}}

\newcommand{\plene}[3]{\put(#1){\punkt{0,0}\linie{0.1,0}{1,0}{0.8}
\put(0.05,0.1){\makebox(-0.1,-0.2)[#2]{\tiny $#3$}}}}

\newcommand{\plzee}[3]{\put(#1){\punkt{0,0}\linie{0.0894,0.0447}{2,1}{0.821}
\put(0.05,0.1){\makebox(-0.1,-0.2)[#2]{\tiny $#3$}}}}

\newcommand{\plzmee}[3]{\put(#1){\punkt{0,0}\linie{0.0894,-0.0447}{2,-1}{
0.821}\put(0.05,0.1){\makebox(-0.1,-0.2)[#2]{\tiny $#3$}}}}

\newcommand{\plvee}[3]{\put(#1){\punkt{0,0}\linie{0.0968,0.0242}{4,1}{0.8064}
\put(0.05,0.1){\makebox(-0.1,-0.2)[#2]{\tiny $#3$}}}}

\newcommand{\plvmee}[3]{\put(#1){\punkt{0,0}\linie{0.0968,-0.0242}{4,-1}{
0.8064}\put(0.05,0.1){\makebox(-0.1,-0.2)[#2]{\tiny $#3$}}}}

\newcommand{\pslene}[1]{\put(#1){\multiput(0.1,0)(0.1,0){9}{\circle*{0.01}}}}

\newcommand{\pslvee}[1]{\put(#1){\multiput(0.0968,0.0242)(0.1,0.025){9}{\circle*{0.01}}}}
\newcommand{\pslvmee}[1]{\put(#1){\multiput(0.0968,-0.0242)(0.1,-0.025){9}{\circle*{0.01}}}}
\newcommand{\pslvde}[1]{\put(#1){\multiput(0.08,0.06)(0.1,0.075){9}{\circle*{0.01}}}}
\newcommand{\pslvmde}[1]{\put(#1){\multiput(0.08,-0.06)(0.1,-0.075){9}{\circle*{0.01}}}}
\newcommand{\plpen}[6]{\put(#1){\punkt{0,0}\linie{0,-0.1}{0,-1}{0.8}%
\punkt{0,-1}%
\put(0.05,0.1){\makebox(-0.1,-0.2)[#2]{#6$#3$}}%
\put(0.05,-0.9){\makebox(-0.1,-0.2)[#4]{#6$#5$}}}}
\newcommand{\plpzn}[6]{\put(#1){\punkt{0,0}\linie{-0.05,-0.1}{0,-1}{0.8}%
\linie{0.05,-0.1}{0,-1}{0.8}%
\punkt{0,-1}%
\put(0.05,0.1){\makebox(-0.1,-0.2)[#2]{#6$#3$}}%
\put(0.05,-0.9){\makebox(-0.1,-0.2)[#4]{#6$#5$}}}}
\newcommand{\plpee}[6]{\put(#1){\punkt{0,0}\linie{0.0447,-0.0894}{1,-2}{0.4103}%
\punkt{0.5,-1}%
\put(0.05,0.1){\makebox(-0.1,-0.2)[#2]{#6$#3$}}%
\put(0.55,-0.9){\makebox(-0.1,-0.2)[#4]{#6$#5$}}}}
\newcommand{\plpeme}[6]{\put(#1){\punkt{0,0}\linie{-0.0447,-0.0894}{-1,-2}{0.4103}%
\punkt{-0.5,-1}%
\put(0.05,0.1){\makebox(-0.1,-0.2)[#2]{#6$#3$}}%
\put(-0.45,-0.9){\makebox(-0.1,-0.2)[#4]{#6$#5$}}}}
\newcommand{\plpze}[6]{\put(#1){\punkt{0,0}\linie{0,-0.1118}{1,-2}{0.4106}%
\linie{0.0894,-0.0671}{1,-2}{0.4106}%
\punkt{0,-1}%
\put(0.05,0.1){\makebox(-0.1,-0.2)[#2]{#6$#3$}}%
\put(0.55,-0.9){\makebox(-0.1,-0.2)[#4]{#6$#5$}}}}
\newcommand{\plpzme}[6]{\put(#1){\punkt{0,0}\linie{-0.0894,-0.0671}{-1,-2}{0.4106}%
\linie{0,-0.1118}{-1,-2}{0.4106}%
\punkt{0,-1}%
\put(0.05,0.1){\makebox(-0.1,-0.2)[#2]{#6$#3$}}%
\put(-0.45,-0.9){\makebox(-0.1,-0.2)[#4]{#6$#5$}}}}
\newcommand{\plped}[6]{\put(#1){\punkt{0,0}\linie{0.0832,-0.0554}{3,-2}{1.3336}%
\punkt{1.5,-1}%
\put(0.05,0.1){\makebox(-0.1,-0.2)[#2]{#6$#3$}}%
\put(1.55,-0.9){\makebox(-0.1,-0.2)[#4]{#6$#5$}}}}
\newcommand{\plpemd}[6]{\put(#1){\punkt{0,0}\linie{-0.0832,-0.0554}{-3,-2}{1.3336}%
\punkt{-1.5,-1}%
\put(0.05,0.1){\makebox(-0.1,-0.2)[#2]{#6$#3$}}%
\put(-1.45,-0.9){\makebox(-0.1,-0.2)[#4]{#6$#5$}}}}
\newcommand{\plpev}[6]{\put(#1){\punkt{0,0}\linie{0.0894,-0.0447}{2,-1}{1.821}%
\punkt{2,-1}%
\put(0.05,0.1){\makebox(-0.1,-0.2)[#2]{#6$#3$}}%
\put(2.05,-0.9){\makebox(-0.1,-0.2)[#4]{#6$#5$}}}}
\newcommand{\plpemv}[6]{\put(#1){\punkt{0,0}\linie{-0.0894,-0.0447}{-2,-1}{1.821}%
\punkt{-2,-1}%
\put(0.05,0.1){\makebox(-0.1,-0.2)[#2]{#6$#3$}}%
\put(-1.95,-0.9){\makebox(-0.1,-0.2)[#4]{#6$#5$}}}}
\newcommand{\plpaen}[8]{\put(#1){\punkt{0,0}\linie{0,-0.1}{0,-1}{0.8}%
\punkt{0,-1}%
\put(0.05,0.1){\makebox(-0.1,-0.2)[#2]{#8$\begin{array}[b]{c}#4\\#3\end{array}$}}%
\put(0.05,-0.9){\makebox(-0.1,-0.2)[#5]{#8$\begin{array}[t]{c}#6\\#7\end{array}$}}}}
\newcommand{\plpaez}[8]{\put(#1){\punkt{0,0}\linie{0.0707,-0.0707}{1,-1}{0.8586}%
\punkt{1,-1}%
\put(0.05,0.1){\makebox(-0.1,-0.2)[#2]{#8$\begin{array}[b]{c}#4\\#3\end{array}$}}%
\put(1.05,-0.9){\makebox(-0.1,-0.2)[#5]{#8$\begin{array}[t]{c}#6\\#7\end{array}$}}}}
\newcommand{\plpaemz}[8]{\put(#1){\punkt{0,0}\linie{-0.0707,-0.0707}{-1,-1}{0.8586}%
\punkt{-1,-1}%
\put(0.05,0.1){\makebox(-0.1,-0.2)[#2]{#8$\begin{array}[b]{c}#4\\#3\end{array}$}}%
\put(-0.95,-0.9){\makebox(-0.1,-0.2)[#5]{#8$\begin{array}[t]{c}#6\\#7\end{array}$}}}}
\begin{document}
\begin{titlepage}
\begin{center}
{\LARGE{\bf On the Gauge Orbit Space Stratification\\[0.5cm](A Review)}}

\vspace{2.5cm}

{\large {\bf G. Rudolph, M. Schmidt and I.P. Volobuev$^1$}}

\vskip 0.5 cm

Institute for Theoretical Physics\\
University of Leipzig\\
Augustusplatz 10\\
04109 Leipzig \& Germany
\\[0.5cm]
$^1$Skobeltsyn Institute of Nuclear Physics\\ 
Moscow State University\\
119992 Moscow \& Russia

\end{center}

\vspace{2.5 cm}

\begin{abstract}
First, we review the basic mathematical structures and results concerning the 
gauge orbit space stratification. This includes general properties of the 
gauge group action, fibre bundle structures induced by this action, basic 
properties of the stratification and the natural Riemannian structures of the 
strata. In the second part, we study the stratification for theories with 
gauge group $\rmSU(n)$ in space time dimension $4$. We develop a general
method for determining the orbit types and their partial ordering, based 
on the $1$-$1$ correspondence between orbit types and holonomy-induced Howe 
subbundles of the underlying principal $\rmSU(n)$-bundle. We show that the 
orbit types are classified by certain cohomology elements of space time 
satisfying two relations and that the partial ordering 
is characterized by a system of algebraic equations. Moreover, operations for 
generating direct successors and  direct predecessors are formulated, which 
allow one to construct the set of orbit types, starting from the principal 
one. Finally, we discuss an application to nodal configurations in 
Yang-Mills-Chern-Simons theory.
\end{abstract}
\vspace{1cm}
\end{titlepage}
\tableofcontents
\newpage

\small\normalsize
\section{Introduction}
\label{Sintro}
One of the basic principles of modern theoretical physics is the
principle of local gauge invariance. Its application to the theory of
particle interactions gave rise to the standard model, which proved to
be a success from both theoretical and phenomenological points of view.
The most impressive results of the model were obtained within the 
perturbation theory, which works well for high energy processes. On
the other hand, the low energy hadron physics, in particular, the
quark confinement, turns out to be dominated by nonperturbative effects,
for which there is no rigorous theoretical explanation yet. To study them, a
variety of different concepts and mathematical methods has been developed. In
particular, for some aspects methods of differential geometry and algebraic
topology seem to be unavoidable. This is certainly true, if one wants to
investigate the structure of the configuration space of a gauge
theory -- the space of gauge group orbits. In general,
this space possesses not only orbits of the so called principal type,
but also orbits of other types, which may give rise to singularities.
This stratified structure of the gauge orbit space is believed to be of
importance for both classical and quantum properties of
non-abelian gauge theories in the nonperturbative approach. 
Let us discuss some aspects indicating its physical relevance. 

First, studying the geometry and topology of the generic (principal)
stratum, one gets an intrinsic topological interpretation of the Gribov-ambiguity 
\cite{Gribov,Singer:Gribov}. We stress that the problem of finding all Gribov copies has been 
discussed within specific models, see e.g. \cite{Langmann}. For a detailed analysis in the case
of $2$-dimensional cylindrical space time (including the Hamiltonian path integral) we refer to
\cite{Shabanov1}. 
Investigating the topology of the determinant line bundle over the generic 
stratum, one gets an understanding of anomalies in terms of the family index 
theorem \cite{Alvarez, Atiyah/Singer}, see also \cite{Carey} for the Hamiltonian 
approach. In particular, one gets anomalies of purely topological type \cite{Witten}, 
which cannot be seen by perturbative quantum field theory. Moreover, there are partial
results and conjectures concerning the relevance of nongeneric
strata. First, generally speaking, nongeneric gauge orbits affect the
classical motion on the orbit space due to boundary conditions
and, in this way, may produce nontrivial contributions to the path
integral. They may lead to localization of certain quantum
states, as it was suggested by finite-dimensional examples
\cite{EmmrichRoemer}. Further, the gauge field configurations
belonging to nongeneric orbits can possess a magnetic charge, i.e.
they can be considered as a kind of magnetic monopole
configurations. Following t`Hooft \cite{tHooft}, these could be responsible for quark confinement. 
The role of these configurations was investigated within the framework of Schr\"odinger 
quantum mechanics on the gauge orbit space of topological Chern-Simons theory in 
\cite{Asorey:Nodes}, see also \cite{Asorey:99} for an approach to 
4-dimensional Yang-Mills theories with $\theta$-term. Within 
t`Hooft`s concept, the idea of abelian projection is of special importance and has been 
discussed by many authors. Recently, this concept was studied within the setting of 
quantum field theory at finite temperature on the $4$-torus \cite{Tok1,Tok2}. 
There, a hierarchy of defects, which should be related to the gauge orbit space
structure, was discovered.
Finally, let us also mention that the existence of additional anomalies corresponding to 
non-generic strata was suggested, see \cite{Heil:Anom}.

Most of the problems mentioned here are still awaiting a
systematic investigation. For that purpose, a deeper insight into the structure 
of the gauge orbit space is necessary. In a series of papers 
\cite{RSV:clfot,RSV:poot,RS:otG} we have made 
a new step in this direction. We have given a complete solution to
the problem of determining the strata that are present in the
gauge orbit space for $\rm SU(n)$ gauge theories in compact Euclidean 
space time of dimension $d=2,3,4$. Our analysis is based on the results of 
Kondracki and Rogulski \cite{KoRo}, where the general structure of 
the full gauge orbit space was investigated for the first time in detail. In particular, 
it was shown that the gauge orbit space is a stratified topological space. 
Moreover, these authors found the basic relation between orbit types
and certain bundle reductions, which we are using. We note that this relation was 
also observed in \cite{Heil:OrbSpa}. 

We mention that there is an approach based upon parameterizing the 
full gauge orbit space by a so called fundamental domain, characterized by 
the fact that, up to identifications on the boundary, it is intersected by 
every gauge orbit exactly once, see 
\cite{Zwanziger1,FSS,vanBaal1,vanBaal2,Zwanziger2} and 
further references therein. However, for the study of the stratified structure of the gauge 
orbit space, this concept seems not to be efficient. 

Finally, we note that the stratification structure for gauge theories within 
the Ashtekar approach has been also clarified, see \cite{Fleisch}.

This review is organized as follows. In the first part, the basic mathematical structures 
and results concerning the gauge orbit space stratification are discussed. 
In Section \rref{Sbasic} we briefly recall the setup and sketch the basic properties  
of the gauge group action, including a slice theorem and an approximation theorem. In Section 
\ref{Sbun}, the fibre bundle structures induced by this action are investigated. 
Next, in Section \ref{Sstrat}, basic properties of the stratification 
are derived and, in Section \ref{SRiemann}, the natural Riemannian structures of the strata 
are discussed. This concludes the general part of the review.
In the remaining part, we specify the gauge group to be $\rmSU(n)$
and space time to be of dimension less than or equal to $4$. Under these
assumptions, the strata can be classified by characteristic classes of certain
reductions of the principal bundle the theory is defined on. This will be
explained in Section \rref{Sclfic}. In Section \rref{Spo}, we show how
the natural partial ordering of strata, which contains information on how
the strata are linked, can be read off from algebraic relations between the
characteristic classes. Finally, we discuss the case of gauge group
$\rmSU(2)$
for some $4$-manifolds in detail and present an application to nodal
configurations in topological Chern-Simons theory. For the convenience of the
reader, we have added two appendices on aspects of bundle theory and algebraic 
topology used in the text, as well as an appendix in which we explain how to
construct the Postnikov towers of the classifying spaces relevant for the
classification of orbit types.
\section{Basics}\label{Sbasic}
\subsection{Setup}
In what follows, we assume that the reader is familiar with the standard
formulation of gauge theories in terms of fibre bundles and connections 
\cite{Daniel,Eguchi,Trautman}. 
Thus, let $M$ be a compact connected orientable Riemannian manifold, let $G$ be
a compact connected linear Lie group with Lie algebra $\mf g$ and let $P$ be 
a smooth locally trivial principal $G$-bundle over $M$. In physical terms, $M$
is a model of space time and $G$ is the gauge group.

For any vector bundle $E$, let $W^k(E)$ denote the Hilbert space of cross sections
of $E$ of Sobolev class $k$. For generalities on such spaces, see 
\cite{Palais}, for the application of these techniques to gauge theories see 
\cite{Mitter}. 
Let $\con$ denote the subspace of $W^k(\rmT^\ast P\otimes\mf g)$ of connection 
forms on $P$ of Sobolev class $k$ and let $\gau$ denote the closure of the
group of smooth $G$-space morphisms $P\rightarrow G$ in $W^{k+1}(P,\rmgl(n,\CC))$. 
Here $n$ is chosen so that $G\subseteq\rmgl(n,\CC)$. In physics, connection
forms represent gauge potentials and $\con$ is the configuration space of the
theory. Elements of $\gau$ represent local gauge transformations acting on
connections by
\begin{equation}\label{Gaction}
A^{(g)} = \Ad(g^{-1})A+g^{-1}\diff g\,.
\end{equation}
The space $\con$ is an affine separable Hilbert space with translational 
vector space
$$
\tsl = W^k(\rmT^\ast M \otimes \Ad P)\,,
$$
where $\Ad P$ denotes the associated bundle $P\times_G\mf g$. 
Throughout the review, we will assume $k>\dim(M)/2+1$. Then the Sobolev
lemma ensures that multiplication of a $W^{k+1}$-function by a $W^l$-function, 
$\dim(M)/2<l\leq k \, ,$ yields a $W^l$-function. It follows that $\gau$ is a
group, acting via \gref{Gaction} on $\con$. In fact, one can prove that
$\gau$ is a Hilbert-Lie group 
with Lie algebra
$$
\rmL\gau = W^{k+1}(\Ad P)
$$
and exponential mapping
\begin{equation}
\label{exp}
\exp_{\gau}(\xi)(p) = \exp_{G}(\xi(p)) \, , \, \forall \xi \in \rmL\gau \, , \, 
p \in P \, ,
\end{equation}
and that the action is smooth \cite{MitterViallet,NaraRama,Singer:Gribov}. 

It should be noted that for both $\tsl$ and $\rmL\gau$, identification of 
sections in associated bundles with the corresponding $G$-equivariant 
horizontal forms on $P$ is understood. We will stick to this 
identification throughout the review.
Also note that the elements of $\con$ and $\gau$ are $C^1$ and $C^2$, 
respectively. In particular, $\gau$ may be viewed as consisting of
vertical automorphisms of $P$ of class $C^2$ or of sections of class $C^2$ 
in the associated fibre bundle $P\times_G G$ \cite{Fischer}.

The gauge orbit space is
$$
\orb := \con/\gau\,,
$$
which is, at this stage, just a topological quotient. It will be equipped
with additional structure later. Note that $\orb$ is the space of classes of gauge
equivalent potentials -- the 'true' configuration space.
 
The scalar products on the Hilbert
spaces $\rmL\gau$ and $\tsl$, respectively, are not intrinsic. Their only
purpose is to define the topology. The geometry of these spaces is defined by
$L^2$-scalar products, induced
from the Riemannian metric on $M$ and an $\Ad(G)$-invariant scalar product
$\langle\cdot,\cdot\rangle$ on $\mf g$ as follows:
$$
(\xi,\eta)_0 := \int_M \langle \xi, \ast \eta \rangle
\,,~~\xi,\eta\in\rmL\gau
\,,~~~~~~
(X,Y)_0 := \int_M\langle X\wedge\ast Y\rangle
\,,~~X,Y\in\tsl\,,
$$
respectively. Here $\ast$ denotes the Hodge duality operator. Both of these
scalar products are invariant under the adjoint action of $\gau$. 

Since $\con$ is affine with translational vector space $\tsl$, we have
\begin{equation}\label{Gidtific}
\rmT\con = \con\times\tsl\,.
\end{equation}
In particular, any smooth assignment of a scalar product in $\tsl$ to the 
elements of $\con$ defines a Riemannian metric on $\con$. Examples are:

(i) The constant assignment $A\mapsto(\cdot,\cdot)_0$ defines the natural 
(weak) $L^2$-metric $\gamma^0$. It is invariant under the induced action
of $\gau$ on $\tsl$, given by 
$$
X^{(g)} = \Ad(g^{-1}) \, X\,.
$$

(ii) The assignment $A\mapsto\gamma^k_A$, induced from 
\begin{equation}
\label{Gdefstrong}
\gamma^k_A(X,Y) := \sum_{l=0}^k ~~\left([\wt\nabla_A]^l \, X,
[\wt\nabla_A]^l \, Y \right)_0\,, 
~~~X,Y\in C^\infty(\rmT^\ast M \otimes\Ad P)\,,
\end{equation}
by prolongation to $\tsl$, defines a natural metric $\gamma^k$. Here 
$$
\wt\nabla_A
:
C^\infty({\rmT^\ast M}^{\otimes l} \otimes \Ad P)
\rightarrow
C^\infty(\rmT^\ast M^{\otimes (l+1)}\otimes\Ad P)
\,,~~~~
\alpha \mapsto \nabla^{LC} \alpha+[A,\alpha]\,,
$$
where $\nabla^{LC}$ is the Levi-Civita connection of the Riemannian metric on $M$
and 
$$
[A,\alpha](X_0,X_1,\dots,X_l)
=
[A(X_0),\alpha(X_1,\dots,X_l)]\,.
$$
The norm on $\tsl$ defined by the scalar products $\gamma^k_A$, $A\in\con$, 
is equivalent to the $W^k$-norm \cite{Ebin}. Therefore, $\gamma^k$ is a strong metric. 
Moreover, due to  
$$
\left(\wt\nabla_{A^{(g)}}\right)^l 
=
\Ad(g^{-1})\left(\wt\nabla_A\right)^l\Ad(g)\,,
$$
it is $\gau$-invariant, $\gamma^k_{A^{(g)}}(X^{(g)},Y^{(g)}) = \gamma^k_A(X,Y)
\, .$

(iii) Let us remark that one can construct further $\gau$-invariant 
metrics using the Laplacian $\square_A =
\nabla_A^\ast\nabla_A+\nabla_A\nabla_A^\ast$ as
\begin{equation}
\label{Gdeflapl}
\eta^k_A(X,Y) 
=
((1+\square_A)^{k/2} X, (1+\square_A)^{k/2} Y)_0\,,
\end{equation}
where $(1+\square_A)^{k/2}$ is defined via functional calculus. For some
specific examples, like the principal $\rmSU(2)$-bundle of second Chern class
('instanton number') $c_2=1$ over $\CC\rmP^2$, the restriction of $\eta^2$ to 
the moduli space of irreducible self-dual connections 
was studied in detail, see \cite{Habermann} and references therein. We do
not comment on this here.

Next, for $A\in\con$, consider the operator of covariant derivative 
w.r.t.~$A$,
$$
\nabla_A: W^{k+1}(\Ad P)\rightarrow W^k(\rmT^\ast M\otimes\Ad P)\,.
$$
Its formal adjoint w.r.t.~the $L^2$-scalar product is the bounded linear operator
$$
\nabla_A^\ast:W^k(\rmT^\ast M \otimes\Ad P) \rightarrow W^{k-1}(\Ad P)\,,
$$
defined by
$$
(\nabla_A\xi,X)_0 = (\xi,\nabla_A^\ast X)_0
\,,~~~~
\forall \xi\in C^\infty(\Ad P), X\in C^\infty(\rmT^\ast M\otimes\Ad P)\,.
$$
Composition then yields a bounded linear operator
$$
\Delta_A = \nabla_A^\ast\nabla_A:W^{k+1}(\Ad P)\rightarrow W^{k-1}(\Ad P)\,.
$$
In the following, instead of $W^l(\Ad P)$ or $W^l(\rmT^\ast M\otimes\Ad P)$
we shall often write $W^l$, because the bundle in which the sections are
taken
can be read off unambiguously from the operators under consideration.
Moreover, the pure symbols $\nabla_A$, $\nabla_A^\ast$, $\Delta_A$ always
stand for the maps $\nabla_A|W^{k+1}$, $\nabla_A^\ast|W^k$, and
$\Delta_A|W^{k+1}$ with $k$ fixed, whereas, for example, $\nabla_A|W^{l+1}$
means that $\nabla_A$ is viewed as an operator $W^{l+1}\rightarrow W^l$
(where $\dim(M)/2<l\leq k$).

Note that the maps 
$$
\con\rightarrow\rmB(W^{k+1},W^k)\, , \, \, A\mapsto \nabla_A, 
~~~
\con\rightarrow\rmB(W^k,W^{k-1})\, , \, \, A\mapsto \nabla_A^\ast, 
$$
are continuous linear. Hence, the map
$$
\con\rightarrow\rmB(W^{k+1},W^{k-1})\, , \, \, A\mapsto\Delta_A, 
$$
is continuous. Since it factorizes into continuous linear maps and
composition of operators, it is even smooth. Moreover, we note the
following equivariance properties:
\begin{equation}\label{GnablaAg}
D_{A^{(g)}} = \Ad(g^{-1}) \, D_A \, \Ad(g)
\,,~~~\forall~A\in\con,g\in\gau \, , 
\end{equation}
where $D$ stands for $\nabla$, $\nabla^\ast$ and $\Delta \, ,$ respectively.
\subsection{Stabilizers}
Recall that the stabilizer (or isotropy subgroup) 
of $A\in\con$ w.r.t.~the action of $\gau$ is the subgroup
$$
\gau_A := \{ g\in\gau~:~A^{(g)} = A \}
$$
of $\gau$. It is determined by the holonomy of $A \, .$ Indeed, 
$g\in\gau_A$ iff $g$ is constant on any curve horizontal with respect to $A$. 
Thus,
\begin{equation} \label{Gstabbunred}
\gau_A = \{ g\in\gau~:~g|_{P_{A,p_0}} = \const \}\,,
\end{equation}
where $P_{A,p_0}$ denotes the holonomy bundle of $A$ based at
$p_0\in P$. Note that $P_{A,p_0}$ is of class $C^2$, because $A$ is
$C^1$.

Let $\xi\in\rmL\gau$. We have 
$$
\nabla_A\xi=0
~~~\LRA~~~
\xi|_{P_{A,p_0}} = \const
~~~\LRA~~~
\exp_\gau(\xi)|_{P_{A,p_0}} = \const\,,
$$
where the second equivalence is due to \gref{exp}. Thus,
$$
\exp_\gau(\rmL\gau)\cap\gau_A = \exp_\gau(\ker(\nabla_A))\,.
$$
Since $\ker(\nabla_A)$ is a closed subspace of the Hilbert space $\rmL\gau$, the r.h.s.~is
a submanifold of $\gau$. Since the l.h.s.~is a neighbourhood of $e$ in $\gau_A$,
it follows that $\gau_A$ is a Lie subgroup of $\gau$ 
with Lie algebra
\begin{equation}\label{GLgau}
\rmL\gau_A=\ker(\nabla_A)=\{ \xi\in\rmL\gau~:~\xi|_{P_{A,p_0}} = \const
\}\,,
\end{equation}
see \cite[\S III.1.3]{Bou:Lie}.
Next, consider the natural group homomorphism
$$
\Phi_{p_0} : \gau\rightarrow G\,,~~~g\mapsto g(p_0)
$$
(the value of $g$ at a point is of course well defined). Since convergence
in $W^{k+1}$, by our choice of $k$, implies pointwise convergence,
$\Phi_{p_0}$ is continuous, hence smooth. Due to \gref{Gstabbunred}, the
restriction of $\Phi_{p_0}$ to the subgroup $\gau_A$ is injective, hence
a Lie group isomorphism onto its image. The image is
$$
\Phi_{p_0}(\gau_A) = \rmC_G(H_{A,p_0})\,,
$$
where $H_{A,p_0}$ denotes the holonomy group of $A$ based at $p_0$. To see
this, recall that $H_{A,p_0}$ is the structure group of $P_{A,p_0}$. Thus,
inclusion from left to right is due to equivariance of the elements
of $\gau$. For the converse inclusion it suffices to note that for
any $a\in\rmC_G(H_{A,p_0})$, the function on $P_{A,p_0}$ with constant
value $a$ is equivariant and, hence, can be equivariantly prolonged to $P$,
thus becoming an element of $\gau_A$.

Let us summarize.
\begin{Theorem}[Stabilizer theorem] \label{Tstab}
$\gau_A$ is a compact Lie subgroup of $\gau$ with Lie algebra given by \gref{GLgau}. 
Through $\Phi_{p_0}$, $\gau_A$ is isomorphic to $\rmC_G(H_{A,p_0})$.
\end{Theorem}
As an immediate consequence of the fact that $\gau_A$ is an (embedded) Lie 
subgroup, the projection 
$\gau \rightarrow \gau/\gau_A $ 
defines a locally trivial principal bundle \cite[\S 6.2.4]{Bou:Lie}.

In \cite{NaraRama} it was shown that the map 
$
\con\times\gau\rightarrow\con\times\con
\,,~
(A,g)\mapsto(A,A^{(g)})\,,
$
is closed. It follows \cite[III,\S4]{Bou:Top}
\begin{Theorem}\label{Tprop}
The action of $\gau$ on $\con$ is proper.
\end{Theorem}
\noindent
Immediate consequences are 

(i) The orbits of the action of $\gau$ on $\con$ are closed.

(ii) The orbit space $\orb$ is Hausdorff.
\vspace{0.2cm}

A different proof of Theorem \ref{Tprop} was given in \cite{KoRo}. 
By assigning to $A\in\con$ a $W^k$-Riemannian metric on $P$,
$$
h_A(u,v) = h_M(\pi_\ast u,\pi_\ast v) + \langle
A(u),A(v)\rangle
\,,~~~ u,v \in \rmT_pP\,, p\in P\,,
$$
where $h_M$ is the Riemannian metric on $M$,
a homeomorphism of $\con$ onto a closed submanifold of the manifold
$\Met^k(P)$ 
of $W^k$-Riemannian metrics on $P$ is constructed (it is even a
diffeomorphism into). $\Met^k(P)$ is acted upon by 
the topological group $\Diff^{k+1}(P)$ of $W^{k+1}$-diffeomorphisms of $P$. 
$\Diff^{k+1}(P)$ is known to be a smooth manifold, but not a Lie group. The 
action is known to be smooth and proper 
\cite{Bourguignon,Ebin,Fischer:Super}. 
It is shown in \cite{KoRo}
that $\gau$ is a closed topological subgroup of $\Diff^{k+1}(P)$ (it is even
a submanifold) and that the embedding $\con\rightarrow\Met^k(P)$ is equivariant. 
Thus, properness carries over from the action of $\Diff^{k+1}(P)$ on 
$\Met^k(P)$ to that of $\gau$ on $\con$. 

Note that compactness of stabilizers is not needed in the second proof.
Rather, it is a consequence of properness of the action.
\subsection{Orbit types}
According to
$
\gau_{A^{(g)}} = g^{-1}\gau_A g\,,
$
the stabilizers along an orbit $x\in\orb$ form a conjugacy class in $\gau$.
This class is called the orbit type of $x$ and is denoted by $\type(x)$.
The set of orbit types carries a natural partial ordering: $\sigma\leq\sigma'$ 
iff there exist
representatives $S$ of $\sigma$ and $S'$ of $\sigma'$ such that $S\supseteq
S'$. Then for any pair of representatives $S$, $S'$ there exists $g\in\gau$
such that $S\supseteq aS'a^{-1}$. One says that $S'$ is subconjugate to $S$.
Note that, although this definition of the partial ordering of orbit types is
the usual one \cite{Bou:Lie,Bredon:CTG}, it is not consistent with
\cite{KoRo}, where the inverse partial ordering is used.  

We are going to characterize orbit types in terms
of certain bundle reductions of $P$, see also \cite{Heil:OrbSpa} for a
similar approach. For that purpose, let us consider, 
for a moment, smooth connections and smooth local gauge transformations.
Recall that a subgroup of $G$ that can be written as a centralizer is usually
called a {\it Howe subgroup}. This is due to the fact that such a subgroup, 
together with its centralizer, forms a reductive dual pair, a notion 
introduced by R.~Howe \cite{Howe1,Howe2,Howe3}. 
According to that, let us call a bundle reduction of
$P$ to
a Howe subgroup of $G$ a {\it Howe subbundle}. 
(All bundle reductions are assumed
to be smooth.) As any subgroup $H\subseteq G$ generates a Howe subgroup 
$\wt H$ (containing $H$) by $\wt H = \rmC_G^2(H)$, any
bundle reduction $Q$ of $P$ to $H$ generates a Howe subbundle $\wt Q$
(containing $Q$) by extending the structure group to $\wt H$,
$$
\wt Q = Q\wt H\,.
$$
In particular, a connection $A$ generates a Howe subbundle $\wt P_{A,p_0}$
through its holonomy bundle. In \cite{KoRo}, $\wt P_{A,p_0}$ was called evolution bundle
of $A$ .
Since an element of $G$ that commutes with $H_{A,p_0}$ still commutes with
$\wt H_{A,p_0}$, a gauge transformation that is constant on $P_{A,p_0}$ is
still constant on $\wt P_{A,p_0}$. Thus,
\begin{equation}\label{GstabHSB}
\gau_A = \{ g\in\gau~:~g|_{\wt P_{A,p_0}}=\const. \}\,.
\end{equation}
We claim that $\wt P_{A,p_0}$ consists of all $p\in P$ obeying 
$$
g(p)=g(p_0)\,,~~~\forall g\in\gau_A\,.
$$
To see this, let $p\in P$ with $g(p)=g(p_0)$, $\forall g\in\gau_A$. There 
exist $p'\in P_{A,p_0}$ and $a\in G$ such that $p=p'a$. Due to equivariance, 
$g(p) = a^{-1}g(p')a$, hence $g(p_0)=a^{-1}g(p_0)a$, $\forall g\in\gau_A$.
Thus, $a$ commutes with $\Phi_{p_0}(\gau_A)$. Now the stabilizer theorem
yields that $a\in\rmC_G^2(H_{A,p_0})\equiv \wt H_{A,p_0}$, hence $p\in\wt P_{A,p_0}$.

It follows that $\wt P_{A,p_0}$ is determined by the subgroup $\gau_A$ rather
than by $A$ itself. Thus, by assigning $\wt P_{A,p_0}$ to $\gau_A$ we obtain
a map from stabilizers to Howe subbundles. Since $\gau_A$ can be recovered
from $\wt P_{A,p_0}$ via \gref{GstabHSB}, the map is injective.
What kind of Howe subbundles arise in this way from stabilizers?
Of course, all of them are generated by a {\em connected} reduction of $P$. 
Howe subbundles with this property will be called holonomy-induced. 
Conversely, let a holonomy-induced Howe subbundle $\wt Q$ with generating
connected bundle reduction $Q$ be given. As is well known \cite{KoNo}, if $\dim M\geq
2$, there exist connections in $P$ which have holonomy bundle $Q$. Then $\wt
Q$ is the Howe subbundle assigned to the stabilizer of any of these
connections. 

To summarize, we have found, within the $C^\infty$-setting, that stabilizers 
are in $1$-$1$ correspondence with holonomy-induced Howe subbundles. 
To carry over 
this characterization to the conjugacy classes, we note that, for gauge
transformations $g$,
\begin{equation}\label{GPAg}
P_{A^{(g)},p_0} = \left(\Theta_g(P_{A,p_0})\right)g(p_0)^{-1}\,,
\end{equation}
where $\Theta_g$ denotes the vertical automorphism of $P$ defined by $g$,
i.e., 
$$
\Theta_g(p) = pg(p)\,,~~~\forall p\in P\,.
$$
Since \gref{GPAg} carries over to the corresponding Howe subbundles, we have
to factorize the holonomy-induced Howe subbundles by vertical automorphisms
of $P$. Since any isomorphism of one bundle reduction of $P$ onto another one
can be extended to a vertical automorphism of $P$, the factorization is
actually by isomorphy. Moreover, in order to make the construction independent 
of the chosen point $p_0$, one must
take Howe subbundles modulo the principal action of $G$ on $P$. Note that
then the corresponding structure
groups are determined up to conjugacy in $G$.

Thus, we have found a characterization of the orbit types of the action of
smooth local gauge transformations on smooth connections. Finally, one can 
prove that the action of $\gau$ on $\con$ has exactly the same orbit types 
\cite{RSV:clfot}. 

Let us summarize.
\begin{Theorem}[Reduction theorem]\label{Totgau}
The orbit types of the action of $\gau$ on $\con$ are in $1$-$1$ 
correspondence with
smooth holonomy-induced Howe subbundles of $P$ modulo isomorphy and modulo the
principal action of $G$ on $P$. The correspondence is given by
\gref{GstabHSB}. 
\end{Theorem}
Note that it is obvious from \gref{GstabHSB} that the partial orderings  
of orbit types and bundle reductions coincide. 
For later use, let us introduce the notation $\con^S$ for the subset of
connections with stabilizer $S$, $\con^\sigma$ for the subset of connections 
of orbit type $\sigma$ and $\orb^\sigma$ for the subset of orbits of type
$\sigma$. Correspondingly, we define
$$
\conleq{S} := \bigcup_{S'\supseteq S} \con^{S'}
\,,~~~
\conleq{\sigma} := \bigcup_{\sigma' \leq \sigma} \con^{\sigma'}
\,,~~~
\orbleq{\sigma} := \bigcup_{\sigma' \leq \sigma} \orb^{\sigma'}\,,
$$
and similarly $\congeq{S}$, $\congeq{\sigma}$, $\orbgeq{\sigma}$.
\subsection{Decomposition theorem}
In what follows we will see that there exists a natural generalization of the
Hodge-de Rham decomposition theorem (w.r.t.~the $L^2$-metric $\gamma_0$) 
to the covariant derivatives $\nabla_A$. 
This has two important consequences. First, 
it ensures that the orbits of the $\gau$-action are submanifolds. Second, it
implies that the two distributions on $\con \, ,$ defined by
\begin{equation}
\label{Gverthoriidtific}
\vert_A = \im(\nabla_A)\,,~~~~\hori_A = \ker(\nabla_A^\ast) \,
,~~~~A\in\con\,,
\end{equation} 
provide a natural orthogonal splitting of the tangent bundle, 
\begin{equation} \label{GdecompTA}
\rmT\con = \vert \oplus \hori\,.
\end{equation}
This splitting is fundamental for all constructions discussed within the rest of 
this and the next three sections. In particular, it is basic for the construction of tubes and
slices, it ensures the (locally trivial) fibre bundle structure on each stratum 
and it induces natural (weak) Riemannian metrics on each stratum of the gauge
orbit space via a Kaluza-Klein construction.

Using the theory of differential operators with $W^l$-coefficients 
\cite{Cantor,Choquet}, one can verify that the following 
decompositions hold, see \cite{Rogulski} for explicit proofs.
\begin{Theorem}[Decomposition theorem]
\label{Tdecomp}
Let $A\in\con$. Then
\begin{eqnarray}\label{Gdecompnabla}
W^k(\rmT^\ast M\otimes\Ad P) & = & \im(\nabla_A)\oplus\ker(\nabla_A^\ast)\,,
\\ \label{GdecompDelta}
W^{k-1}(\Ad P) & = & \im(\Delta_A)\oplus\ker(\Delta_A)\,,
\end{eqnarray}
where the sums are orthogonal w.r.t. the corresponding $L^2$-scalar products.
\end{Theorem}

\noindent
{\bf Remarks:}\\
1. The decompositions still hold if one replaces $\nabla_A$, $\nabla_A^\ast$,
$\Delta_A$ by $\nabla_A|W^{l+1}$, $\nabla_A^\ast|W^l$, $\Delta_A|W^{l+1}$,
respectively, with $\dim(M)/2<l\leq k$.
\\
2. As an immediate consequence of \gref{Gdecompnabla},
\begin{eqnarray} \label{GkerDAkernA}
\ker(\Delta_A) & = & \ker(\nabla_A)\,,
\\ \label{GimDAimnA}
\im(\Delta_A) & = & \im(\nabla_A^\ast)\,.
\end{eqnarray}
3. In the decomposition \gref{GdecompDelta}, the subspace $\ker(\Delta_A)$
of $W^{k+1}$ is viewed as a subspace of $W^{k-1}$. Actually, there should
occur $\ker(\Delta_A|W^{k-1})$ instead. However, by virtue of point 1 above,
formula \gref{GkerDAkernA} holds also in degree $\dim(M)/2<l\leq k$.
Since $\diff$ is elliptic, $\ker(\nabla_A|W^{l+1}) = \ker(\nabla_A)$, for any
$\dim(M)/2<l\leq k$. Hence, \gref{GkerDAkernA} implies $\ker(\Delta_A|W^{k-1}) 
= \ker(\Delta_A)$.
\vspace{0.3cm}

\noindent
As an important consequence of the decomposition theorem one has 
\begin{Theorem}\label{Torb}
For any $A\in\con$, the orbit of $A$ under the action of $\gau$ is a 
submanifold of $\con$, naturally diffeomorphic to $\gau/\gau_A$.
\end{Theorem}
This was proved in \cite{KoRo}. Since the orbits are closed due to
properness of the action and since 
the topology of $\con$ is second countable (recall that $\con$ is separable), 
it suffices to show that the map
\begin{equation}\label{Gorbitmap}
\iota_A:\gau\rightarrow\con\,, ~~~g\mapsto A^{(g)}\,,
\end{equation}
is a subimmersion \cite[\S 5.12.5]{Bou:Lie}. The map $\iota_A$ factors through
$\gau/\gau_A$,
$$
\gau\rightarrow\gau/\gau_A\stackrel{\tw\iota_A}{\rightarrow}\con\,.
$$
Since the first mapping is the projection in a locally trivial principal bundle, it 
is a submersion. We claim that $\tw\iota_A$ is a smooth immersion (so that
\gref{Gorbitmap} is  a subimmersion, indeed). 

Smoothness follows from the fact that, due to local triviality of the
principal bundle $\gau\rightarrow\gau/\gau_A$, $\gau/\gau_A$ can be covered
by smooth local sections $\gau/\gau_A\supseteq U\rightarrow\gau$.
Namely, locally, $\tw\iota_A$ factors through such a section and $\iota_A$. 

To prove that $\tw\iota_A$ is an immersion, it suffices to show that 
it is an immersion at $[e]\in\gau/\gau_A$, the class of the identity of 
$\gau$. Given a closed subspace $\calY$ 
of $\rmL\gau$ complementary to $\rmL\gau_A$,
one can find an appropriate local section $(U,s)$ about $[e]$ such that its
tangent map $(s_\ast)_{[e]}$ maps $\rmT_{[e]}\gau/\gau_A$ isomorphically 
onto $\calY$. Then 
$$
({{\tilde\iota}_A}{}_\ast)_{[e]}\circ\left((s_\ast)_{[e]}\right)^{-1}
=
({\iota_A}_\ast)_e |_{\calY}\,. 
$$
Since $(s_\ast)_{[e]}$ is an isomorphism, it suffices to show that the r.h.s.
is injective and has closed image. 
For that purpose, recall that the Killing field at $A$ generated by $\xi$ is 
\begin{equation}\label{GT1iotaA}
{\iota_A}_\ast\,\,\xi = \nabla_A\xi\,.
\end{equation}
Now, injectivity is obvious from \gref{GLgau}.
Moreover, $\im \left(({\iota_A}_\ast)_e |_{\calY} \right) = \im(\nabla_A)$
and, due to the decomposition theorem, the image is closed and
admits a closed complement.  
\vspace{0.2cm}

As a second important consequence of the decomposition theorem we note that
the tangent bundle splitting (\ref{GdecompTA}) holds and is orthogonal 
w.r.t.~the $L^2$-metric $\gamma^0 \,.$ Due to \gref{GnablaAg}, 
the distributions $\vert$ and $\hori$ are equivariant,
\begin{equation}\label{Gevrdtb}
\vert_{A^{(g)}} = (\vert_A)^{(g)}
\,,~~~
\hori_{A^{(g)}} = (\hori_A)^{(g)}\,.
\end{equation}
Geometrically, $\vert$ consist of the subspaces tangent to the orbits. 
We stress that, in general, neither $\vert$ nor $\hori$ are smooth or
locally trivial. However, as we will see later, restrictions to strata 
will be so.

Let us determine the projectors 
$$
\bfv,\bfh:\rmT\con\rightarrow\rmT\con
$$
onto $\vert$ and $\hori$, respectively. They are given by maps
$$
\con\rightarrow\rmB(\tsl)
\,,~~~~
A\mapsto\bfv_A,\bfh_A\,,
$$
where $\bfv_A$ and $\bfh_A$ denote the projectors associated to the 
decomposition \gref{Gdecompnabla}.
Since $\ker(\Delta_A)\subseteq W^{k+1}$, the decomposition
\gref{GdecompDelta} implies 
\begin{equation}\label{GdecompWk+1}
W^{k+1} = \ker(\Delta_A)\oplus\ker(\Delta_A)^{\perp_0}\,,
\end{equation}
where $\ker(\Delta_A)^{\perp_0} = W^{k+1} \cap \im(\Delta_A)$.
Thus, by restriction, $\Delta_A$ induces a bounded operator 
$\ker(\Delta_A)^{\perp_0}\rightarrow\im(\Delta_A)$ which is invertible,
hence has bounded inverse by the open mapping theorem. The inverse can be
prolonged to a bounded operator 
\begin{equation}\label{GdefGA}
\rmG_A : W^{k-1}(\Ad P) \rightarrow W^{k+1}(\Ad P)\,,
\end{equation}
the Green's operator associated to $\Delta_A$, 
by setting $\rmG_A|\ker(\Delta_A) = 0$. Note that 
$\rmG_A\Delta_A:W^{k+1}\rightarrow W^{k+1}$ is the $L^2$-orthogonal 
projector onto $\ker(\Delta_A)^{\perp_0}$. Hence,
\begin{equation}\label{Gweglass}
\nabla_A\rmG_A\Delta_A = \nabla_A
\,,~~~~
\Delta_A\rmG_A\nabla_A^\ast = \nabla_A^\ast\,.
\end{equation}
Note, in particular, that $\rmG_A$ is not the inverse of $\Delta_A$, 
unless $\gau_A$ is discrete, as in the case of the 
principal stratum for semisimple structure group \cite{NaraRama}.

Now consider the composition $\nabla_A\rmG_A\nabla_A^\ast$, which is a
bounded operator on $\tsl$. Using \gref{Gweglass} one can check that it is
a projector and that it acts trivially on $\hori_A$ and identically on
$\vert_A$. Thus,
\begin{equation} \label{GvA,hA}
\bfv_A = \nabla_A\rmG_A\nabla_A^\ast
\,,~~~
\bfh_A = \II-\bfv_A\,.
\end{equation}
From \gref{GnablaAg} we infer 
\begin{equation} \label{GevrGA}
\rmG_{A^{(g)}} = \Ad(g^{-1})~\rmG_A~\Ad(g)\,.
\end{equation}
It follows 
\begin{equation} \label{GevrvA}
\bfv_{A^{(g)}} = \Ad(g^{-1})~\bfv_A~\Ad(g)
\,,~~~~
\bfh_{A^{(g)}} = \Ad(g^{-1})~\bfh_A~\Ad(g)
\end{equation}
which is consistent with \gref{Gevrdtb}.
\subsection{Slice theorem} 
\label{SSslice}
We assume the reader to be familiar with the notions of tube and slice
\cite{Bredon:CTG}. They
are generalizations of the notions of local trivialization and local
section,
respectively, which apply to group actions with a single orbit type.

Following \cite{KoRo}, the normal distribution $\hori$ can be used to construct
tubes and slices for the action of $\gau$ on $\con$.
For $x\in\orb$, the normal bundle of the orbit $\pi^{-1}(x)$ is given by 
$$
N_x := \hori|_{\pi^{-1}(x)}\,.
$$
According to \gref{Gevrdtb}, $N_x$ is equivariant. We claim that it is a
smooth locally trivial vector subbundle of $\rmT\con|_{\pi^{-1}(x)}$. To
see this, observe that for given
$A\in\pi^{-1}(x)$, due to local triviality
of the principal bundle $\gau\rightarrow\gau/\gau_{A}$, there exists a 
neighbourhood $U_{A}$ of $A$ in $\pi^{-1}(x)$ and a smooth map 
$\theta:U_{A}\rightarrow\gau$ such that
$A'=A^{(\theta(A'))}$, for any $A'\in U_A$. 
The map 
$$
U_{A}\times\tsl\rightarrow\rmT\con|_{U_A}
\,,~~~~
(A',X)\mapsto(A',X^{(\theta(A'))})\,,
$$
is easily seen to be a diffeomorphism. Due to equivariance of $N_x$, 
the pre-image of $N_x|_{U_A}$ under this map is 
$U_A\times\hori_A$. This proves the assertion .
Let us note that the argument shows that any equivariant vector subbundle of
$\rmT\con|_{\pi^{-1}(x)}$ which has closed fibres is smooth and locally trivial.

For $\ve>0$, define 
$$
\hori_{A,\ve} := \{X\in\hori_A~:~\sqrt{\gamma^k_A(X,X)}\leq\ve\}\,,
$$
where the $W^k$-metric $\gamma^k$ was defined in \gref{Gdefstrong}.
Consider the smooth subbundle 
$$
N_{x,\ve} := \{(A,X)\in N_x ~:~ X\in\hori_{A,\ve}\}
$$
of $N_x$. Note that $N_{x,\ve}$ is not just the $\ve$-disk bundle of 
$N_x$, because orthogonality and length are taken w.r.t.~different
metrics.
Due to $\gau$-invariance of $\gamma^k$, $N_{x,\ve}$ is equivariant.

As $\gau$-spaces, $N_x$ and $N_{x,\ve}$ are equivariantly diffeomorphic
through the rescaling map 
$$
\vr_\ve:N_x\rightarrow N_{x,\ve}\,,~~(A,X)\mapsto
\left(A,\frac{\ve}{\sqrt{\gamma_A^k(X,X)+1}}X\right)\,.
$$
By restriction, the map 
$$
\exp:\rmT\con\rightarrow\con\,,~~(A,X)\mapsto A+X\,,
$$
which is in fact the exponential map w.r.t.~the $L^2$-metric 
$\gamma^0$, defines a smooth $\gau$-equivariant map 
$N_{x,\ve}\rightarrow\con$. The image is   
\begin{equation}\label{Gdeftube}
\tube_{x,\ve} = \{A+X~:~\pi(A)=x \, , \, X\in\hori_{A,\ve}\}\,.
\end{equation}
It is an open invariant neighbourhood of $\pi^{-1}(x)$ in $\con$ 
(called 'tubular neighbourhood').
Using that $\pi^{-1}(x)$ is a submanifold, one can show \cite{KoRo} that 
there exists $\ve>0$ such that the restriction of $\exp$ to 
$N_{x,\ve}\subseteq\rmT\con$ is injective. Consequently, the composition 
\begin{equation}\label{Gtube}
\exp\circ\vr_\ve:N_x\rightarrow\con\,,
\end{equation}
is an equivariant diffeomorphism onto $\tube_{x,\ve}$, i.e., it is a tube. 
(Note that already $\exp|_{N_{x,\ve}}$ alone is a tube.)

From \gref{Gdeftube} we can easily read off the slice about $A\in\pi^{-1}(x)$ 
associated to $\tube_{x,\ve}$. It is the subset
$$
\slice_{A,\ve}:=\{A+X~:~X\in\hori_{A,\ve}\}
$$
of $\tube_{x,\ve}$. By construction, $\slice_{A,\ve}$ obeys the defining 
properties of a slice:
\vspace{0.2cm}

(i) $\tube_{x,\ve}=\left(\slice_{A,\ve}\right)^{(\gau)}$,
\vspace{0.2cm}

(ii) $\slice_{A,\ve}$ is closed in $\tube_{x,\ve}$,
\vspace{0.2cm}

(iii) $\slice_{A,\ve}$ is invariant under the stabilizer $\gau_A$,
\vspace{0.2cm}

(iv) For any $g\in\gau$, $\left(\slice_{A,\ve}\right)^{(g)}\cap\slice_{A,\ve}\neq\emptyset$
implies $g\in\gau_A$.
\vspace{0.2cm}

\noindent
We conclude:
\begin{Theorem}[Slice theorem]
\label{Tslice}
For any $x\in\orb$ there exists $\ve>0$ such that \gref{Gtube} is a tube
about $x$. For any $A\in\con$ there exists $\ve>0$ such that
$\slice_{A,\ve}$ is a slice about $A$. In particular, the action of $\gau$
on $\con$ admits a slice at any point.
\end{Theorem}
In the following, whenever we write $\tube_{x,\ve}$ or
$\slice_{A,\ve}$, it is understood that $\ve$ is small enough to make the
subset a tubular neighbourhood or a slice, respectively.

The authors of \cite{KoRo} actually prove more: they show that for any $x\in\orb$ and any
open invariant neighbourhood $U$ of $x$ in $\con$ there exists $\ve>0$ such that
$\overline{\tube_{x,\ve}}\subseteq U$ and
$U\setminus\overline{\tube_{x,\ve}}\neq\emptyset$. They call this the 'local
slice theorem'. As a consequence, $\orb$ is a regular topological
space, meaning that whenever one has a closed subset $V$ and a point $x \notin V$ 
then there exists a neighbourhood of $x \, ,$ whose closure in $\orb$ does not
intersect $V$. According to Urysohn's metrization theorem, regularity
in combination with second countability (which
is due to separability of $\con$) then implies that $\orb$ is a metrizable space.

As an application, let us note an immediate consequence of the 
slice theorem. Property (iv) of slices implies that for any 
$x\in\orb^\sigma$ and any $A\in\con^S$,
\begin{equation}\label{GSincongeq}
\tube_{x,\ve}\subseteq\congeq{\sigma}
\,,~~~~
\slice_{A,\ve} \subseteq \congeq{S}\,.
\end{equation}
It follows that for any stabilizer $S$ and orbit type $\sigma$ the following
subsets are open: 
$$
\con^S \mbox{ in }\conleq{S}
\,,~~~~
\con^\sigma \mbox{ in } \conleq{\sigma}
\,,~~~~
\orb^\sigma \mbox{ in } \orbleq{\sigma}\,.
$$
To see this, let $A\in\con^S$. Since $\tube_{\pi(A),\ve}$ is a 
neighbourhood of $A$ in
$\con$, its intersection with $\conleq{S}$ is a neighbourhood of $A$ in
$\conleq{S}$. Due to \gref{GSincongeq}, the intersection is contained in 
$$
\congeq{S}\cap\conleq{S}=\con^S\,.
$$
The argument applies without change to $\con^\sigma$. For $\orb^\sigma$ it
suffices to note that $\tube_{x,\ve}$ projects to a neighbourhood of $x$ in $\orb$.

\subsection{Approximation theorem}
It is well known that connections with trivial stabilizer under $\gau$-action
are dense in $\con$, see \cite{Singer:Gribov}. More generally, the question 
arises, whether
$\con^\sigma$ is dense in $\conleq{\sigma}$, in other words, whether a 
connection
with nontrivial stabilizer can be approximated by connections with a
prescribed, strictly smaller stabilizer. In \cite{KoRo}, the following is
proved.
\begin{Theorem}[Approximation theorem]
\label{Tapprox}
Assume $\dim M\geq 2 \, .$ Let $A\in\con$ and let $Q$ be a connected bundle
reduction of $P$ to a (not necessarily closed) Lie subgroup. Assume that
$Q$ contains a holonomy bundle of $A$. Then there exists $X\in\tsl$ such
that all $A+tX$, $t\in\RR\setminus\{0\}$, have holonomy bundle $Q$.
\end{Theorem}

By virtue of the characterization of stabilizers by bundle reductions of $P$, 
see \gref{Gstabbunred}, the approximation theorem implies that the following 
subsets are dense: 
\begin{equation}\label{Gdense}
\con^S\subseteq\conleq{S}\,,~
\con^\sigma\subseteq\conleq{\sigma}\,,~ 
\orb^\sigma\subseteq\orbleq{\sigma}\,. 
\end{equation}
Namely, let $A\in\conleq{S}$. Then $S\subseteq\gau_A$. Hence, according to
\gref{Gstabbunred}, the bundle reduction $Q_S$ associated to
$S$, based at some $p_0$, contains the holonomy bundle of $A$, based at
$p_0$. Of course, so does already the connected component $Q_{S,p_0}\subseteq
Q_S$ of $p_0$. Thus, Theorem \rref{Tapprox} yields that $A$ can be
approximated by connections with holonomy bundle $Q_{S,p_0}$. By
construction, such connections have stabilizer $S$. Hence, $\con^S$ is dense 
in $\conleq{S}$. Then denseness of $\con^\sigma$ in $\conleq{\sigma}$
and of $\orb^\sigma$ in $\orbleq{\sigma}$ follows.

One can combine openness, found above, and denseness by saying that $\con^S$, 
$\con^\sigma$,
$\orb^\sigma$ are generic sets in $\conleq{S}$, $\conleq{\sigma}$, and
$\orbleq{\sigma}$, respectively. 

Combining the approximation theorem with the slice theorem one arrives at the
following closure formulae: for any orbit type $\sigma$,
\begin{equation}\label{Gclosure}
\overline{\con^\sigma} = \conleq{\sigma}
\,,~~~~
\overline{\orb^\sigma} = \orbleq{\sigma}
\end{equation}
Indeed, the inclusions from right to left are obvious from \gref{Gdense}. 
The converse inclusions follow from the slice theorem: let
$A\in\overline{\con^\sigma}$. Consider
$\tube_{A,\ve}  \cap  \overline{\con^\sigma}$. Since this is a neigh\-bour\-hood of $A$ in
$\overline{\con^\sigma}$, it contains some $B\in\con^\sigma$. According to
\gref{GSincongeq}, then $\sigma\geq\type(A)$. Thus, $A\in\conleq{\sigma}$.
The inclusion for $\orb^\sigma$ then follows by noting that for saturated
sets like $\con^\sigma$, closure and projection commute.

We remark that for stabilizers $S$ one has a similar formula:
\begin{equation}\label{GclosureS}
\overline{\con^S} = \conleq{S}\,.
\end{equation}
While $\supseteq$ is again due to \gref{Gdense}, $\subseteq$ can
be proved without the slice theorem by the following simple argument. For
any $g\in\con$, consider the map
$$
\Phi_g:\con\rightarrow\tsl\,, ~~~A\mapsto A^{(g)}-A\,.
$$
As the $\Phi_g$ are continuous, the subsets $\Phi_g^{-1}(0)$ are closed in
$\con$. Then $\conleq{S} = \bigcap_{g\in S}\Phi_g^{-1}(0)$ is closed. Hence,
$\overline{\con^S}\subseteq\conleq{S}$.
\section{Smooth fibre bundle structure of strata}
\label{Sbun}
In this section, we shall explain how the projections
$$
\pi^\sigma:\con^\sigma\rightarrow\orb^\sigma
$$ 
induced from $\pi:\con\rightarrow\orb$ can be equipped with the structure of 
smooth locally trivial fibre bundles. As a result, in a sense, $\pi$ fibres 
over the set of orbit types into such bundles. 
\subsection{Submanifold structure of the configuration space strata}
To prove that $\con^\sigma$ is a submanifold of $\con$, it suffices to show 
that for any $x\in\orb^\sigma$ the subset
$$
\tube^\sigma_{x,\ve} := \tube_{x,\ve}\cap\con^\sigma\,,
$$
which is a neighbourhood of the orbit $\pi^{-1}(x)$ in $\con^\sigma$, 
is a submanifold of $\tube_{x,\ve}$. For any  
$A\in\con^\sigma$, define 
\begin{eqnarray}\nonumber
\slice^\sigma_{A,\ve} & := & \slice_{A,\ve}\cap\con^\sigma\,,
\\ \label{GdefhAs}
\hori^\sigma_{A} & := & \{X\in\hori_A~:~\gau_X\supseteq\gau_A\}\,,
\\ \nonumber
\hori^\sigma_{A,\ve} & := & \hori_{A,\ve}\cap\hori_A^\sigma\,.
\end{eqnarray}
Due to \gref{GSincongeq}, 
\begin{equation}\label{GGA0}
\gau_{A'} = \gau_A\,,~~~~\forall A'\in\slice^\sigma_{A,\ve}\,.
\end{equation}
Hence,
$
\slice^\sigma_{A,\ve} 
=
\{A+X~:~X\in\hori_{A,\ve}\,,~\gau_{A+X} = \gau_{A}\}\,.
$~
Since $\gau_{A+X} = \gau_{A}$ iff $\gau_X\supseteq\gau_{A}$,
\begin{equation}\label{GSAs}
\slice^\sigma_{A,\ve}
=
\{A+X~:~X\in\hori^\sigma_{A,\ve}\}\,.
\end{equation}
Then 
$$
\tube^\sigma_{x,\ve} =
\{A+X~:~A\in\pi^{-1}(x)\,,~X\in\hori^\sigma_{A,\ve}\}\,.
$$
Therefore, the pre-image of $\tube^\sigma_{x,\ve}$ under the equivariant 
diffeomorphism \gref{Gtube} is the vector subbundle
%
%
%
$$
N^\sigma_x := \bigcup_{A\in\pi^{-1}(x)}\hori^\sigma_{A}
$$
%
%
%
of $N_x$. As we have argued in Subsection \rref{SSslice}, 
since $N^\sigma_x$ is equivariant and since 
its fibres are closed subspaces of $\tsl$, it is a smooth subbundle of 
$\rmT\con|_{\pi^{-1}(x)}$, hence of $N_x$. It follows that 
$\tube^\sigma_{x,\ve}$ is a smooth
submanifold of $\tube_{x,\ve}$, for any $x\in\orb^\sigma$, as asserted.

For later purposes, let us note that the vector subbundle 
$N^\sigma_x$ is in fact trivial, where a smooth trivialization is given by 
$$
\gau/\gau_{A}\times\hori^\sigma_{A}
\rightarrow
N^\sigma_x
\,,~~~~
([g],X)\mapsto(A^{(g)},X^{(g)})\,,
$$
for some $A\in\pi^{-1}(x)$. Note that this map is well defined precisely
because $\gau_X\supseteq\gau_{A}$. It follows that $\tube^\sigma_{x,\ve}$ also
has a direct product structure. This can be made explicit by introducing 
maps
\begin{equation}\label{Gloctriv}
\chi^\sigma_{A,\ve}
:
\slice^\sigma_{A,\ve}\times\gau/\gau_{A}
\rightarrow
\tube^\sigma_{\pi(A),\ve}
\,,~~~~
(A',[g])\mapsto {A'}^{(g)}\,,
\end{equation}
which are easily seen to be diffeomorphisms. Note that, for obvious reasons, the 
roles of fibre and base have changed here.
Also for later purposes, let us note that
\begin{equation}\label{GTSAs} 
\rmT\slice^\sigma_{A,\ve} 
=
\slice^\sigma_{A,\ve}\times\hori^\sigma_{A}\,,
\end{equation}
for any $A\in\con^\sigma$, which is obvious from \gref{GSAs}. 
\subsection{Manifold structure of the orbit space strata}
We shall construct an atlas of the stratum $\orb^\sigma$ using the partial 
slices $\slice^\sigma_{A,\ve}$, $A\in\con^\sigma$. 
For any $x\in\orb^\sigma$, define
$$
V^\sigma_{x,\ve}
:=
\pi(\tube^\sigma_{x,\ve})\,.
$$
By restriction in domain and range, for any $A\in\pi^{-1}(x)$, $\pi$ 
defines a map
\begin{equation}\label{GdefpisAe} 
\pi^\sigma_{A,\ve}
:
\slice^\sigma_{A,\ve}
\rightarrow
V^\sigma_{x,\ve}
\end{equation}
We prove:

(i) $\pi^\sigma_{A,\ve}$ is bijective: Due to \gref{GGA0} and property (iv)
of slices, none of the elements of $\slice^\sigma_{A,\ve}$ has a gauge copy 
in $\slice^\sigma_{A,\ve}$.

(ii) $\pi^\sigma_{A,\ve}$ is a homeomorphism onto $V^\sigma_{x,\ve}$: 
It suffices to check
that $\pi$ maps open subsets of $\slice^\sigma_{A,\ve}$ to open subsets of
$V^\sigma_{x,\ve}$. Let $U\subseteq\slice^\sigma_{A,\ve}$ be open. Then
$U=\slice^\sigma_{A,\ve}\cap U'$, where $U'\subseteq\slice_{A,\ve}$ is open. 
Using a local trivialization of the normal bundle 
$N_x$, one can show that the saturation $\tw
U'={U'}^{(\gau)}$ is open in $\con$. Since $\slice^\sigma_{A,\ve}$ does not
contain gauge copies, $U=\slice^\sigma_{A,\ve}\cap\tw U'$. Since $\tw U'$ is
saturated,  
$$
\pi(U) = \pi(\slice^\sigma_{A,\ve})\cap\pi(\tw U')
=
V^\sigma_{x,\ve}\cap\pi(\tw U')\,.
$$
Here $\pi(\tw U')$ is open in $\orb$. Hence, $\pi(U)$ is open in
$V^\sigma_{x,\ve}$.

(iii) $V^\sigma_{x,\ve}$ is open in $\orb^\sigma$: obviously,
$V^\sigma_{x,\ve}=\orb^\sigma\cap\pi(\tube_{\pi(A),\ve})$, where
$\tube_{\pi(A),\ve}$ is open in $\con$.

Since the partial slices $\slice^\sigma_{A,\ve}$ are open subsets of
closed affine subspaces of $\con$, see \gref{GSAs}, the family
$(V^\sigma_{\pi(A),\ve},(\pi^\sigma_{A,\ve})^{-1})$, $A\in\con^\sigma$,
provides a covering of $\orb^\sigma$ by local charts (one can make this more
explicit by further mapping $\slice^\sigma_{A,\ve}\rightarrow\hori^\sigma_{A,\ve}$).
We finally have to check whether the transition maps between these charts 
are smooth. Due to
\gref{Gloctriv}, for $A_1,A_2\in\con^\sigma$ we have a diffeomorphism
%
%
%
$$
\slice^\sigma_{A_1,\ve_1}\cap\tube^\sigma_{\pi(A_2),\ve_2} 
\times
\gau/\gau_{A_1}
\stackrel{\chi^\sigma_{A_1,\ve_1}}{\longrightarrow}
\tube^\sigma_{\pi(A_1),\ve_1}\cap\tube^\sigma_{\pi(A_2),\ve_2}
\stackrel{(\chi^\sigma_{A_2,\ve_2})^{-1}}{\longrightarrow}
\slice^\sigma_{A_2,\ve_2}\cap\tube^\sigma_{\pi(A_1),\ve_1}
\times
\gau/\gau_{A_2}\,.
$$
%
%
%
The transition map $(\pi^\sigma_{A_2,\ve_2})^{-1}\circ\pi^\sigma_{A_1,\ve_1}$
is given by the composition of the embedding $A'\mapsto(A',[e])$, the above
diffeomorphism, and projection to the first component. Hence, it is smooth.
Thus, the atlas we have constructed equips $\orb^\sigma$ with the structure of 
a smooth Hilbert manifold.
\subsection{Smooth fibre bundle structure}
\label{smfbstr}
Using the local diffeomorphisms $\chi^\sigma_{A,\ve}$, we obtain local
diffeomorphisms 
$$
V^\sigma_{\pi(A),\ve}\times\gau/\gau_{A}
\stackrel{(\pi^\sigma_{A,\ve})^{-1}\times\id}{\longrightarrow}
\slice^\sigma_{\pi(A),\ve}\times\gau/\gau_{A}
\stackrel{\chi^\sigma_{A,\ve}}{\longrightarrow}
\tube^\sigma_{\pi(A),\ve}
$$
which provide a covering of $\con^\sigma$ by local trivializations of the
projection $\pi^\sigma:\con^\sigma\rightarrow\orb^\sigma$. Thus, the latter
is a smooth locally trivial fibre bundle with standard fibre
$\gau/\gau_{A}$, for some $A\in\con^\sigma$. In particular, $\pi^\sigma$
is a submersion, because locally it is the projection onto the first component.

Let us consider, in particular, the principal orbit type $\sigma=\sigma_{\rm p}$, which is 
the conjugacy class consisting of the subgroup $\tilde{\rmZ}(G)$ of constant functions
$P \rightarrow \rmZ(G)$, where $\rmZ(G)$ denotes the center of $G$. Since 
$\tilde{\rmZ}(G)$ is normal in $\gau$, the smooth locally trivial fibre bundle 
\begin{equation}\label{Gprinc}
\pi^{\rm p}:\con^{\rm p}\rightarrow\orb^{\rm p}
\end{equation}
is in fact principal, with structure group 
$\wt\gau := \gau/ \tilde{\rmZ}(G)$. This bundle has been studied intensively
\cite{Mitter,MitterViallet,NaraRama,Singer:Gribov}.
An important aspect is that nontriviality of this bundle is an obstruction to the existence of
smooth (or even continuous) gauges. An elegant argument to
show nontriviality, i.e., nonexistence  of smooth gauges, is due to Singer
\cite{Singer:Gribov}. Namely, assume that the bundle were trivial, i.e.,
$\con^{\rm p}\cong\orb^{\rm p}\times\wt\gau$. Since $\con^{\rm p}$ is contractible, then the
homotopy groups were $\pi_i(\wt\gau)=0$, $i\geq 1$. Since, in many cases this
is not true, one concludes that in these cases \gref{Gprinc} is nontrivial.
For $G=\rmSU(n)\, ,$ examples of this situation  are: 
space time manifolds $M=\rmS^3$ and $\rmS^4$ 
\cite{Singer:Gribov}, $\rmT^4$ and $\rmS^2\times\rmS^2$
\cite{Killingback:Gribov} and others. This explains the Gribov ambiguity 
\cite{Gribov} for the corresponding models.
\vspace{0.3cm}

\noindent
{\bf Remark:} For the other orbit types, 
representatives $S$ are not normal in $\gau$. In 
order to have a similar picture as in the case of the principal stratum, 
one would have to take the submanifold $\con^S$ of connections with
stabilizer $S$. $\con^S$ is acted upon freely by $N/S$, where
$N$ denotes the normalizer of $S$ in $\gau$. Provided one could show
that $N$ is a Lie subgroup of $\gau$ -- a problem which, to our
knowledge, is not settled yet -- the projection
$\pi^S:\con^S\rightarrow\orb^\sigma$ would be a smooth locally trivial
principal fibre bundle and $\pi^\sigma:\con^\sigma\rightarrow\orb^\sigma$
would be associated to this bundle via the action of $N/S$ on
$\gau/S$.
\section{The stratification of the gauge orbit space $\orb$}
\label{Sstrat}
A stratification of a topological space $X$ is a countable disjoint
decomposition into smooth manifolds $X_i$, $i  \in I$, (so-called
strata) such that the 'frontier condition' is satisfied:
$$
X_i \cap \overline{X_{i'}} \neq \emptyset
~~~\RA~
X_i \subseteq \overline{X_{i'}}
\,,~~~~~
\forall i,i'\in I\,.
$$
As this notion is rather weak, one usually adds additional assumptions 
about the linking between the strata, thus
arriving at special types of stratification.
According to \cite{KoRo}, the type of stratification appropriate for our 
purposes is called 'regular' and is defined by the property
$$
X_i \cap \overline{X_{i'}} \neq \emptyset
~~~\RA~
X_i \mbox{ closed in }X_i \cup X_{i'}
\,,~~~~~
\forall i, i' \in I\,.
$$
The following is due to Kondracki and Rogulski \cite{KoRo}.
\begin{Theorem}[Stratification theorem]
\label{Tstrat}
The decomposition of $\orb$ by orbit types is a regular stratification. 
\end{Theorem}
To prove it, one has to check countability of orbit types and the frontier
and regularity conditions.  
\subsection{Countability of orbit types}
Due to the reduction theorem, orbit types are in $1$-$1$ correspondence with 
certain 
reductions of $P$ to Howe subgroups, modulo isomorphy of the reductions and
modulo conjugacy of the subgroups. We note the following facts:

(i) Howe subgroups are closed.

(ii) There are at most countably many conjugacy classes of closed subgroups in 
a compact group \cite{KoNo}.

(iii) There are at most countably many isomorphism classes of principal 
bundles with a given structure group over a compact manifold. The classes
are in $1$-$1$ correspondence with arch-wise connected components of the space of 
continuous maps from the base space of the bundle to the classifying space. 
General arguments ensure that there are at most countably many such components.

It follows from (i)--(iii) that the number of orbit types is at most countable.

Let us note that the number of Howe subgroups in a compact Lie group is
actually finite. This follows from the fact that any centralizer in a compact
Lie group is generated by finitely many elements \cite[ch.~9]{Bou:Lie} and 
that a 
compact group action on a compact manifold has a finite number of orbit types 
\cite{Bredon:CTG}. 
\subsection{Frontier and regularity conditions}
Let $\sigma$, $\sigma'$ be orbit types such that
$\overline{\orb^\sigma}\cap\orb^{\sigma'}\neq\emptyset$. According to the
closure formula \gref{Gclosure}, $\overline{\orb^\sigma}$ is a union of
strata. If $\orb^{\sigma'}$ intersects the union, it must in fact
coincide with one of these strata. Then
$\orb^{\sigma'}\subseteq\overline{\orb^\sigma}$. Thus, the decomposition
by orbit types satisfies the frontier condition.

On the other hand, we know from the slice theorem that $\orb^\sigma$ is open
in $\orbleq{\sigma}$, hence in $\overline{\orb^\sigma}$. Then $\orb^\sigma$
is open in $\orb^\sigma\cup\orb^{\sigma'}$, because the latter is a subset
of $\overline{\orb^\sigma}$ due to the frontier condition. Then
$\orb^{\sigma'}$, being the complement, is closed. Hence, the decomposition
by orbit types is a regular stratification.

(This actually shows that if all strata are open in their closures, the
frontier condition implies regularity.)
\vspace{0.3cm}

\noindent
{\bf Remarks:}\\
1.~Consider the relation
$$
\orb^\sigma\leq\orb^{\sigma'}
\Leftrightarrow
\overline{\orb^\sigma}\cap\orb^{\sigma'}\neq\emptyset\,.
$$
For any stratification, this relation is reflexive and transitive, i.e., a
quasi-ordering (the 'natural quasi-ordering' of the stratification). If the
stratification is regular, the relation is also antisymmetric, hence a
partial ordering. As for the stratification of $\orb$ by orbit types,
\gref{Gclosure} implies that the natural partial ordering of the strata is
just inverse to that of the corresponding orbit types. \\
2.~Instead of using Sobolev techniques one can also stick to
smooth connection forms and gauge transformations. Then one obtains
essentially analogous results about the stratification of the corresponding 
gauge orbit space where, roughly speaking, one has to replace
'Hilbert manifold' and 'Hilbert Lie group' by 'tame Fr\'echet manifold' and
'tame Fr\'echet Lie group', see \cite{Abbati:Gau,Abbati:OrbSpa}.
\section{$L^2$-Riemannian structure on strata}\label{SRiemann}
The $L^2$-metric $\gamma^0$ on $\con$ induces a weak
Riemannian metric on each stratum $\orb^\sigma$. This was discussed
for the case of the principal stratum in \cite{BabelonViallet:Rm,Singer:Geo} 
and for the general case in \cite{BerziReni}. The basic idea consists in
restricting the tangent bundle splitting (\ref{GdecompTA}) to strata. This
yields a smooth connection in each bundle which allows to lift tangent
vectors, thus projecting $\gamma^0$ to a metric on each stratum.
\subsection{A natural connection}
By restriction, the distribution $\vert$, made up by the tangent spaces of the orbits,
induces a distribution $\vert^\sigma$
on $\con^\sigma$. Contrary to $\vert$, $\vert^\sigma$ is smooth and locally
trivial, because 
$\vert^\sigma = \ker({\pi^\sigma}_\ast)$ and $\pi^\sigma$ is a 
smooth submersion. Let $\hori^{\sigma}$ denote the normal distribution 
associated to $\vert^\sigma$ w.r.t.~the $L^2$-metric $\gamma^0 \, .$ By 
construction,
$$
\hori^\sigma := \hori\cap\rmT\con^\sigma\,.
$$
Due to \gref{GdecompTA} and $\vert^\sigma \subseteq  \rmT\con^\sigma \, ,$ 
\begin{equation}\label{GdecompTAsigma}
\rmT\con^\sigma = \vert^\sigma\oplus\hori^\sigma\,,
\end{equation}
where the sum is orthogonal w.r.t.~$\gamma^0$. Moreover, $\hori^\sigma$ is 
$\gau$-equivariant,
$$
\hori^\sigma_{A^{(g)}} = \left(\hori^\sigma_A\right)^{(g)} \,.
$$

We draw the attention of the reader to the fact that we had already
introduced the notation $\hori^\sigma_A$ for the subspace of $\hori_A$
consisting of elements invariant under $\gau_A$, see \gref{GdefhAs}. This
notation suggests that $\hori^\sigma_A$ is in fact the fibre at $A$ of the
distribution $\hori^\sigma$. To see that this holds indeed, recall that 
$\hori_A = \rmT_A\slice_{A,\ve}$. Hence, the fibre of $\hori^\sigma$ 
is
$$
\rmT_A\slice_{A,\ve}\cap\rmT_A\con^\sigma
=
\rmT_A\slice^\sigma_{A,\ve}\,.
$$
According to \gref{GTSAs}, the r.h.s.~is given by $\hori^\sigma_A$. 

In the remaining part of this subsection we shall prove that the
distribution $\hori^\sigma$ is smooth and locally trivial (viewed as a subbundle 
of $\rmT \con^\sigma $). 
Note that, due to
weakness of $\gamma^0$, this is not obvious from smoothness and local
triviality of $\vert^\sigma$. 
It follows then that $\hori^\sigma$ is a smooth connection in the $\gau$-bundle 
$\pi^\sigma:\con^\sigma\rightarrow\orb^\sigma$.

Smoothness of $\hori^\sigma$ would follow from smoothness of either one of the
corresponding $\gamma^0$-orthogonal projectors $\bfh|_{\rmT\con^\sigma}$ or
$\bfv|_{\rmT\con^\sigma}$ which, in turn, would follow from smoothness of 
the restrictions of $\bfh$ or $\bfv$, respectively, to $\rmT\con|_{\con^\sigma}$. Recall from 
\gref{GvA,hA} that the restriction of $\bfv$ is given by the map 
$$
\con^\sigma\rightarrow\rmB(\tsl)\,,~~~~A\mapsto\nabla_A\rmG_A\nabla_A^\ast\,.
$$
This map decomposes as
$$
\con^\sigma
\stackrel{\mbox{\footnotesize diag}}{\longrightarrow}
\con^\sigma\times\con^\sigma\times\con^\sigma
\stackrel{\nabla_\cdot\times\rmG_\cdot\times\nabla_\cdot^\ast}{
\longrightarrow}
\rmB(W^{k+1},W^k)\times\rmB(W^{k-1},W^{k+1}) \times \rmB(W^k,W^{k-1})
\stackrel{\mbox{\footnotesize comp.}}{\longrightarrow}
\rmB(W^k)\,.
$$
Since diagonal embedding, $\nabla_\cdot$, $\nabla_\cdot^\ast$ and composition
of bounded operators are continuous (multi-) linear maps, it suffices to prove
smoothness of the map
\begin{equation}\label{GmapGA}
\con^\sigma\rightarrow\rmB(W^{k-1},W^{k+1})\,,~~~A\mapsto G_A\,.
\end{equation}
Pulling it back with a local trivialization $\chi^\sigma_{A_0,\ve}$,
$A_0\in\con^\sigma$, see \gref{Gloctriv}, we obtain a map
%
%
%
$$
\slice^\sigma_{A_0,\ve}\times\gau/\gau_{A_0}
\rightarrow\rmB(W^{k-1},W^{k+1})\,,~~~(A,[g])\mapsto G_{A^{(g)}}\,,
$$
%
%
%
which is well defined, because $\gau_A=\gau_{A_0}$, $\forall
A\in\slice^\sigma_{A_0,\ve}$. Due to \gref{GevrGA}, this map is smooth along
$\gau/\gau_{A_0}$. Thus, what we actually have to show is that the
restrictions of the map \gref{GmapGA} to the partial slices $\slice^\sigma_{A_0,\ve}$, 
$A_0\in\con^\sigma$, are smooth. For that purpose, recall that
$\rmG_A$ is constructed from the (bounded) inverse of the operator
\begin{equation} \label{GwtDelta}
\wt\Delta_A:\ker(\Delta_A)^{\perp_0}\rightarrow\im(\Delta_A)
\end{equation}
induced by $\Delta_A$. Due to $\gau_A=\gau_{A_0}$, equation
\gref{GkerDAkernA} and the decomposition theorem, we have
\begin{equation} \label{GkerDelta}
\ker(\Delta_A) = \ker(\Delta_{A_0})
\,,~~~~
\im(\Delta_A) = \im(\Delta_{A_0})\,.
\end{equation}
Hence, \gref{GwtDelta} reads
$$
\wt\Delta_A:\ker(\Delta_{A_0})^{\perp_0}\rightarrow\im(\Delta_{A_0})
\,,~~~~
\forall A\in\slice^\sigma_{A_0,\ve}\,.
$$
Thus, the map under consideration decomposes into
$$
\slice^\sigma_{A_0,\ve}
\stackrel{\wt\Delta_\cdot}{\longrightarrow}
\Inv\left(\ker(\Delta_{A_0})^{\perp_0},\im(\Delta_{A_0})\right)
\stackrel{\mbox{\footnotesize inv}}{\longrightarrow}
\Inv\left(\im(\Delta_{A_0}),\ker(\Delta_{A_0})^{\perp_0}\right),
$$
followed by prolongation to a bounded operator $W^{k-1}\rightarrow W^{k+1}$.
Here $\Inv(\cdot,\cdot)\subseteq\rmB(\cdot,\cdot)$ denotes the open subset
of invertible bounded operators, whereas 'inv' stands for the inversion map,
which is smooth. Since the first step factorizes into continuous linear
maps and composition of bounded operators, it is smooth, too.

This concludes the proof of smoothness of the projectors $\bfv|_{\rmT\con^\sigma}$
and $\bfh|_{\rmT\con^\sigma}$ and, hence, of the distribution $\hori^\sigma$.

Next, let us construct local trivializations of $\hori^\sigma$. To this end,
for $A_0\in\con^\sigma$, consider the distribution
$\mf D^\sigma_{A_0,\ve}$ on
$\slice^\sigma_{A_0,\ve}\times\gau/\gau_{A_0}$, made up by the subspaces
tangent to $\slice^\sigma_{A_0,\ve}$. Due to \gref{GTSAs}, it is trivial.
We claim that the map
\begin{equation}\label{Glocbuniso}
\mf D^\sigma_{A_0,\ve}
\rightarrow
\rmT(\slice^\sigma_{A_0,\ve}\times\gau/\gau_{A_0})
\stackrel{(\chi^\sigma_{A_0,\ve})_\ast}{\rightarrow}
\rmT\tube^\sigma_{\pi(A_0),\ve}
\stackrel{\bfh}{\rightarrow}
\hori^\sigma|_{\tube^\sigma_{A_0,\ve}}
\end{equation}
is a smooth vector bundle isomorphism and, thus, provides a local
trivialization of $\hori^\sigma$. To see this, note that
$(\chi^\sigma_{A_0,\ve})_\ast$ maps
$\mf D^\sigma_{A_0,\ve}$ isomorphically on the
equivariant distribution 
$$
\bigcup_{[g]\in\gau/\gau_{A_0}}
\rmT\slice^\sigma_{A_0^{(g)},\ve}\,.
$$
Hence, due to equivariance of $\hori^\sigma$ and $\bfh$, it suffices to show
that the map
\begin{equation}\label{Ghslice}
\rmT\slice^\sigma_{A_0,\ve}\stackrel{\bfh}{\rightarrow}\hori^\sigma|_{\slice^\sigma_{A_0,\ve}}
\end{equation}
is a smooth vector bundle isomorphism. We shall construct a smooth inverse.

Recall that $\slice_{A_0,\ve}$ is transversal to
any orbit it meets. Hence, 
$$
\hori_{A_0}\cap\vert_A
=
\ker(\nabla_{A_0}^\ast)\cap\im(\nabla_A)
=
\{0\}
\,,~~~~
\forall A\in\slice^\sigma_{A_0,\ve}\,.
$$
Then $\Delta_{A_0A} := \nabla_{A_0}^\ast\nabla_A$ has kernel
$\ker(\nabla_A) = \ker(\Delta_A)$ and image
$\im(\nabla_{A_0}^\ast)=\im(\Delta_{A_0})$. In particular, for any element
$A$ of the partial slice $\slice^\sigma_{A_0,\ve}$, $\ker(\Delta_{A_0A}) =
\ker(\Delta_{A_0})$. Thus, we can construct a partial inverse $\rmG_{A_0A}$ similar to
$\rmG_{A_0}$ and $\rmG_A$. By construction,
\begin{equation}\label{GA0A}
\rmG_{A_0A}\Delta_{A_0A} = \rmG_{A_0}\Delta_{A_0} = \rmG_A\Delta_A
\,,~~~~
\Delta_{A_0A}\rmG_{A_0A} = \Delta_{A_0}\rmG_{A_0} = \Delta_A\rmG_A\,.
\end{equation}
Define $\bfh_{A_0A} := \id_\tsl - \nabla_A\rmG_{A_0A}\nabla_A^\ast$.
Using \gref{Gweglass} and \gref{GA0A}, one can check that
\begin{equation}\label{GhA0A}
\bfh_{A_0A}\bfh_A = \bfh_{A_0}\bfh_{A_0A} = \bfh_{A_0A}
\,,~~~~
\bfh_{A_0A}\bfh_{A_0} = \bfh_{A_0}
\,,~~~~
\bfh_A\bfh_{A_0A} = \bfh_A\,,
\end{equation}
for any $A\in\slice^\sigma_{A_0,\ve}$. It follows that $\bfh_{A_0A}$ maps
$\hori_A$ to $\hori_{A_0}$. Since, due to \gref{GnablaAg}, 
$$
\bfh_{A_0^{(g)}A^{(g)}} = \Ad(g^{-1})~\bfh_{A_0A}~\Ad(g)\,,
$$
$\bfh_{A_0A}$ maps $\hori^\sigma_A$ onto $\hori^\sigma_{A_0}$. 
Formulae \gref{GhA0A} imply 
%
%
%
$$
\bfh_{A_0A}\bfh_A|\hori^\sigma_{A_0} = \id_{\hori^\sigma_{A_0}}
\,,~~~~
\bfh_A\bfh_{A_0A}|\hori^\sigma_A = \id_{\hori^\sigma_A}
\,,~~~~
\forall A\in\slice^\sigma_{A_0A}\,.
$$
%
%
%
Since the map $\slice^\sigma_{A_0A}\rightarrow\rmB(\tsl)$, 
$A\mapsto\bfh_{A_0 A}$, is smooth, which can be shown in a similar way as for 
the map $A\mapsto\bfh_A$, it provides the desired inverse of \gref{Ghslice},
thus proving that \gref{Glocbuniso} is a local trivialization of
$\hori^\sigma$.

We remark that the operators
$\bfh_{A_0A}$
and $\bfv_{A_0A} := \nabla_A\rmG_{A_0A}\nabla_{A_0}^\ast$, where
$A\in\slice^\sigma_{A_0A}$, $A_0\in\con^\sigma$, are the projectors
associated to the (not necessarily $L^2$-orthogonal) decomposition
$$
\tsl = \vert_A\oplus\hori_{A_0}\,.
$$
This can be checked using \gref{Gweglass} again.

Finally, we note that, with the above connection, there is associated an
equivariant differential form with values in $\rmL\gau\, ,$ given by
\begin{equation}
\label{connecform}
\Omega_A (A,X) := \rmG_A \nabla_A^\ast X \, ,
\end{equation}
for all $(A,X) \in \con \times \tsl = \rmT\con \, .$ 
For the principal stratum $\orb^\rmp$ , we have
\begin{equation}
\label{connecform1}
\Omega_A (A,\nabla_A \xi) = \xi \, , \forall \xi \in \rmL \gau \, ,
\end{equation}
showing that $\Omega$ is an ordinary connection form in the 
principal fibre bundle over $\orb^\rmp$ with structure group $\gau$ factorized
by its center. For the other strata, however, $\Omega_A$ maps the Killing field
generated by $\xi$ to the projection of $\xi$ onto the $L^2l$-orthogonal
complement of $\rmL \gau_A $ in $\rmL \gau \, .$ We further comment on this
below.

\subsection{The metric}
The natural connection $\hori^\sigma$ and the Riemannian metric 
$\gamma^0$ induce a Riemannian metric $\gamma^{0,\sigma}$ on $\orb^\sigma$ as
follows. Due to the open mapping theorem, restriction of ${\pi^\sigma}_\ast$ 
to a fibre $\hori^\sigma_A$, $A\in\con^\sigma$, induces a Banach space 
isomorphism onto $\rmT_{\pi(A)}\orb^\sigma$. This allows to lift tangent 
vectors at $x\in\orb^\sigma$ to horizontal tangent vectors at
$A\in\pi^{-1}(x)$ and evaluate their scalar product w.r.t.~$\gamma^0$. 
Due to equivariance of $\hori^\sigma$ and invariance of $\gamma^0$, the 
result does not depend on the choice of the representative 
$A$. Due to smoothness of $\hori^\sigma$, the Riemannian metric
$\gamma^{0,\sigma}$ on $\orb^\sigma$ so constructed is smooth. 

Let us determine the local representatives of $\gamma^{0,\sigma}$
w.r.t.~the charts $(\pi^\sigma_{A_0,\ve})^{-1}$, $A_0\in\con^\sigma$,
see \gref{GdefpisAe}. Let $A\in\slice^\sigma_{A_0,\ve}$. For tangent 
vectors $(A,X_i)\in\rmT_A\slice^\sigma_{A_0,\ve} =
\slice^\sigma_{A_0,\ve}\times\hori^\sigma_{A_0}$, we have
$$
(\pi^\sigma_{A_0,\ve})^\ast\gamma^{0,\sigma}((A,X_1),(A,X_2))
=
\gamma^{0,\sigma}\left(({\pi^\sigma_{A_0,\ve}})_\ast(A,X_1),
({\pi^\sigma_{A_0,\ve}})_\ast(A,X_2)\right)\,.
$$
Horizontal lift of $({\pi^\sigma_{A_0,\ve}})_\ast(A,X_i)$ to $A$ yields
$(A,\bfh_A X_i)$. Hence, 
\begin{equation}\label{Gmetricincharts}
({\pi^\sigma_{A_0,\ve}})^\ast\gamma^{0,\sigma}((A,X_1),(A,X_2))
= 
(X_1,\bfh_A X_2)_0\,.
\end{equation}
where we have used $\bfh_A^\ast = \bfh_A$ and $\bfh_A^2=\bfh_A$. In
this formula, we can replace $\bfh_A$ by $\bfh_{A_0}\bfh_A$. Since the
latter maps $\hori^\sigma_{A_0}$ to itself, the operator which represents
the scalar product \gref{Gmetricincharts} on $\hori^\sigma_{A_0}$ is 
$$
\bfh_{A_0}\bfh_A|_{\hori^\sigma_{A_0}}\,.
$$
Thus, w.r.t.~the charts $(\pi^\sigma_{A_0,\ve})^{-1}$, $\gamma^{0,\sigma}$ 
is given by the smooth map
$$
\slice^\sigma_{A_0,\ve}\rightarrow\rmB(\hori^\sigma_{A_0})
\,,~~~A\mapsto\bfh_{A_0}\bfh_A|_{\hori^\sigma_{A_0}}\,.
$$
Using \gref{GhA0A}, one can check that the inverse of
$\bfh_{A_0}\bfh_A|_{\hori^\sigma_{A_0}}$ is given by
$\bfh_{A_0A}\bfh_{A_0A}^\ast$. In particular, 
$\bfh^\sigma_{A_0}\bfh^\sigma_A|_{\hori^\sigma_{A_0}}$ 
is indeed a Banach space isomorphism.
\vspace{0.3cm}

\noindent
{\bf Remarks:}
1. It can be easily seen that the $\gau$-invariant $L^2$-metric $\gamma^0$ on the 
bundle space $\con^\sigma$ is uniquely characterized  by the triple 
$(\gamma^{0,\sigma}, \Omega, (\cdot,\cdot)_0) \, ,$ where $(\cdot,\cdot)_0$ 
denotes the $L^2$-scalar product on $\rmL \gau \, .$ 
This is a structure similar to that in Kaluza-Klein theory, where $G$-invariant metrics 
$\eta$ on a $G$-bundle $Q$ with fibre $G/H$ over space time $M$ are in
$1$-$1$ correspondence 
with triples $(\eta_M, \omega, \langle \cdot,\cdot \rangle) \, .$ Here $\eta_M$ is a
metric on $M \, ,$ $\omega$ is a connection form in the principal bundle $P$ with structure
group $N/H$ associated with $Q$ and $\langle \cdot,\cdot \rangle$ is a $\Ad(G)$-
invariant scalar product on the Lie algebra of $G \, .$ Moreover, $N$ denotes
the normalizer of $H$ in $G \, .$ According to the remark at the end of Subsection 
\ref{smfbstr}, in our case it is unclear whether the normalizer of a given 
stabilizer 
$\gau_A$ in $\gau$ is a Lie subgroup. Thus, we cannot construct the above associated
principal bundle and give an interpretation of $\Omega$ as a connection form in
this bundle.
\\
2. In a similar way one can project $W^k$-metrics, like $\gamma^k$, see
\gref{Gdefstrong}, or $\eta^k$, see \gref{Gdeflapl}, to metrics on the strata.
To our knowledge this has not been investigated yet, see, however, 
\cite{Habermann} for results on the restriction of $\eta^2$ to some instanton
spaces.
\subsection{Curvature}
The same tedious but straightforward computation as in the case of the
principal stratum \cite{BabelonViallet:Rm} yields for the local
representative of the Riemannian curvature tensor
\begin{eqnarray}\nonumber
R & : & \slice^\sigma_{A_0,\ve} \rightarrow \rmB(\hori^\sigma_{A_0}\otimes 
\hori^\sigma_{A_0}\otimes
\hori^\sigma_{A_0},\hori^\sigma_{A_0})\,,
\\ \label{GRmcurv}
R_A(X,Y)Z 
& = &
\bfh^\sigma_{A_0} \left(-2\rmK_Z\rmG_A\rmK_X^\ast Y - \rmK_Y\rmG_A\rmK_X^\ast Z +
\rmK_X\rmG_A\rmK_Y^\ast Z\right)\,,
\end{eqnarray}
where $X,Y,Z\in\hori^\sigma_{A_0}$, $A\in\slice^\sigma_{A_0,\ve}$ and 
$\rmK_X:W^{k+1}(\Ad P)\rightarrow W^k(\rmT^\ast M\otimes\Ad P)$ denotes
taking the commutator with $X$ and 
$\rmK_X^\ast : W^k(\rmT^\ast M\otimes\Ad P)\rightarrow W^k(\Ad P)$ 
its formal adjoint. 

From \gref{GRmcurv} one obtains for the local representative of the sectional 
curvature $\mf R$ of a $2$-plane $\mf P\subseteq \hori^\sigma_{A_0}$
$$
\mf R_A(\mf P) = 3(\rmK_X^\ast Y,\rmG_A\rmK_X^\ast Y)_0\,,
$$
where $X,Y\in \hori^\sigma_{A_0}$ are orthonormal vectors spanning $\mf P$. We 
claim that
the sectional curvature is nonnegative, as in the case of the principal
stratum \cite{BabelonViallet:Rm,Singer:Geo}. To see this, denote 
$\xi=\rmK^\ast_X Y$.
Since $\xi\in W^{k-1}(\Ad P)$, one can decompose it according to the
decomposition theorem
$
\xi = \xi_{\im}+\xi_{\ker}\,.
$
By construction of $\rmG_A$, $\xi_\im = \Delta_A\rmG_A\xi$ and 
$\im(\rmG_A)\perp_0\ker(\Delta_A)$. It follows
$$
(\xi,\rmG_A\xi)_0
=
(\xi_\im,\rmG_A\xi)_0
=
(\Delta_A\rmG_A\xi,\rmG_A\xi)_0
=
(\nabla_A\rmG_A\xi,\nabla_A\rmG_A\xi)_0\,.
$$
\subsection{Formal volume element}
For the case of the principal stratum $\orb^{\rm p}$, a formal expression for
the volume element
of the metric $\gamma^{0,{\rm p}}$ was derived in \cite{BabelonViallet:FP}:
\begin{equation}\label{Gvolelm}
\det\left(\bfh_{A_0}\bfh_A|_{\hori_{A_0}}\right)^{1/2}
=
\frac{\det(\Delta_{A_0A})}{\det(\Delta_{A_0})^{1/2}\det(\Delta_A)^{1/2}}
\,,~~A \in \slice^{\rm p}_{A_0,\ve}, \, \, A_0 \in \con^{\rm p} \, ,
\end{equation}
(recall that $\hori^{\rm p}_{A_0}=\hori_{A_0}$). The function 
$A\mapsto\det(\Delta_{A_0A})$ is known as the Faddeev-Popov determinant 
in the background potential $A_0$. It follows that the functional
integral derived by the Faddeev-Popov procedure \cite{FaddeevPopov},
can be geometrically interpreted as the formal integral defined by the
natural $L^2$-Riemannian structure on $\orb^{\rm p}$ \cite{BabelonViallet:FP}.
Schr\"odinger quantum mechanics on the gauge orbit space has been discussed in
this context, see e.g. \cite{Gaw} and references therein.

It is easy to see that \gref{Gvolelm} extends to the other strata.
Namely, for $A_0\in\con^\sigma$ and $A\in\slice^\sigma_{A_0,\ve}$
we have seen that $\Delta_A$, $\Delta_{A_0}$, and $\Delta_{A_0A}$ have 
common kernel $\ker(\Delta_{A_0})$ and image $\im(\Delta_{A_0})$. By 
defining their determinant as that of the restricted operators
$$
\ker(\Delta_{A_0})^{\perp_0}\rightarrow\im(\Delta_{A_0})
$$
(i.e., by 'removing zero modes'),
one can establish \gref{Gvolelm} by essentially the same proof as in the case
of the principal stratum.

In particular, one can use \gref{Gvolelm} to formally define an integral
for each stratum. However, as for the physical interpretation, the mere sum
of such integrals would certainly not be a reasonable
extension of the Faddeev-Popov procedure from the principal stratum to the
whole orbit space, because it does not take into account any
'interaction'
between strata.
\subsection{Geodesics}
In \cite{BerziReni}, the following was proved.
\begin{Theorem}\label{Tgeodesics}
Let $A\in\con^\sigma$ and $X\in \hori^\sigma_A$. Let $I$ denote the connected component
of $0$ in $\{t\in\RR~:~A+tX\in\con^\sigma\}$. Then $I$ is non-empty, open, and 
$$
I\rightarrow\orb^\sigma\,,~~~t\mapsto \pi^\sigma(A+tX)\,,
$$
is a geodesic in $\orb^\sigma$. Conversely, any geodesic in $\orb^\sigma$ is
of this form.
\end{Theorem}
Note that 
\begin{equation}\label{Ggeodperp}
\nabla_{A+tX}^\ast X = \nabla_A^\ast X = 0\,,~~~\forall \,  A \in\con \, , \, 
X\in \hori^\sigma_A \, , \, t\in\RR\,,
\end{equation}
so that the straight line $A+tX$ is perpendicular to any orbit it meets.
Thus, the theorem says that the geodesics in $\orb^\sigma$ are given by 
projections of segments of straight lines inside $\con^\sigma$ which are 
perpendicular to orbits.

Note also that the theorem, in particular, shows that the charts
$(\pi^\sigma_{A_0,\ve})^{-1}$ provide normal coordinates.

In \cite{BerziReni}, the above characterization of orbits is used to prove
that the principal stratum, in general, is not geodesically complete. In
fact, the argument given there can be extended to prove
\begin{Theorem}\label{Tgeodcomplete}
$\orb^\sigma$ is geodesically complete if and only if there does not exist
$\sigma'$ such that $\sigma<\sigma'$.
\end{Theorem}                                         
Indeed, for $A\in\con^\sigma$ and $X\in \hori^\sigma_A$, we have
$
\gau_{A+tX} \supseteq \gau_A\cap\gau_X = \gau_A\,.
$
Therefore,
\begin{equation}\label{Ggeodingeq}
A+tX\in\conleq{\sigma}\,,~~~\forall t\in\RR\,.
\end{equation}
In particular, if there is no $\sigma'$ with $\sigma<\sigma'$, the geodesic
associated to $A$ and $X$ is defined for all values $t\in\RR$. 

Now assume that $\sigma<\sigma'$ for some $\sigma'$. Choose
$x'\in\orb^{\sigma'}$ and a tube $\tube_{x',\ve}$ about the orbit
$\pi^{-1}(x')$. Since $\tube_{x',\ve}$ is a neighbourhood of $\pi^{-1}(x')$ in $\con$,
the denseness properties \gref{Gdense} imply 
$
\tube_{x',\ve}\cap\con^\sigma \neq \emptyset\,.
$
Since $\tube_{x',\ve} = \bigcup_{A'\in\pi^{-1}(x')} \slice^{\sigma'}_{A',\ve}$ one finds
$A'$ such that $\slice^{\sigma'}_{A',\ve}\cap\con^\sigma\neq\emptyset$. Choose $A$
from the intersection and let $X\in\tsl$ such that $A'=A+X$. Since $X\in
\hori^{\sigma'}_{A'}$, \gref{Ggeodperp} implies that $\nabla_A^\ast X=0$. Since
$A\in\slice^{\sigma'}_{A',\ve}$, $\gau_A \subseteq \gau_{A'}$. It follows that 
$X \in \hori^\sigma_A$. Thus, $A$ and $X$ define a geodesic in $\orb^\sigma$ 
that cannot be prolonged to values $t\geq 1$.

The following theorem was stated for the principal stratum in
\cite{BerziReni}.
\begin{Theorem}\label{Tnowheredense}
Let $A\in\con^\sigma$, $X\in \hori^\sigma_A$. The set of values $t\in\RR$ for which
$A+tX\notin\con^\sigma$ is discrete.
\end{Theorem}
To see this, denote $C(t)=A+tX$. According to
\gref{Ggeodingeq}, $C^{-1}(\con^\sigma)$ is open in $\RR$, because
$\con^\sigma$ is open in $\conleq{\sigma}$. Hence, $\RR\setminus
C^{-1}(\con^\sigma)$ is closed in $\RR$.

Let $t_0\in\RR\setminus C^{-1}(\con^\sigma)$. According to \gref{Ggeodperp},
$X\in\ker(\nabla_{C(t_0)}^\ast)$, so that the slice theorem implies 
$C(t)=C(t_0)+(t-t_0)X\in\slice_{C(t_0),\ve}$ for $t$ close to $t_0$. 
If $t_0$ was an accumulation point of $\RR\setminus C^{-1}(\con^\sigma)$,
there would exist $t_1\neq t_0$ such that
$C(t_1)\in\slice_{C(t_0),\ve}\cap\con^{\sigma'}$ for some $\sigma'>\sigma$. 
By the properties of the slice, $\gau_{C(t_1)}\subseteq\gau_{C(t_0)}$. 
Since $C(t_1) = C(t_0) + (t_1-t_0)X$, then $\gau_X \supseteq \gau_{C(t_1)}$. 
Writing 
$
A = C(t_1) - t_1 X
$
one sees that then $\gau_A \subseteq \gau_{C(t_1)}$ (contradiction). Hence,
$\RR\setminus C^{-1}(\con^\sigma)$ consists of isolated points. Due to
closedness, it is then discrete.
%
%
\section{Classification of gauge orbit types for $G=\rmSU(n)$}
\label{ClassGOT}\label{Sclfic}
%
%

Until now, complete classification results for the set of orbit types are
known only for gauge group $\rmSU(n)$ and base manifolds of dimension up to
$4$ \cite{RSV:clfot,RSV:poot}, see also \cite{RS:otG} for the discussion of a coarser 
stratification. In the following two sections these results 
will be reviewed. 
According to the reduction theorem, to determine the set of orbit types one 
has to work out the following programme:

1. Classification of Howe subgroups of $\rmSU(n)$ up to conjugacy,

2. Classification of Howe subbundles of $P$ up to isomorphy,

3. Specification of Howe subbundles which are holonomy-induced,

4. Factorization by $\rmSU(n)$-action,

5. Determination of the natural partial ordering of Howe subbundles.
\subsection{Howe subgroups of $\rmSU(n)$}
\label{Howe}
%
%
%
General references for the determination of Howe subgroups of classical Lie
groups are \cite{Przebinda,Schmidt}, see also \cite{Rubenthaler} for the
case of complex semisimple Lie algebras. For $\rmSU(n)$, however, it is not
necessary to apply the general theory, because one can show, using the 
double commutant theorem, that the Howe subgroups of $\rmSU(n)$ are in 
$1$-$1$ correspondence to unital $\ast$-subalgebras of $\rmM_n(\CC)$, 
the algebra of complex $n\times n$ matrices. The relation is given by 
intersecting the subalgebras with $\rmSU(n)$. 

The unital $\ast$-subalgebras of $\rmM_n(\CC)$ can be described 
as follows. Let $\rmK(n)$ denote the set of pairs 
$J=(\bfk,\bfm)$ of 
sequences $\bfk=(k_1,\dots, k_r)$, $\bfm =(m_1,\dots, m_r)$, 
$r=1,\dots,n$, consisting of positive integers such that 
\begin{equation}\label{Gsumcond}
\bfk\cdot\bfm = \sum_{i=1}^r k_im_i = n\,. 
\end{equation}
Any $J\in\rmK(n)$ defines a decomposition
\begin{equation}\label{Gdecomp}
\CC^n=\bigoplus_{i=1}^r \CC^{k_i}\otimes\CC^{m_i}
\end{equation}
and an embedding
\begin{equation}\label{GdefMJ}
\prod_{i=1}^r    
\rmM_{k_i}(\CC)\rightarrow\rmM_n(\CC)
\,,~~~~          
(D_1,\dots,D_r)  
\mapsto          
\bigoplus_{i=1}^r
D_i\otimes\II_{m_i}
\,.              
\end{equation}
We denote the image of this embedding by $\rmM_J(\CC)$, its intersection
with $\rmU(n)$ by $\UJ$ and its intersection with $\rmSU(n)$ by $\SUJ$.
By construction, $\rmM_J(\CC)$
is a unital $\ast$-subalgebra of $\rmM_n(\CC)$. Conversely,
it is not hard to show that any unital $\ast$-subalgebra of $\rmM_n(\CC)$ is
conjugate to $\rmM_J(\CC)$ for some $J\in\rmK(n)$. Hence, up to conjugacy,
the Howe subgroups of $\rmSU(n)$ are given by the subgroups $\SUJ$,
$J\in\rmK(n)$. Finally, it is evident that $\SUJ$ and $\SUJp$ are conjugate 
iff $J'$ can be obtained from $J$ by a simultaneous permutation of $\bfk$ 
and $\bfm$.
\vspace{0.3cm}

\noindent
{\bf Remark:} $\UJ$ is the image of the restriction of \gref{GdefMJ} to
$\rmU(k_1)\times\cdots\times\rmU(k_r)$. If we identify
$\CC^{k_i}\otimes\CC^{m_i}\cong\CC^{k_im_i}$, 
$(c_1,\dots,c_{k_i})\otimes (d_1,\dots,d_{m_i})\mapsto(c_1d_1,\dots,
c_{k_i}d_1,\dots,c_1d_{m_i},\dots,c_{k_i}d_{m_i})$, the elements of $\UJ$
are given by matrices
$$
\vveckmatrix{\wt D_1}{0}{0}{0}{\wt D_2}{0}{0}{0}{\wt D_r}
\,,~~~~
\wt D_i = \vveckmatrix{D_i}{0}{0}{0}{D_i}{0}{0}{0}{D_i}
\,,
$$
where $D_i\in\rmU(k_i)$ and $\wt D_i$ has dimension $m_i$. 
Then $\SUJ$ consists of all such matrices which 
have determinant $1$.
\vspace{0.3cm}

For later purposes, we introduce the following notation:
$$
\begin{array}{cccccl}
j_J & : & \SUJ & \longrightarrow & \UJ & \mbox{(embedding),}
\\
i_J & : & \UJ & \longrightarrow & \rmU(n) & \mbox{(embedding),}
\\
\prM{J}{i} & : & \MJ & \longrightarrow & \MC{k_i} & \mbox{(projection onto
the $i$th factor),}
\\
\prU{J}{i} & : & \UJ & \longrightarrow & \rmU(k_i) & \mbox{(projection onto
the $i$th factor).}
\end{array}
$$
Let $\gcd$ denote the greatest common divisor of $\bfm$ and let 
$\twbfm=(\twm_1,\dots,\twm_r)$ be defined by 
$m_i=\gcd\twm_i \, ,$ $\forall i$. For any $D\in\UJ$, 
$$
\det\nolimits_{\rmU(n)}(D) 
= 
\prod_{i=1}^r \left[
\det\nolimits_{\rmU(k_i)}\left(\prU{J}{i}(D)\right) \right]^{m_i}\,.
$$
We can extract the $g$-th root of the determinant by defining the Lie group 
homomorphism
$$
\lUJ: \UJ \longrightarrow  \rmU(1)
\,,~~
D \mapsto
\prod_{i=1}^r \left[\det\nolimits_{\rmU(k_i)}\left(\prU{J}{i}(D)\right) \right]^{\twm_i}\,. 
$$
Then 
$$
\det\nolimits_{\rmU(n)}(D)
=
\left[\lUJ(D)\right]^g
\,,~~~~\forall D\in\UJ\,.
$$
Since $\lUJ(\SUJ) = \ZZ_\gcd\subseteq\rmU(1)$, $\lUJ$ induces a homomorphism 
$\lSJ:\SUJ \rightarrow \ZZ_\gcd$. We have the commutative
diagram
\begin{equation}\label{GctvdgrlSJ}
\begin{CD}
\SUJ @>j_J>> \UJ
\\
@V{\lSJ}VV 
@VV{\lUJ}V
\\
\ZZ_g @>>j_g> \rmU(1)
\end{CD}
\end{equation}  
where $j_g$ denotes natural embedding.  

Below we shall need the low dimensional homotopy groups of $\SUJ$. In
dimension $k\geq 1$, they can be derived in a standard way from the
corresponding homotopy groups of
$\UJ\cong\rmU(k_1)\times\cdots\times\rmU(k_r)$ by means of the exact
homotopy sequence of the $\SUJ$-bundle $\det_{\rmU(n)}:\UJ\rightarrow\rmU(1)$.
In dimension $k=0$ we have, by definition, $\pi_0(\SUJ) = \SUJ/\SUJ_0$,
where $\SUJ_0$ denotes the connected component of the identity of $\SUJ$. 
One can show $\SUJ/\SUJ_0\cong\ZZ_g\,,$ with the isomorphism being induced
by $\lSJ$, see \cite[Lemma 5.2]{RSV:clfot}. Thus,
\begin{equation}\label{Ghtpgr}
\pi_k(\SUJ) = \left\{\begin{array}{ccl}
\ZZ_g & | & k=0
\\
\ZZ^{\oplus(r-1)} & | & k=1
\\
\pi_k(\rmU(k_1))\oplus\cdots\oplus\pi_k(\rmU(k_r)) & | & k>1\,.
\end{array}\right.
\end{equation}
%
%


\subsection{Howe subbundles of $\rmSU(n)$-bundles}
\label{SUJBun}


In this subsection, let  
$J\in\rmK(n)$ be arbitrary but fixed. We are going to derive a classification, 
up to isomorphy, of principal $\SUJ$-bundles over $M$ 
in terms of appropriately chosen characteristic classes. Recall that we
assume $\dim(M)\leq 4$. Then, on the level of these classes, we shall obtain 
a characterization of those $\SUJ$-bundles which are reductions of a given 
$\rmSU(n)$-bundle $P$. In the following we use some facts from bundle theory
as well as from algebraic topology. For a brief account, see Appendices A
and B. 

Generally, each isomorphism class of principal $\SUJ$-bundles over $M$
is in $1$-$1$ correspondence to a homotopy class of maps from $M$ to the classifying 
space $\BSUJ$ of $\SUJ$, its so-called classifying map. As usual, we denote
the set of all homotopy classes by $[M,\BSUJ]$. Due to the
potentially complicated structure of the space $\BSUJ$, $[M,\BSUJ]$
is hardly tractable in full generality. However, we can use three major
inputs from algebraic topology to get control of it under our specific
assumption $\dim(M)\leq 4$. 

First, assume that we are able to find a simpler space $\BSUJ_n$ and a map
$f_n:\BSUJ\rightarrow\BSUJ_n$ such that the homomorphism induced by $f_n$ on
homotopy groups is an isomorphism in dimension $k<n$ and surjective in
dimension $n$. Then composition with $f_n$ defines a bijection from
$[M,\BSUJ]$ onto $[M,\BSUJ_n] \, ,$ see \cite[Chapter VII]{Bredon:Top}. 
We remark that $\BSUJ_n$ is called an
$n$-equivalent approximation of $\BSUJ$ and $f_n$ is called an 
$n$-equivalence.

Second, algebraic topology provides a method to successively construct
$n$-equivalent approximations, starting from $n=1$: the method of Postnikov
tower. It renders $\BSUJ_n$ as an $n$-stage fibration over a point, where
the fibre at stage $k$ is given by the Eilenberg-MacLane space
$\rmK(\pi_k(\BSUJ),k)$. This space is defined as a $CW$ complex, up to
homotopy equivalence, by the property that its only nonvanishing homotopy 
group is $\pi_k(\BSUJ)$ in dimension $k$. Recall that 
$\pi_k(\BSUJ)\cong\pi_{k-1}(\SUJ) \, .$ For the precise formulation of the 
method see \rref{AlgTop}. For a detailed explanation as well as an
application to standard groups, we refer to \cite{AvisIsham}.

Applying the method of Postnikov tower to $\BSUJ$ up to stage $5$ 
we obtain, see \cite[Theorem 5.4]{RSV:clfot},
\begin{equation} 
\label{GBG5}
\BSUJ_5 = K(\ZZ_g,1)\times \prod_{j=1}^{r-1}K(\ZZ,2)\times 
\prod_{j=1}^{r^\ast}K(\ZZ,4)\,,
\end{equation}
where $r^\ast$ denotes the number of members $k_i>1$. For the convenience of 
the reader we give the proof of \gref{GBG5} in \rref{Postnikov}. 
We note that the successive fibrations mentioned above turn out to 
be trivial here, i.e., they are just direct products. As a consequence, we 
have a bijection 
\begin{eqnarray}\nonumber
[M,\BSUJ] 
& ~\rightarrow~ & 
[M,K(\ZZ_g,1)]
\times
\prod_{i=1}^{r-1} [M,K(\ZZ,2)]
\times
\prod_{i=1}^{r^\ast} [M,K(\ZZ,4)]
\\ \label{GMBSUJbij}
\phantom{[M,\BUJ]}f 
& ~\mapsto~ &
\left(
\pr_1\circ f_5\circ f,
\{
\pr_{2i}\circ f_5\circ f
\}_{i=1}^{r-1},
\{
\pr_{4i}\circ f_5\circ f
\}_{i=1}^{r^\ast}
\right),
\end{eqnarray}
where $f_5: \BSUJ \rightarrow \BSUJ _5$ is a $5$-equivalence and the
$\pr_{ij}$ are the projections from $\BSUJ_5$ onto
its factors.  

To treat the factors on the rhs.~we use a third input from algebraic
topology. We will explain it for $[M,K(\ZZ_g,1)]$. Namely, the theory of 
Eilenberg-MacLane spaces provides the
following relation between homotopy and cohomology, see \rref{AlgTop}. 
There exists $\gamma_1\in H^1(\rmK(\ZZ_g,1),\ZZ_g)$ (the first $\ZZ_g$-valued 
cohomology group) such that the assignment 
\begin{equation}\label{GEMLScohom}
[M,K(\ZZ_g,1)]\rightarrow H^1(M,\ZZ_g)
\,,~~~~
\pr_1\circ f_5\circ f\mapsto(\pr_1\circ f_5\circ f)^\ast\gamma_1\,,
\end{equation}
is a bijection. Here $(\pr_1\circ f_5\circ f)^\ast$ denotes the homomorphisms
induced in cohomology. Writing $(\pr_1\circ f_5\circ f)^\ast\gamma_1 =
f^\ast(\pr_1\circ f_5)^\ast\gamma_1$, we observe that the bijection
\gref{GEMLScohom} is characterized by the image under $f^\ast$ of the fixed
element $(\pr_1\circ f_5)^\ast\gamma_1$ of $H^1(\BSUJ,\ZZ_g)$. Thus, if for
given maps $f,f':M\rightarrow\BSUJ$ the induced homomorphisms
$f^\ast,{f'}^\ast:H^1(\BSUJ,\ZZ_g)\rightarrow H^1(M,\ZZ_g)$ coincide then
the maps $\pr_1\circ f_5\circ f$ and $\pr_1\circ f_5\circ f'$ are homotopic.
Analogously, one finds for $k=2,4$ that if the induced homomorphisms
$f^\ast,{f'}^\ast:H^k(\BSUJ,\ZZ)\rightarrow H^k(M,\ZZ)$ coincide then 
$\pr_{ki}\circ f_5\circ f$ and $\pr_{ki}\circ f_5\circ f'$ are homotopic,
for all admissible $i$. Thus, using that \gref{GMBSUJbij} is a bijection, we
arrive at the following result: 
Two maps $f,f':M\rightarrow\BSUJ$ are
homotopic if they induce the same homomorphisms on the cohomology groups 
$H^1(\BSUJ,\ZZ_g)$, $H^2(\BSUJ,\ZZ)$, and  $H^4(\BSUJ,\ZZ)$.
Thus, to characterize homotopy classes of maps $M \rightarrow \BSUJ$, as usual, 
we have to determine a set of generators for these
cohomology groups and and to evaluate $f^\ast$ on them. In this way, 
a set of characteristic classes is associated to any
element of $[M,\BSUJ]$, hence to any $\SUJ$-bundle through its classifying
map. This set is complete in
the sense that coincidence of characteristic classes implies isomorphy of
the corresponding bundles. 

To construct a set of generators, we use the commutative diagram
\gref{GctvdgrlSJ}, which on the level of classifying spaces reads
\begin{equation}
\label{GctvdgrlBSJ}
\begin{CD}
\BSUJ @>{\rmB j_J}>> \BUJ
\\
@V{\rmB\lSJ}VV @VV{\rmB\lUJ}V
\\
\rmB\ZZ_g @>>{\rmB j_g}> \rmB\rmU(1)
\end{CD}
\end{equation}
First, consider the $\ZZ$-valued cohomology. Recall that the cohomology 
algebra $H^\ast(\rmB\rmU(k_i),\ZZ)$ is generated freely over $\ZZ$ by elements
$\gamma_{\rmU(k_i)}^{(2j)}\in H^{2j}(\rmB\rmU(k_i),\ZZ)$, $j=1,\dots, k_i$, 
see \cite{Borel}. We denote
\begin{equation}\label{GdefgUk}
\gamma_{\rmU(k_i)}
=
1+\gamma_{\rmU(k_i)}^{(2)}+\cdots+\gamma_{\rmU(k_i)}^{(2k_i)}\,.
\end{equation}
The generators $\gamma_{\rmU(k_i)}^{(2j)}$ define elements
\begin{eqnarray}  \label{GdeftwgJi2j}
\twgamma_{J,i}^{(2j)} 
& = &
\left(\rmB\prU{J}{i}\right)^\ast\gamma_{\rmU(k_i)}^{(2j)}\,,
\\ \label{GdefgJi2j}
\gamma_{J,i}^{(2j)} 
& = &
\left(\rmB j_J\right)^\ast\twgamma_{J,i}^{(2j)}
\end{eqnarray}
of $H^{2j}(\BUJ,\ZZ)$ and $H^{2j}(\BSUJ,\ZZ)$, respectively. 
We denote
\begin{eqnarray} \label{GdeftwgJi}
\twgamma_{J,i} 
& = &
1+\twgamma_{J,i}^{(2)}+\cdots+\twgamma_{J,i}^{(2k_i)}
\,,~~~~
\twgamma_J = (\twgamma_{J,1},\dots,\twgamma_{J,r})\,,
\\ \label{GdefgJi}
\gamma_{J,i}
& = &
1+\gamma_{J,i}^{(2)}+\cdots+\gamma_{J,i}^{(2k_i)}
\,,~~~~
\gamma_J = (\gamma_{J,1},\dots,\gamma_{J,r})\,.
\end{eqnarray}
It is a direct consequence of the K\"unneth Theorem for cohomology that 
the cohomology algebra $H^\ast(\BUJ,\ZZ)$ is freely generated over $\ZZ$ 
by the elements $\twgamma_{J,i}^{(2j)}$, $j=1,\dots,k_i$, $i=1,\dots,r$.
Moreover, using that $\rmB j_J:\BSUJ \rightarrow \BUJ$ is a
$\rmU(1)$-bundle and, therefore, induces a Gysin sequence one can show that
$\left(\rmB j_J\right)^\ast$ is surjective, see \cite[Lemma 5.7]{RSV:clfot}.
Thus, the cohomology algebra $H^\ast(\BSUJ,\ZZ)$ is generated over 
$\ZZ$ by the elements $\gamma_{J,i}^{(2j)}$, $j=1,\dots, k_i$, 
$i=1,\dots, r$. We remark that the generators $\gamma_{J,i}^{(2)}$ of 
$H^\ast(\BSUJ,\ZZ)$ are subject to a relation, 
which is however irrelevant for our purposes,
because it follows from another relation to be derived below. 

Next, we have to consider $H^1(\BSUJ,\ZZ_g) \, .$ We notice the following
facts:

(i) The induced homomorphism 
$
\left(\rmB\lambda^\rmS_J\right)^\ast
:
H^1(\rmB\ZZ_g,\ZZ_g)\rightarrow H^1(\BSUJ,\ZZ_g)
$ 
is an isomorphism. This follows by virtue of the Hurewicz and universal
coefficient theorems from the obvious fact that $\lSJ$ induces an isomorphism
of homotopy groups $\pi_0(\SUJ)\rightarrow\pi_0(\ZZ_g)$. 

(ii) From the (long) exact sequence induced by the short exact sequence of
coefficient groups 
$
0
\rightarrow
\ZZ
\rightarrow
\ZZ
\rightarrow
\ZZ_g
\rightarrow
0
$
one can read
off that the associated Bockstein homomorphism 
$
\beta_\gcd:H^1\left(\rmB\ZZ_\gcd,\ZZ_\gcd\right)
\rightarrow
H^2\left(\rmB\ZZ_\gcd,\ZZ\right)
$   
is an isomorphism.

(iii) The surjectivity of $\rmB j_J$, mentioned above, implies, in particular, 
surjectivity of the homomorphism 
$
\left(\rmB j_g\right)^\ast
:
H^2(\rmB\rmU(1),\ZZ)
\rightarrow
H^2(\rmB\ZZ_g,\ZZ)\,.
$

\noindent
It follows that $H^1(\BSUJ,\ZZ_g)$ is generated by the single element
\begin{equation} \label{genHBSUJZg}
\delta_J
:=
\left(\rmB\lSJ\right)^\ast \beta_\gcd^{-1} \left(\rmB j_\gcd\right)^\ast
\gamma_{\rmU(1)}^{(2)}\,.
\end{equation}
Finally, the commutative diagram (\ref{GctvdgrlBSJ}) induces a
relation between the generators $\gamma_{J,i}^{(2)}$ and $\delta_J$. 
To formulate it, we introduce the following
notation. For any topological space $X$ and any 
sequence of nonnegative integers $\bfb = (b_1,\dots,b_s)$,
define a polynomial function
\begin{equation}\label{GdefEm}
E_{\bfb}:\prod_{i=1}^s H^\rmeven(X,\ZZ)\rightarrow H^\rmeven(X,\ZZ)\,,~~
(\alpha_1,\dots,\alpha_s)
\mapsto
\alpha_1^{b_1}\dots\alpha_s^{b_s}\,.
\end{equation}
One can check the following formulae for  
the components of $E_{\bfb}$ in degree $2$ and $4$:
\begin{eqnarray}\label{GEJ2}
\hspace*{-0.5cm}E_{\bfb}^{(2)}(\alpha)
& = &
\sum_{i=1}^s b_i\alpha_i^{(2)}\,,
\\ \label{GEJ4}
\hspace*{-0.5cm}E_{\bfb}^{(4)}(\alpha)
& = &
\sum_{i=1}^s 
b_i\alpha_i^{(4)} 
+ 
\sum_{i=1}^s
\frac{b_i(b_i-1)}{2}\alpha_i^{(2)}\alpha_i^{(2)} 
+
\sum_{i<j=2}^s 
b_ib_j\alpha_i^{(2)}\alpha_j^{(2)}\,.
\end{eqnarray}
A straightforward computation, see \cite[Lemma 5.12]{RSV:clfot}, yields
\begin{equation}\label{GBiJast}
\left(\rmB\lUJ\right)^\ast\gamma_{\rmU(1)}^{(2)}
= 
E_{\twbfm}^{(2)}\left(\twgamma_J\right)\,,
\end{equation}
Then the commutative diagram \gref{GctvdgrlBSJ} implies
$$
E^{(2)}_{\twbfm}\left(\gamma_J\right)
= 
\left(\rmB j_J\right)^\ast E^{(2)}_{\twbfm}\left(\twgamma_J\right)
=
\left(\rmB j_J\right)^\ast\left(\rmB\lUJ\right)^\ast
\gamma_{\rmU(1)}^{(2)}
=
\left(\rmB\lambda^\rmS_J\right)^\ast\left(\rmB j_g\right)^\ast
\gamma_{\rmU(1)}^{(2)}\,.
$$
Thus, by definition of $\delta_J$, the relation is
\begin{equation}\label{Grel}
E^{(2)}_{\twbfm}\left(\gamma_J\right)
=
\beta_g(\delta_J)\,.
\end{equation}

The generators $\gamma_{J,i}^{(2j)}$, $\delta_J$ constructed above define the 
following characteristic classes for $\SUJ$-bundles $Q$ over $M$:
\begin{eqnarray*} 
%
\xi_J(Q)
& := &
\left(\ka{Q}\right)^\ast\delta_J\,.
\\
\alpha^{(2j)}_{J,i}(Q) 
& := & 
\left(\ka{Q}\right)^\ast\gamma^{(2j)}_{J,i}
\,,~~~j=1,\dots,k_i\,,~i=1,\dots,r\,.
\end{eqnarray*}
Here $\ka{Q}$ denotes the classifying
map of $Q$. We denote 
$
\alpha_{J,i} 
= 
1+\alpha_{J,i}^{(2)}+\cdots+\alpha_{J,i}^{(2k_i)}
$
and 
$
\alpha_J = (\alpha_{J,1},\dots,\alpha_{J,r})
$.
Due to \gref{Grel}, $\alpha_J$ and $\xi_J$ are subject to the
relation 
\begin{equation}\label{Grelcc}
E_{\twbfm}^{(2)}\left(\alpha_J(Q)\right)
=
\beta_{g}\left(\xi_J(Q)\right)\,.
\end{equation}

By construction, the characteristic classes so defined have the following
interpretation in terms of ordinary characteristic classes of certain 
bundles naturally associated to $Q$. First, by extending the structure group
of $Q$ to $\UJ$ we obtain a $\UJ$-bundle $\twQ$. Since
$\UJ\cong\rmU(k_1)\times\cdots\times\rmU(k_r)$, $\twQ$ decomposes into a Whitney
product of $\rmU(k_i)$-bundles $\twQ_i$. Formally, $\twQ_i$ is
given by the associated bundle $Q\times_{\SUJ}\rmU(k_i)$, where $\SUJ$ acts
via $\prU{J}{i}\circ j_J$ by left multiplication on $\rmU(k_i)$. Hence, its
classifying map is $\rmB\prU{J}{i}\circ\rmB j_J\circ\ka{Q}$, see
formula \gref{Gclfmapassbun} in \rref{bunth}. Using this, a standard
calculation yields
\begin{equation}
\label{Gaclass}
\alpha_{J,i}(Q) = c(\twQ_i)\,,
\end{equation}
where $c$ denotes the total Chern class. Second, factorizing $Q$ by
$\SUJ_0$, the connected component of the identity of $\SUJ$, we obtain a 
$\ZZ_g$-bundle $Q_0$. It is given by the associated bundle
$Q\times_{\SUJ}\ZZ_g$, where $\SUJ$ acts on $\ZZ_g$ via the homomorphism
$\lSJ$. Then formula \gref{Gclfmapassbun} implies that $Q_0$ has classifying
map $\rmB\lSJ\circ f_Q$. This allows to calculate
\begin{equation}
\label{Gxclass}
\xi_J(Q) = \chi_g(Q_0)\,,
\end{equation}
where $\chi_g$ is a (suitably chosen) generating characteristic class for
$\ZZ_g$-bundles over $M$. 

We remark that the commutative diagram \gref{GctvdgrlSJ} implies that
extension of $Q_0$ to structure group $\rmU(1)$ and factorization of $\twQ$
by $\SUJ_0$ yield isomorphic $\rmU(1)$-bundles. In this way, the relation
\gref{Grelcc} expresses itself on the level of the associated bundles.

So far, we have found that the classes $\alpha_J$ and $\xi_J$ assign to 
any $\SUJ$-bundle $Q$ over $M$ an element of the set
$$
\rmK(M,J)
= 
\left\{
(\alpha,\xi)
\in 
\prod_{i=1}^r \prod_{j=1}^{k_i} H^{2j}(M,\ZZ)
\times 
H^1(M,\ZZ_g)
~\left|~
E_{\twbfm}^{(2)}(\alpha) = \beta_{g}(\xi)
\right.
\right\}\,.
$$
We already know that $\alpha_J(Q) = \alpha_J(Q')$ and $\xi_J(Q) = \xi_J(Q')$
imply $Q\cong Q'$. Thus, for $\rmK(M,J)$ to be a classifying set for
$\SUJ$-bundles it remains to prove that for any of its elements a bundle
with the corresponding characteristic classes exists. Thus, let
$(\alpha,\xi)$ be given. There exist

(i) $\rmU(k_i)$-bundles $\twQ_i$ such that $c(\twQ_i) = \alpha_i$. Their
Whitney product defines a $UJ$-bundle $\twQ$. 

(ii) a $\ZZ_g$-bundle $Q_0$ such that $\chi_g(Q_0) = \xi$.

\noindent
The defining relation of $\rmK(M,J)$ ensures that $Q_0$ is a reduction of
the quotient bundle $\twQ/\SUJ_0$, see \cite[Lemma 5.15]{RSV:clfot}. 
Then the pre-image $Q$ of $Q_0$ in $\twQ$ is
an $\SUJ$-bundle. By construction, \gref{Gaclass} and \gref{Gxclass} hold. 
Hence, we have $\alpha_J(Q) = \alpha$ and $\xi_J(Q) = \xi$.

We summarize.
\begin{Theorem}\label{TBunMSUJ}
Let $M$ be a manifold, $\dim M\leq 4$, and let $J\in\rmK(n)$. Then the
characteristic classes $\alpha_J$ and $\xi_J$ define a bijection from
isomorphism classes of principal $\SUJ$-bundles over $M$ onto $\rmK(M,J)$.
\end{Theorem}

Next, we have to characterize the $\SUJ$-bundles $Q$ that are reductions of
a given $\rmSU(n)$-bundle $P$. Evidently, $Q\subseteq P$ iff $P$ can be
obtained from $Q$ by extending the structure group to $\rmSU(n)$, or iff the
extensions of $P$ and $\twQ$ to structure group $\rmU(n)$ coincide. A
standard calculation yields that the total Chern class of the
extension of $\twQ$ is given by $E_\bfm(\alpha_J(Q))$. Thus, using the
notation 
$$
\rmK(P,J)
=
\left\{
\left.
(\alpha,\xi)\in\rmK(M,J) 
~\right|~
E_{\bfm}(\alpha)=c(P)
\right\}\,,
$$
we have 
\begin{Theorem}\label{TRedPSUJ}
Let $P$ be a principal $\rmSU(n)$-bundle over a manifold $M$, $\dim M\leq
4$, and let $J\in\rmK(n)$. Then the characteristic classes 
$\alpha_J$, $\xi_J$ define a bijection from isomorphism classes of
reductions of $P$ to the subgroup $\SUJ$ onto $\rmK(P,J)$.
\end{Theorem}
The equation $E_{\bfm}(\alpha)=c(P)$ actually contains the two equations
$E_{\bfm}^{(2)}(\alpha) = 0$ and $E_{\bfm}^{(4)}(\alpha)=c_2(P)$. However,
under the assumption $(\alpha,\xi)\in\rmK(M,J)$, the first one is redundant,
because due to \gref{GEJ2},
$E_{\bfm}^{(2)}(\alpha)=g\,E_{\twbfm}^{(2)}(\alpha)=
g\,\beta_g(\xi)=0$.
Thus, the relevant equations are
\begin{eqnarray}
\label{GKMJ}
E_{\twbfm}^{(2)}(\alpha)
& = &
\beta_g(\xi)\,,
\\ \label{GKPJ}
E_{\bfm}^{(4)}(\alpha)
& = &
c_2(P)\,.
\end{eqnarray}
The set of solutions of \gref{GKMJ} yields $\rmK(M,J)$, the set of solutions 
of both equations \gref{GKMJ} and \gref{GKPJ} yields $\rmK(P,J)$.

This concludes the classification of Howe subbundles of $P$, i.e., step 2 of
our programme. We have found that, up to the principal action of $\rmSU(n)$,
the Howe subbundles are given by triples $(J;\alpha,\xi)$, where
$J\in\rmK(n)$ and $(\alpha,\xi)\in\rmK(P,J)$. For further use, let us denote 
the set of all such triples by $\rmK(P)$. It may be viewed as the
disjoint union of all $\rmK(P,J)$, $J\in\rmK(n)$. 
Moreover, for given $L\in\rmK(P)$,
$L=(J;\alpha,\xi)$, let $Q_L$ denote the corresponding Howe subbundle. That
is, $Q_L$ is the reduction of $P$ to $\SUJ$ which has characteristic classes
$\alpha_J(Q_L)=\alpha$ and $\xi_J(Q_L)=\xi$. It is unique up to 
isomorphy.
\subsection{Examples}
\label{SSex}
We determine $\rmK(P,J)$ for several specific values of $J$ and for base 
manifolds $M=\sphere{4},\sphere{2}\times\sphere{2},\torus{4}$, and
$\lens{p}{3}\times\sphere{1}$. Here $\lens{p}{3}$ denotes the
$3$-dimensional lens space which is defined to be
the quotient of the restriction of the natural action of $\rmU(1)$ on the
sphere $\sphere{3}\subset\CC^2$ to the subgroup $\ZZ_p$. Note that
$\lens{p}{3}$ is orientable.

Let us derive the respective Bockstein homomorphisms
$\beta_g:H^1(M,\ZZ_g)\rightarrow H^2(M,\ZZ)$. Since the Abelian group
$H^1(M,\ZZ_g)$ has vanishing free part and since for products of spheres the
integer-valued second cohomology is free Abelian, the Bockstein homomorphism is
trivial here. For  $M=\lens{p}{3}\times\sphere{1}$, on the other hand, 
let $\gamma_{\lens{p}{3};\ZZ_g}^{(1)}$ and $\gamma_{\sphere{1}}^{(1)}$ 
be generators of $H^1(\lens{p}{3},\ZZ_g)$ and $H^1(\sphere{1},\ZZ)$,
respectively. One has 
$
H^1(\lens{p}{3}\times\sphere{1},\ZZ_g)
=
\ZZ_{\langle p,g\rangle}\oplus\ZZ_g
$,
where $\langle p,g\rangle$ denotes the greatest common divisor of $p$ and
$g$. Here the first factor is generated by $\gamma_{\lens{p}{3};\ZZ_g}^{(1)}
\!\times\! 1_{\sphere{1}}$ and  the second one by
$1_{\lens{p}{3};\ZZ_g}\!\times\!\gamma_{\sphere{1}}^{(1)}$. 
In terms of these generators and an appropriately chosen generator 
$\gamma_{\lens{p}{3};\ZZ}^{(2)}$ of $H^2(\lens{p}{3},\ZZ)\cong\ZZ_p$, 
the Bockstein homomorphism is
\begin{equation} \label{GbgLdpS}
\beta_g\left(\gamma_{\lens{p}{3};\ZZ_g}^{(1)}\!\times\!1_{\sphere{1}}\right) 
= 
\frac{p}{\langle p,g \rangle}\gamma_{\lens{p}{3};\ZZ}^{(2)}\!\times\!
1_{\sphere{1}}
\,,~~~~
\beta_g\left(1_{\lens{p}{3};\ZZ_g}\!\times\!\gamma_{\sphere{1}}^{(1)}\right)
= 0\,.
\end{equation}
Now we discuss specific $J$. We write them in the form
$J=(k_1,\dots,k_r|m_1,\dots,m_r)$.
\\
\\
{\it Example} 1. $J=(1|n)\in\rmK(n)$.
Here $\SUJ=\ZZ_n$, the center of $\rmSU(n)$. Moreover,
$g=n$. Variables are $\xi\in H^1(M,\ZZ_n)$ and $\alpha=1+\alpha^{(2)}$,
$\alpha^{(2)}\in H^2(M,\ZZ)$. The system of equations \gref{GKMJ} and
\gref{GKPJ} reads
\begin{eqnarray}\label{G(1|n)KMJ}
\alpha^{(2)} & = & \beta_n(\xi)
\\ \label{G(1|n)KPJ}
\frac{n(n-1)}{2}\alpha^{(2)}\smile\alpha^{(2)} & = & c_2(P)\,.
\end{eqnarray}
Equation \gref{G(1|n)KMJ} yields $n\,\alpha^{(2)}=0$, so that equation 
\gref{G(1|n)KPJ}
requires $c_2(P)=0$. Thus, $\rmK(P,J)$ is nonempty iff $P$ is trivial and
is then parametrized by $\xi$. This coincides with what is known about
$\ZZ_n$-reductions of $\rmSU(n)$-bundles.
\\
\\
{\it Example} 2. $J=(n|1)\in\rmK(n)$.
Here $\SUJ=\rmSU(n)$, the whole group. Due to $g=1$, the only
variable is $\alpha=1+\alpha^{(2)}+\alpha^{(4)}$. Equations
\gref{GKMJ} and \gref{GKPJ} read $\alpha^{(2)} = 0$ and 
$\alpha^{(4)} = c_2(P)\,, $ respectively. Thus, $\rmK(P,J)$ consists of $P$
itself.
\\
\\
{\it Example} 3. $J=(1,1|2,2)\in\rmK(4)$. Here $g=2$.
One can check that $\SUJ$ has connected components
$
\{\diag(z,z,z^{-1},z^{-1})|z\in\rmU(1)\}
$
and
$
\{\diag(z,z,-z^{-1},-z^{-1})|z\in\rmU(1)\}
\,.$
It is therefore isomorphic to $\rmU(1)\times\ZZ_2$. Variables are $\xi\in
H^1(M,\ZZ_2)$ and $\alpha_i=1+\alpha_i^{(2)}$, $i=1,2$. The system of 
equations under consideration is
\begin{eqnarray}\label{G(1,1|2,2)KMJ}
\alpha_1^{(2)}+\alpha_2^{(2)} & = & \beta_2(\xi)
\\ \label{G(1,1|2,2)KPJ}
\left(\alpha_1^{(2)}\right)^2 + \left(\alpha_2^{(2)}\right)^2
+ 4 \alpha_1^{(2)}\alpha_2^{(2)} & = & c_2(P)\,.
\end{eqnarray}
We solve equation \gref{G(1,1|2,2)KMJ}  w.r.t.~$\alpha_2^{(2)}$ and
insert it into equation \gref{G(1,1|2,2)KPJ}. Since, due to compactness and
orientability of $M$, $H^4(M,\ZZ)$ is torsion-free, 
products including $\beta_2(\xi)$ vanish. Thus, we obtain that $\xi$ can be
chosen arbitrarily, whereas $\alpha_1^{(2)}$ must solve the equation
\begin{equation} \label{G(1,1|2,2)}
-2\left(\alpha_1^{(2)}\right)^2 
= 
c_2(P)\,.
\end{equation}
Let us discuss the result for the different base manifolds.
 
(i) $M=\sphere{4}$: Due to $H^1(M,\ZZ_2) = 0$ and $H^2(M,\ZZ) = 0$, 
$\rmK(P,J)$ is nonempty iff $c_2(P)=0$,
in which case it contains the (necessarily trivial) 
$\rmU(1)\times\ZZ_2$-bundle over $\sphere{4}$.

(ii) $M=\lens{p}{3}\times\sphere{1}$: We have $H^1(M,\ZZ_2)
\cong\ZZ_{\langle 2,p \rangle}\oplus\ZZ_2$ and $H^2(M,\ZZ)\cong\ZZ_p$.
In particular, $(\alpha_1^{(2)})^2=0$. Hence, if $c_2(P)=0$,
$\rmK(P,J)=(\ZZ_{\langle 2,p \rangle}\oplus\ZZ_2)\times\ZZ_p$. 
Otherwise, $\rmK(P,J)=\emptyset$.

(iii) $M=\sphere{2}\times\sphere{2}$: We have $H^1(M,\ZZ_2)=0$ and
$H^2(M,\ZZ)\cong\ZZ\oplus\ZZ$. The latter is generated by
$\gamma_{\sphere{2}}^{(2)}\!\times\! 1_{\sphere{2}}$ and  
$1_{\sphere{2}}\!\times\!\gamma_{\sphere{2}}^{(2)}$, where 
$\gamma_{\sphere{2}}^{(2)}$ is a generator of $H^2(\sphere{2},\ZZ)$. 
Then $H^4(M,\ZZ)$ is generated by 
$\gamma_{\sphere{2}}^{(2)}\!\times\!\gamma_{\sphere{2}}^{(2)}$. 
Writing 
\begin{equation}\label{Ga1S2S2p}
\alpha_1^{(2)}=a~\gamma_{\sphere{2}}^{(2)}\!\times\! 1_{\sphere{2}}
+b~1_{\sphere{2}}\!\times\!\gamma_{\sphere{2}}^{(2)}
\end{equation}
with $a,b\in\ZZ$, equation \gref{G(1,1|2,2)} becomes
\begin{equation}\label{GS2S2ex}
-4ab~\gamma_{\sphere{2}}^{(2)}\!\times\!\gamma_{\sphere{2}}^{(2)}
= 
c_2(P)\,.
\end{equation}
If $c_2(P)=0$, there are two series of solutions: $a=0$ and $b\in\ZZ$ as well
as $a\in\ZZ$ and $b=0$. Here $\rmK(P,J)$ is infinite. If
$c_2(P)=4l~\gamma_{\sphere{2}}^{(2)}\!\times\!\gamma_{\sphere{2}}^{(2)}$,
$l\neq 0$, then $a=q$ and $b=-l/q$, where $q$ runs through the (positive and
negative) divisors of $l$. Hence, in this case, the cardinality of
$\rmK(P,J)$
is twice the number of divisors of $l$. If $c_2(P)$ is not
divisible by $4$ then $\rmK(P,J) = \emptyset$.

(iv) $M=\torus{4}$: Here $H^1(M,\ZZ_2)\cong\ZZ_2^{\oplus 4}$ and 
$H^2(M,\ZZ)\cong\ZZ^{\oplus 6}$. The latter is generated by elements 
$\gamma^{(2)}_{\torus{4};ij}$, $1 \leq i < j \leq 4$, where 
$
\gamma_{\torus{4};12}^{(2)} 
= 
\gamma_{\sphere{1}}^{(1)}\!\times\!
\gamma_{\sphere{1}}^{(1)}\!\times\!
1_{\sphere{1}}\!\times\! 
1_{\sphere{1}}
\,,$~
$
\gamma_{\torus{4};13}^{(2)} 
= 
\gamma_{\sphere{1}}^{(1)}\!\times\! 
1_{\sphere{1}}\!\times\!
\gamma_{\sphere{1}}^{(1)}\!\times\!
1_{\sphere{1}}
$
etc. $H^4(M,\ZZ)$ is generated by
$\gamma_{\torus{4}}^{(4)} = \gamma_{\sphere{1}}^{(1)}\!\times\! 
\gamma_{\sphere{1}}^{(1)}\!\times\! \gamma_{\sphere{1}}^{(1)}\!\times\! 
\gamma_{\sphere{1}}^{(1)}$. One can check 
$
\gamma^{(2)}_{\torus{4};ij}\smile\gamma^{(2)}_{\torus{4};kl}
=
\epsilon_{ijkl}~\gamma^{(4)}_{\torus{4}}\,,
$
where $\epsilon_{ijkl}$ denotes the totally antisymmetric tensor in $4$
dimensions. Writing 
$
\alpha_1^{(2)}=\sum_{1\leq i<j\leq 4} a_{ij}\gamma^{(2)}_{\torus{4};ij}\,,
$
equation \gref{G(1,1|2,2)} becomes 
$$
-4\left(a_{12}a_{34}-a_{13}a_{24}+a_{14}a_{23}\right)\gamma_{\torus{4}}^{(4)}
=
c_2(P)\,.
$$
Hence, again $\rmK(P,J)\neq\emptyset$ iff $c_2(P)$ is divisible by $4$, 
in which case now it always has infinitely many elements.
\\
\\
{\it Example} 4. $J=(1,1|2,3)\in\rmK(5)$.
The subgroup $\SUJ$ of $\rmSU(5)$ consists of matrices of the form
$\diag(z_1,z_1,z_2,z_2,z_2)$, where $z_1,z_2\in\rmU(1)$ such that $z_1^2z_2^3=1$. 
We can
parametrize $z_1=z^3$, $z_2=z^{-2}$, $z\in\rmU(1)$.
Hence, $\SUJ$ is isomorphic to $\rmU(1)$. Variables are
$\alpha_i=1+\alpha_i^{(2)}$, $i=1,2$. The equations to be solved read
\begin{eqnarray}\label{G(1,1|2,3)KMJ}
2\alpha_1^{(2)}+3\alpha_2^{(2)} & = & 0\,,
\\ \label{G(1,1|2,3)KPJ}
\left(\alpha_1^{(2)}\right)^2
+ 3 \left(\alpha_2^{(2)}\right)^2
+ 6 \alpha_1^{(2)}\smile\alpha_2^{(2)}
& = & c_2(P)\,.
\end{eqnarray}
Equation \gref{G(1,1|2,3)KMJ} can be parametrized by $\alpha_1^{(2)} = 3\eta$,
$\alpha_2^{(2)}=-2\eta$, where $\eta\in H^2(M,\ZZ)$. Then 
\gref{G(1,1|2,3)KPJ} becomes $-15 \eta^2 = c_2(P)$.
The discussion of this equation is analogous to that of equation
\gref{G(1,1|2,2)} above. For example, in case
$M=\sphere{2}\times\sphere{2}$, $\rmK(P,J)\neq\emptyset$ iff $c_2(P)$ is
divisible by 15.
\\
\\
{\it Example} 5. $J=(2,3|1,1)\in\rmK(5)$.
Here $\SUJ \cong \rmS[\rmU(2)\!\times\!\rmU(3)]$. This is the symmetry
group of the standard model. In the grand unified $\rmSU(5)$-model
it is the subgroup to which $\rmSU(5)$ is broken by the heavy
Higgs field. Moreover, it is the centralizer of the subgroup
discussed in Example 4. 

Since $g=1$, variables are
$\alpha_i=1+\alpha_i^{(2)}+\alpha_i^{(4)}$, $i=1,2$. 
Equations \gref{GKMJ}
and \gref{GKPJ} read
\begin{eqnarray}\label{G(2,3|1,1)KMJ}
\alpha_1^{(2)}+\alpha_2^{(2)}
& = &
0\,,
\\ \label{G(2,3|1,1)KPJ}
\alpha_1^{(4)}+\alpha_2^{(4)}+\alpha_1^{(2)}\smile\alpha_2^{(2)}
& = &
c_2(P)\,.
\end{eqnarray}
Using \gref{G(2,3|1,1)KMJ} to replace $\alpha_2^{(2)}$ in
\gref{G(2,3|1,1)KPJ} we obtain for the latter
$
\alpha_2^{(4)}=c_2(P)-\alpha_1^{(4)}
+
\left(\alpha_1^{(2)}\right)^2.
$
Thus, $\rmK(P,J)$ can be parametrized by $\alpha_1$ (or $\alpha_2$), i.e., 
by the Chern class of one of the factors $\rmU(2)$ or $\rmU(3)$. This is
well known \cite{Isham}.
\\
\\
{\it Example} 6. $J=(2|2)$. We have $g=2$.
The subgroup $\SUJ$ of $\rmSU(4)$ consists of matrices $D\oplus D$, where 
$D\in\rmU(2)$ such that $(\det D)^2=1$. Hence, it has connected components 
$\{D\oplus D | D\in\rmSU(2)\}$ and $\{(iD)\oplus(iD) | D\in\rmSU(2)\}$.
One can check that $\SUJ\cong(\rmSU(2)\times\ZZ_4)/\ZZ_2$.
Variables are $\xi\in
H^1(M,\ZZ_2)$ and $\alpha=1+\alpha^{(2)}+\alpha^{(4)}$. The equations under
consideration are
\begin{eqnarray} \label{G(2|2)KMJ}
\alpha^{(2)} & = & \beta_2(\xi)\,,
\\ \label{G(2|2)KPJ}
\left(\alpha^{(2)}\right)^2 + 2\alpha^{(4)} & = & c_2(P)\,.
\end{eqnarray}
Equation \gref{G(2|2)KMJ} fixes $\alpha^{(2)}$ in terms of $\xi$. For
example, in case $M=\lens{p}{3}\times\sphere{1}$, by expanding 
$
\xi 
= 
\xi_\rmL~\gamma_{\lens{p}{3};\ZZ_2}^{(1)}\!\times\!1_{\sphere{1}} 
+
\xi_\rmS~1_{\lens{p}{3};\ZZ_2}\!\times\!\gamma_{\sphere{1}}^{(1)}
$,
equations \gref{GbgLdpS} and \gref{G(2|2)KMJ} imply
$$
\alpha^{(2)} = \left\{\begin{array}{ccl}
q\xi_L~\gamma_{\lens{p}{3};\ZZ}^{(2)}\!\times\!1_{\sphere{1}}
& | & p=2q \\
0 & | & p=2q+1.
\end{array}\right.
$$
For general $M$, due to \gref{G(2|2)KMJ}, equation \gref{G(2|2)KPJ} becomes 
$2 \alpha^{(4)} =
c_2(P)$. Thus, $\rmK(P,J)$ is nonempty iff $c_2(P)$ is even and is then 
parametrized by $\xi\in H^1(M,\ZZ_2)$. 
%
%
%

\subsection{Holonomy-induced Howe subbundles and factorization by
$\rmSU(n)$-action}
\label{HolonIndHSB}


In this subsection, we will accomplish steps 3 and 4 of our programme.

In step 3, we have to specify those reductions $Q\subseteq P$ to $\SUJ$, 
$J\in\rmK(n)$, which are holonomy-induced, i.e., which possess a connected
reduction to some subgroup $H$ such that $\rmC^2_{\rmSU(n)}(H) = \SUJ$. 
Let $Q$ be given and consider a connected component of $Q$. This is a
connected reduction of $Q$ to some subgroup $H\subseteq\SUJ$ which has the
same dimension as $\SUJ$. Then so has the Howe subgroup 
$\twH := \rmC^2_{\rmSU(n)}(H)$ generated by $H$, because 
$H\subseteq\twH\subseteq\SUJ$. Then the Howe subgroups 
$\rmC^2_{\rmU(n)}(\twH)$ and $\rmC^2_{\rmU(n)}(\SUJ)$ of $\rmU(n)$ have the 
same dimension and obey 
$\rmC^2_{\rmU(n)}(\twH)\subseteq\rmC^2_{\rmU(n)}(\SUJ)$. Since they are
closed and connected (recall that they are conjugate to $\UJ$ for some
$J\in\rmK(n)$), they coincide. It follows $\twH = \SUJ$. We conclude that
any Howe subbundle of an $\rmSU(n)$-bundle is holonomy-induced, so that the
condition is redundant here.

We remark that, in general, Howe subbundles exist which are not 
holonomy-induced. A simple example is provided by the Howe subgroup
$H=\{\II_3,\diag(-1,-1,1)\}$ of $\rmSO(3)$. 
While the reduction $Q=M\times H\subseteq M\times\rmSO(3)$ is a Howe
subbundle, any connected reduction of $Q$ has the center 
$\{\II_3\}$ as its structure group, hence is a Howe subbundle itself. Thus,
$Q$ is not holonomy-induced.

In step 4, we have to factorize the set of Howe subbundles by the principal
action of $\rmSU(n)$. That is, we have to identify elements $L,L'$ of
$\rmK(P)$ for which $D\in\rmSU(n)$ exists such that 
\begin{equation}\label{GisoQL}
Q_{L'} \cong Q_L\cdot D\,.
\end{equation}
First, assume that such $D$ exists. Then $\SUJp=D^{-1}\,\SUJ\,D$, hence
$\rmM_{J'}(\CC) = D^{-1}\,\rmM_J(\CC)\,D$. It follows that $r=r'$ and there
exists a permutation $\sigma$ such that 
\begin{equation}\label{Gkpsk}
\bfk' = \sigma\bfk\,,~~~~\bfm' = \sigma\bfm\,.
\end{equation}
A straightforward calculation, see \cite[Lemma 7.1]{RSV:clfot}, yields 
\begin{equation}\label{GccQLD}
\alpha_{J'}(Q_L\cdot D) = \sigma\alpha
\,,~~~~
\xi_{J'}(Q_L\cdot D) = \xi\,.
\end{equation}
Hence, \gref{GisoQL} implies 
\begin{equation}\label{Gapsa}
\alpha' = \sigma\alpha\,,~~~~\xi' = \xi\,.
\end{equation}
Conversely, assume that $r=r'$ and that a permutation exists such that
\gref{Gkpsk} and \gref{Gapsa} hold. Due to \gref{Gkpsk} one can construct
$D\in\rmSU(n)$ such that conjugation of $\rmM_J(\CC)$ by $D^{-1}$ yields
$\rmM_{J'}(\CC)$, where the factors are permuted according to $\sigma$, see
\cite[Lemma 4.2]{RSV:clfot}. Then \gref{GccQLD} and \gref{Gapsa} imply 
$\alpha_{J'}(Q_L\cdot D) = \alpha'$ and $\xi_{J'}(Q_L\cdot D) = \xi'$.
Hence, \gref{GisoQL} holds. Note that \gref{GisoQL} is actually a special 
case of a more general situation discussed in Subsection \rref{parord}.

Thus, on the level of $\rmK(P)$, factorization by the principal 
$\rmSU(n)$-action on Howe subbundles amounts to the identification of
elements which can be transformed to each other by a simultaneous
permutation of $\bfk$, $\bfm$, and $\alpha$. The set of equivalence classes 
so obtained will be denoted by $\hat\rmK(P)$ and its elements will be 
denoted by $[L]$.


\subsection{Summary}
\label{Summary}


Before we proceed, we summarize the results of this section. 
The set of Howe subbundles of $P$ modulo isomorphy and the principal 
$\rmSU(n)$-action, which classifies the orbit types of the action of 
$\gau$ on $\con$ by virtue of the reduction
theorem, can be described as follows. Its elements are labelled by symbols 
$[J;\alpha,\xi]$, where 

(i) 
$
J=((k_1,\dots,k_r),(m_1,\dots, m_r))
$
is a pair of sequences of positive integers obeying 
$
\sum_{i=1}^r k_im_i=n\,,
$

(ii) $\alpha=(\alpha_1,\dots,\alpha_r)$ is a sequence of cohomology elements 
$\alpha_i\in H^\ast(M,\ZZ)$, which are admissible values of the total Chern 
class of $\rmU(k_i)$-bundles over $M$, 

(iii) $\xi\in H^1(M,\ZZ_g)$ with $g$ being the greatest common divisor
of $(m_1,\dots, m_r)$. 

\noindent
The cohomology elements $\alpha_i$ and $\xi$ are subject to the relations
\begin{eqnarray}\nonumber
\sum_{i=1}^r \frac{m_i}{g} \alpha_i^{(2)} & = & \beta_g(\xi)\,,
\\ \nonumber
\alpha_1^{m_1}\smile\dots\smile\alpha_r^{m_r} & = & c(P)\,,
\end{eqnarray}
where $\beta_g: H^1(M,\ZZ_g)\rightarrow H^2(M,\ZZ)$ is the connecting
homomorphism associated to the short exact sequence 
$0\rightarrow\ZZ\rightarrow\ZZ\rightarrow\ZZ_g\rightarrow 0$ of coefficient groups in
cohomology. For any permutation $\sigma$ of $r$ elements, the symbols
\begin{eqnarray*}
\Big[\Big((k_1,\dots,k_r),(m_1,\dots,m_r)\Big)
;
(\alpha_1,\dots,\alpha_r),\xi\Big]\,,
\\
\Big[\Big((k_{\sigma(1)},\dots, k_{\sigma(r)})
,
(m_{\sigma(1)},\dots,m_{\sigma(r)})\Big)
;
(\alpha_{\sigma(1)},\dots, \alpha_{\sigma(r)}),\xi
\Big]
\end{eqnarray*}
have to be identified.


\section{Partial ordering of gauge orbit types for $G=\rmSU(n)$}
\label{PO}\label{Spo}


\subsection{Characterization of the partial ordering relation}
\label{parord}


In this subsection we are going to characterize the natural partial ordering of
Howe subbundles in terms of the classifying set $\hat{\rmK}(P)$. 
Thus, let $L,L'\in\rmK(P)$. By definition, $[L]\leq[L']$ iff
$D\in\rmSU(n)$ exists such that $Q_{L}\cdot D\subseteq Q_{L'}$, where 
inclusion is understood up to isomorphy. We say that $Q_L$ is subconjugate
to $Q_{L'}$. 

First, we observe that $Q_{L}\cdot D\subseteq Q_{L'}$ implies
$D^{-1}\,\SUJ\, D\subseteq\SUJp$, i.e., subconjugacy of the structure
groups. Then also $D^{-1}\,\MJ\, D\subseteq \MJp$. We have an 
associated embedding 
$$
\hMD : \MJ\longrightarrow\MJp\,,~~C\mapsto D^{-1}CD\,,
$$
and, derived from that, embeddings $\hUD : \UJ\longrightarrow\UJp$ and 
$\hSD : \SUJ\longrightarrow\SUJp$. Since $\MJ$ and $\MJp$ are
finite-dimensional unital $\rmC^\ast$-algebras, the embedding $\hMD$ is 
characterized by a so-called {\it inclusion matrix} $\Delta$. This is an 
$(r'\times r)$-matrix with nonnegative integer entries, defined as follows:
$\Delta_{i'i}$ is the number of fundamental irreducible representations
contained in the representation
$$
\MC{k_i}\longrightarrow
\MJ\stackrel{\hMD}{\longrightarrow}
\MJp\stackrel{\prM{J'}{i'}}{\longrightarrow}
\MC{k'_{i'}}\,.
$$
Here the first map is the canonical embedding to the $i$th factor of $\MJ$.
Since the embedding $\hMD$ is unital, $\sum_i\Delta_{i'i}k_i = k'_{i'}$, for
all $i'$.
Since conjugation of $\MJ$ by $D^{-1}$ preserves the total number of
fundamental irreducible representations of the factor $\rmM_{k_i}(\CC)$ in 
$\rmM_n(\CC)$, $\sum_{i'}\Delta_{i'i}m'_{i'} = m_i$, for all $i$. Thus,
$\Delta$ solves the system of equations
\begin{eqnarray}\label{GeqkD}
\Delta\,\bfk & = & \bfk'
\\ \label{GeqmD}
\bfm & = & \bfm'\Delta\,,
\end{eqnarray}
where $\bfm$ and $\bfm'$ are viewed as row vectors. Conversely, assume that
a solution $\Delta$ of \gref{GeqkD} and \gref{GeqmD} is given. Then
the decompositions \gref{Gdecomp} associated to $J$ and $J'$ admit
subdecompositions
\begin{eqnarray*}
\CC^n
& = &
\bigoplus_{i=1}^r\CC^{k_i}
\otimes
\left(\bigoplus_{i'=1}^{r'}\CC^{\Delta_{i'i}}\otimes\CC^{m'_{i'}}\right)
\,,
\\
\CC^n
& = &
\bigoplus_{i'=1}^{r'}
\left(\bigoplus_{i=1}^r\CC^{k_i}\otimes\CC^{\Delta_{i'i}}\right)
\otimes\CC^{m'_{i'}}\,,
\end{eqnarray*}
respectively, which differ by  a permutation of the factors
$\CC^{k_i}\otimes\CC^{\Delta_{i'i}}\otimes\CC^{m'_{i'}}$. From this
permutation, $D\in\rmSU(n)$ can be constructed which obeys
$D^{-1}\,\MJ\,D\subseteq\MJp$ and which has inclusion matrix $\Delta$, see
\cite[Lemma 3.1]{RSV:poot}. It follows that
$\SUJ$ is subconjugate to $\SUJp$, or $\MJ$ is subconjugate to $\MJp$, iff the 
system of equations \gref{GeqkD}, \gref{GeqmD} has a solution $\Delta$.

Second, let $Q_L^D$ denote the extension of $Q_{L}\cdot D$ to structure
group $\SUJp$. We observe that $Q_{L}\cdot D\subseteq Q_{L'}$ implies 
$Q_L^D\cong Q_{L'}$. This provides a relation between the characteristic 
classes $\alpha$, $\xi$ and $\alpha'$, $\xi'$. To derive it, we have to 
compute the
characteristic classes of $Q_L^D$. Let us sketch how this can be done. For a
detailed computation, purely on the level of cohomology, we refer to
\cite[Lemma 3.2]{RSV:poot}. 

To compute $\alpha_{J'}(Q_L^D)$ we may form
the extension $\wt{Q_L^D}$~of $Q_L^D$ to structure group $\UJp$ and 
compute the total Chern class of the Whitney factors.
To do so, we use that $\wt{Q_L^D}$ coincides with the extension 
$\twQ_L^D$ of 
$\twQ_L\cdot D$ to structure group $\UJp$. A close look at how $\hUD$ embeds 
the factors of $\UJ$ into those of $\UJ'$ reveals that the $i'$th Whitney 
factor of $\twQ_L^D$ contains the Whitney product $(\twQ_L)_1^{\Delta_{i'1}}
\times\cdots\times (\twQ_L)_r^{\Delta_{i'r}}$ as a subbundle. Hence, the
total Chern class of this factor is
$
\alpha_1^{\Delta_{i'1}}\cdots\alpha_r^{\Delta_{i'r}}\,.
$
Using the notation
$$
E_\Delta(\alpha)
=
\left(
\alpha_1^{\Delta_{11}}\cdots\alpha_r^{\Delta_{1r}}
,\dots,
\alpha_1^{\Delta_{r'1}}\cdots\alpha_r^{\Delta_{r'r}}
\right)\,,
$$
which is a generalization of \gref{GdefEm}, we can write
\begin{equation}\label{GaJassoc}
\alpha_{J'}(Q_L^D)
=
E_\Delta(\alpha)\,.
\end{equation}

To determine $\xi_{J'}(Q_L^D)$, we can compute the class $\chi_{g'}$ of the
quotient $Q_L^D/\SUJp_0$. The latter is given by the associated bundle
$Q_L\times_{\SUJ}\ZZ_{g'}$, where $\SUJ$ acts on $\ZZ_{g'}$ via the
homomorphism $\lSJp\circ\hSD$. A straightforward computation yields
$\lSJp\circ\hSD = \vr_{g'}\circ\lSJ$, where $\vr_{g'}$ denotes reduction
modulo $g'$. Note that \gref{GeqmD} implies that $g'$ divides $g$, hence
$\vr_{g'}$ is a well defined homomorphism. Moreover, one can check that the
characteristic class of the mod$\,g'$-reduction of a $\ZZ_g$-bundle is given
by the mod$\,g'$-reduction of the characteristic class of this bundle.
Hence,
\begin{equation}\label{GxJassoc}
\xi_{J'}(Q_L^D)
=
\vr_{g'}(\xi)\,.
\end{equation}
Thus, $Q_L\cdot D\subseteq Q_{L'}$ implies
\begin{eqnarray}\label{Geqa}
E_\Delta(\alpha) & = & \alpha'
\\ \label{Geqx}
\vr_{g'}(\xi) & = & \xi'\,.
\end{eqnarray}
Let us introduce the following notation. If \gref{Geqx} holds,   
let $\rmN(L,L')$ be the set of solutions of the combined system of equations
\gref{GeqkD}, \gref{GeqmD}, \gref{Geqa} in the indeterminate $\Delta$. 
If \gref{Geqx} does not hold, let $\rmN(L,L')=\emptyset$.
So far, we have found that if $Q_L$ is subconjugate to $Q_{L'}$ then 
$\rmN(L,L')\neq\emptyset$. Now assume that, conversely, $\rmN(L,L')$
contains an element $\Delta$. We have seen above that
due to \gref{GeqkD}, \gref{GeqmD} 
there exists $D\in\rmSU(n)$, obeying $D^{-1}\,\MJ\,D\subseteq\MJp$, 
which has inclusion matrix $\Delta$.
Consider $Q_L^D$, i.e., the extension of $Q_L\cdot D$ to structure group
$\SUJp$. Due to \gref{GaJassoc} and \gref{Geqa}, $\alpha_{J'}(Q_L^D) =
\alpha'$. Due to \gref{GxJassoc} and \gref{Geqx}, $\xi_{J'}(Q_L^D) =
\xi'$. It follows $Q_L^D\cong Q_{L'}$, hence $Q_L\cdot D\subseteq Q_{L'}$.
Thus, we have shown that $Q_L$ is subconjugate to $Q_{L'}$ iff
$\rmN(L,L')\neq\emptyset$. Consequently, on the level of $\hat\rmK(P)$, the 
partial ordering of Howe subbundles is given by  
\begin{Theorem}\label{Tpo}
Let $L,L'\in\rmK(P)$. Then $[L]\leq[L']$ if and only if 
$\rmN(L,L')\neq\emptyset$.
\end{Theorem}

\noindent
{\bf Example:}
Let $P=M\times\rmSU(4)$. Consider elements $L$, $L'$ with $J=((1,1),(2,2))$ 
and $J'=((2,2),(1,1))$, respectively. Recall that
$\SUJ\cong\rmU(1)\times\ZZ_2$. The subgroup
$\SUJp$ can be parametrized as follows:
$$
\SUJp 
= 
\left\{\left.
\zzmatrix{z A}{0}{0}{z^{-1}B}
\right|
z\in\rmU(1), A,B\in\rmSU(2)
\right\}\,.
$$
It is therefore isomorphic to 
$\left[ \rmU(1) \times\rmSU(2)\times\rmSU(2) \right]/ \ZZ_2 $. 
To determine $\rmN(L,L')$, we first consider equations \gref{GeqkD} and 
\gref{GeqmD}:
$$
\zzmatrix{\Delta_{11}}{\Delta_{12}}{\Delta_{21}}{\Delta_{22}}
\zematrix{1}{1}
=
\zematrix{2}{2}\,,
\hspace{1cm}
\begin{array}{c} \ezmatrix{1}{1} \\ \phantom{2}\end{array}
\zzmatrix{\Delta_{11}}{\Delta_{12}}{\Delta_{21}}{\Delta_{22}}
\begin{array}{c} = \\ \phantom{2}\end{array}
\begin{array}{c} \ezmatrix{2}{2}\,. \\ \phantom{2}\end{array}
$$
The solutions are
$
\Delta^a
=
\zzmatrix{1}{1}{1}{1}
$,~
$
\Delta^b
=
\zzmatrix{2}{0}{0}{2}
$,~
$
\Delta^c
=
\zzmatrix{0}{2}{2}{0}
$.
For $\alpha=(\alpha_1,\alpha_2)$, they yield
$
E_{\Delta^a}(\alpha)
=
(\alpha_1\alpha_2,\alpha_1\alpha_2)
$,~
$
E_{\Delta^b}(\alpha)
=
(\alpha_1^2,\alpha_2^2)
$,~
$
E_{\Delta^c}(\alpha)
=
(\alpha_2^2,\alpha_1^2)\,.
$~
Condition \gref{Geqx} is trivially satisfied due to $g'=1$. Thus,
$\rmN(L,L')\neq\emptyset$, i.e., $Q_L$ is subconjugate to $Q_{L'}$ or
$[L]\leq[L']$, precisely
in one of the following cases: (a) $\alpha_1' = \alpha_2' =
\alpha_1\alpha_2$, (b) $\alpha_1' = \alpha_1^2$, $\alpha_2' = \alpha_2^2$,
and (c) $\alpha_1' = \alpha_2^2$, $\alpha_2' = \alpha_1^2$.
\vspace{0.3cm} 

\noindent
{\bf Remark:} Any inclusion matrix can be visualized by a diagram
consisting of a series of upper vertices, labelled by
$i=1,\dots,r$, and a series of lower vertices, labelled by $i'=1,\dots,r'$.
For each combination of $i$ and $i'$ the corresponding vertices are
connected by $\Delta_{i'i}$ edges. For example, the matrices $\Delta^a$,
$\Delta^b$, and $\Delta^c$ in the above example give rise to the
following diagrams:
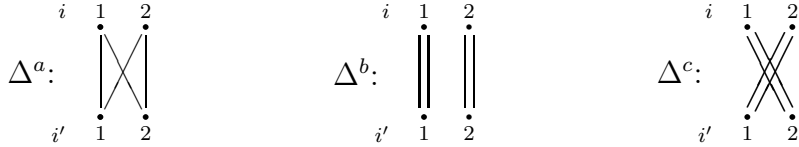
\begin{center}
~\hfill
\unitlength1.2cm
\begin{picture}(1,2.6)
\put(0,1.3){\makebox(0,0)[cr]{$\Delta^a$:}}
\put(0.5,1.8){
\plpen{0,0}{br}{i~~~~}{tr}{i'~~~~}{\scriptsize}
\plpen{0,0}{bc}{1}{tc}{1}{\scriptsize}
\plpee{0,0}{}{}{tc}{2}{\scriptsize}
\plpeme{0.5,0}{bc}{2}{}{}{\scriptsize}
\plpen{0.5,0}{}{}{}{}{}
}
\end{picture}
\hfill
\begin{picture}(1,2.6)
\put(0,1.3){\makebox(0,0)[cr]{$\Delta^b$:}}
\put(0.5,1.8){
\plpzn{0,0}{br}{i~~~~}{tr}{i'~~~~}{\scriptsize}
\plpzn{0,0}{bc}{1}{tc}{1}{\scriptsize}
\plpzn{0.5,0}{bc}{2}{tc}{2}{\scriptsize}
}
\end{picture}
\hfill
\begin{picture}(1,2.6)
\put(0,1.3){\makebox(0,0)[cr]{$\Delta^c$:}}
\put(0.5,1.8){
\plpze{0,0}{br}{i~~~~}{}{}{\scriptsize}
\plpzme{0.5,0}{}{}{tr}{i'~~~~}{\scriptsize}
\plpze{0,0}{bc}{1}{tc}{2}{\scriptsize}
\plpzme{0.5,0}{bc}{2}{tc}{1}{\scriptsize}
}
\end{picture}
\hfill~
\end{center}
The diagrams associated in this way to the elements of $\rmN(J,J')$, 
$J,J'\in\rmK(n)$, are special cases of so-called {\it Bratteli diagrams}
\cite{Bratteli}. The latter have, in general, several stages
picturing the subsequent inclusion matrices associated to an ascending
sequence of finite dimensional von-Neumann algebras ${\bf A}_1\subseteq{\bf
A}_2\subseteq{\bf A}_3\subseteq\cdots$ . For this
reason, we refer to the diagram associated to $\Delta\in \rmN(J,J')$ as the
Bratteli diagram of $\Delta$. We remark that, due to equation \gref{GeqkD}, 
$\Delta$ cannot have a zero row. Due to \gref{GeqmD}, it cannot have a zero
column either. Accordingly, each vertex  of the Bratteli diagram of
$\Delta$ is cut by at least one edge.
Since equations \gref{GeqkD}, \gref{GeqmD}, \gref{Geqa} have an obvious
reformulation on the level of Bratteli diagrams, these diagrams can be used 
to simplify calculations. Furthermore, 
some of the arguments in the sequel are easier to formulate on the level of
Bratteli diagrams than on the level of the corresponding matrices.
 

\subsection{Direct successors}
\label{SSopn}


In this subsection we derive a characterization of direct successors. 
For a detailed discussion we refer to \cite[\S 5]{RSV:poot}.

Let $L,L'\in\rmK(P)$ such that $[L]\leq[L']$. It is not hard to see that 
under this assumption $[L']$
is a direct successor of $[L]$ iff $[\SUJ']$ is a direct successor of $[\SUJ]$
in the set of conjugacy classes of Howe subgroups of $\rmSU(n)$, or iff
$[\MJp]$ is a direct successor of $[\MJ]$ in the set of conjugacy classes of
unital $\ast$-subalgebras of $\rmM_n(\CC)$. It is known by 'folklore' -- and 
can be proved using the notion of the level of an inclusion matrix, see 
\cite{RSV:poot} -- 
that $[\MJp]$ is a direct successor of $[\MJ]$ iff the following holds: 
There exists $D\in\rmSU(n)$
obeying $D^{-1}\,\MJ\, D\subseteq\MJp$, where the Bratteli diagram of the 
corresponding inclusion matrix has either one of the following shapes with
arbitrary $i_0$ and $i_1<i_2$:
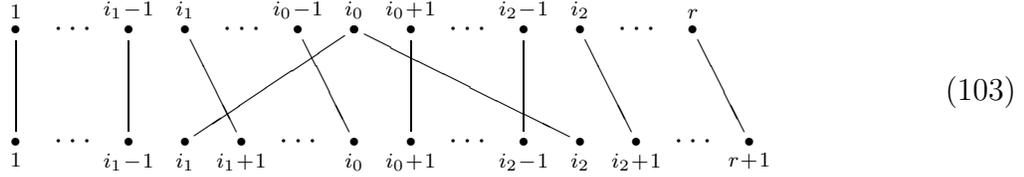
\begin{eqnarray} \label{Gdgrspl}
\unitlength1.5cm
\begin{array}{c}
\begin{picture}(6.5,2)
\put(0,1.5){
\plpen{0,0}{bc}{1}{tc}{1}{\scriptsize}
\put(0.5,0){\makebox(0,0)[cc]{$\cdots$}}
\put(0.5,-1){\makebox(0,0)[cc]{$\cdots$}}
\plpen{1,0}{bc}{i_1\!-\!1}{tc}{i_1\!-\!1}{\scriptsize}
\plpee{1.5,0}{bc}{i_1}{tc}{i_1\!+\!1}{\scriptsize}
\put(2,0){\makebox(0,0)[cc]{$\cdots$}}
\put(2.5,-1){\makebox(0,0)[cc]{$\cdots$}}
\plpee{2.5,0}{bc}{i_0\!-\!1}{tc}{i_0}{\scriptsize}
\plpemd{3,0}{bc}{i_0}{tc}{i_1}{\scriptsize}
\plpev{3,0}{}{}{tc}{i_2}{\scriptsize}
\plpen{3.5,0}{bc}{i_0\!+\!1}{tc}{i_0\!+\!1}{\scriptsize}
\put(4,0){\makebox(0,0)[cc]{$\cdots$}}
\put(4,-1){\makebox(0,0)[cc]{$\cdots$}}
\plpen{4.5,0}{bc}{i_2\!-\!1}{tc}{i_2\!-\!1}{\scriptsize}
\plpee{5,0}{bc}{i_2}{tc}{i_2\!+\!1}{\scriptsize}
\put(5.5,0){\makebox(0,0)[cc]{$\cdots$}}
\put(6,-1){\makebox(0,0)[cc]{$\cdots$}}
\plpee{6,0}{bc}{r}{tc}{r\!+\!1}{\scriptsize}
}
\end{picture}
\end{array}
\\ \label{Gdgrmrg}
\begin{array}{c}
\unitlength1.5cm
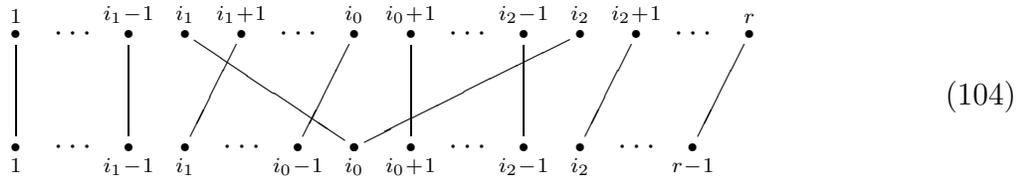
\begin{picture}(6.5,2)
\put(0,1.5){
\plpen{0,0}{bc}{1}{tc}{1}{\scriptsize}
\put(0.5,0){\makebox(0,0)[cc]{$\cdots$}}
\put(0.5,-1){\makebox(0,0)[cc]{$\cdots$}}
\plpen{1,0}{bc}{i_1\!-\!1}{tc}{i_1\!-\!1}{\scriptsize}
\plped{1.5,0}{bc}{i_1}{tc}{i_0}{\scriptsize}
\plpeme{2,0}{bc}{i_1\!+\!1}{tc}{i_1}{\scriptsize}
\put(2.5,0){\makebox(0,0)[cc]{$\cdots$}}
\put(2,-1){\makebox(0,0)[cc]{$\cdots$}}
\plpeme{3,0}{bc}{i_0}{tc}{i_0\!-\!1}{\scriptsize}
\plpen{3.5,0}{bc}{i_0\!+\!1}{tc}{i_0\!+\!1}{\scriptsize}
\put(4,0){\makebox(0,0)[cc]{$\cdots$}}
\put(4,-1){\makebox(0,0)[cc]{$\cdots$}}
\plpen{4.5,0}{bc}{i_2\!-\!1}{tc}{i_2\!-\!1}{\scriptsize}
\plpemv{5,0}{bc}{i_2}{}{}{\scriptsize}
\plpeme{5.5,0}{bc}{i_2\!+\!1}{tc}{i_2}{\scriptsize}
\put(6,0){\makebox(0,0)[cc]{$\cdots$}}
\put(5.5,-1){\makebox(0,0)[cc]{$\cdots$}}
\plpeme{6.5,0}{bc}{r}{tc}{r\!-\!1}{\scriptsize}
}
\end{picture}
\end{array}
\end{eqnarray}
Thus, if $[L']$ is a direct successor of $[L]$ then $\rmN(L,L')$ contains an
element with Bratteli diagram \gref{Gdgrspl} or \gref{Gdgrmrg}. Conversely,
if $\rmN(L,L')$ contains such an element $\Delta$ then $[L]\leq[L']$. 
As noted above,
there exists $D\in\rmSU(n)$, obeying $D^{-1}\,\MJ\, D\subseteq\MJp$, which 
has inclusion matrix $\Delta$. Since the Bratteli diagram of $\Delta$ 
is of the form \gref{Gdgrspl} or \gref{Gdgrmrg}, $[\MJp]$ is a direct 
successor of $[\MJ]$. Thus, $[L']$ is a direct successor of $[L]$. It
follows
\begin{Theorem}\label{Tds}
Let $L,L'\in\rmK(P)$. Then $[L']$ is a direct successor of $[L]$ if and only
if $\rmN(L,L')$ contains an element with Bratteli diagram \gref{Gdgrspl} or
\gref{Gdgrmrg} for some $i_0$ and $i_1<i_2$.
\end{Theorem}
\subsection{Generation of direct successors and direct predecessors}
\label{generation}
In this subsection, we sketch how to derive operations to create the 
direct successors and the direct predecessors of a given element of 
$\hat\rmK(P)$. Again, for a detailed discussion we refer to \cite{RSV:poot},
Sections 5 and 6.
 
In view of Theorem \rref{Tds}, to determine all direct successors of a given
element $[L]$ of $\hat\rmK(P)$, we have to go through all the diagrams
\gref{Gdgrspl} and \gref{Gdgrmrg} and find all $L'$ that obey \gref{Geqx} as
well as the system of
equations \gref{GeqkD}, \gref{GeqmD}, \gref{Geqa} with $L$ being some 
representative of $[L]$ and $\Delta$ being given by the corresponding
diagram. Of course, the amount of work can be reduced by observing that

(i) consideration of one representative $L$ is sufficient,

(ii) diagrams that differ only by a permutation of the lower vertices yield
equivalent $L'$, hence identical direct successors.

\noindent
It follows that the diagrams to be considered are 
\begin{eqnarray}\label{Gspli0}
\unitlength1.5cm
\begin{array}{c}
\begin{picture}(4,2)
\put(0,1.5){
\plpen{0,0}{bc}{1}{tc}{1}{\scriptsize}
\put(0.5,0){\makebox(0,0)[cc]{$\cdots$}}
\put(0.5,-1){\makebox(0,0)[cc]{$\cdots$}}
\plpen{1,0}{bc}{i_0\!-\!1}{tc}{i_0\!-\!1}{\scriptsize}
\plpen{1.5,0}{bc}{i_0}{tc}{i_0}{\scriptsize}
\plpee{1.5,0}{}{}{tc}{i_0\!+\!1~~}{\scriptsize}
\plpee{2,0}{bc}{i_0\!+\!1}{tc}{~~i_0\!+\!2}{\scriptsize}
\put(2.5,0){\makebox(0,0)[cc]{$\cdots$}}
\put(3,-1){\makebox(0,0)[cc]{$\cdots$}}
\plpee{3,0}{bc}{r}{tc}{r\!+\!1}{\scriptsize}
}
\end{picture}
\end{array}
\\ \label{Gmrgi1i2}
\unitlength1.5cm
\begin{array}{c}
\begin{picture}(5,2)
\put(0,1.5){
\plpen{0,0}{bc}{1}{tc}{1}{\scriptsize}
\put(0.5,0){\makebox(0,0)[cc]{$\cdots$}}
\put(0.5,-1){\makebox(0,0)[cc]{$\cdots$}}
\plpen{1,0}{bc}{i_1\!-\!1}{tc}{i_1\!-\!1}{\scriptsize}
\plpen{1.5,0}{bc}{i_1}{tc}{i_1}{\scriptsize}
\plpen{2,0}{bc}{i_1\!+\!1}{tc}{i_1\!+\!1}{\scriptsize}
\put(2.5,0){\makebox(0,0)[cc]{$\cdots$}}
\put(2.5,-1){\makebox(0,0)[cc]{$\cdots$}}
\plpen{3,0}{bc}{i_2\!-\!1}{tc}{i_2\!-\!1}{\scriptsize}
\plpemv{3.5,0}{bc}{i_2}{}{}{\scriptsize}
\plpeme{4,0}{bc}{i_2\!+\!1}{tc}{i_2}{\scriptsize}
\put(4.5,0){\makebox(0,0)[cc]{$\cdots$}}
\put(4,-1){\makebox(0,0)[cc]{$\cdots$}}
\plpeme{5,0}{bc}{r}{tc}{r\!-\!1}{\scriptsize}
}  
\end{picture}
\end{array}
\end{eqnarray}
for arbitrary $i_0$ and $i_1<i_2$, respectively.
Taking this into account it can be easily seen that all necessary $L'$ are
generated from $L$ by the following two kinds of operations:
\vspace{0.2cm}

\noindent
{\it Splitting:} Choose $i_0$ such that $m_{i_0}\neq 1$. Choose a
decomposition $m_{i_0}=m_{{i_0},1}+m_{{i_0},2}$ with strictly positive
integers $m_{{i_0},1},m_{{i_0},2}$. Define $J'=(\bfk',\bfm')$ and $\alpha'$
by
\begin{eqnarray*}\label{Gsplk}
\bfk'
& = &
\left(k_1,\dots, k_{{i_0}-1},k_{i_0},k_{i_0},k_{{i_0}+1},\dots, k_r\right)\,,
\\ \label{Gsplm}
\bfm'
& = &
\left(
m_1,\dots, m_{{i_0}-1},m_{{i_0},1},m_{{i_0},2},m_{{i_0}+1},\dots, m_r
\right)\,,
\\ \label{Gspla}
\alpha'
& = &
\left(\alpha_1,\dots,
\alpha_{i_0-1},\alpha_{i_0},\alpha_{i_0},\alpha_{{i_0}+1},
\dots,\alpha_r\right)\,.
\end{eqnarray*}
Since the greatest common divisor $g'$ of $\bfm'$ divides $g$, we
can furthermore define $\xi' = \varrho_{g'}(\xi)$.
We have to check whether $L'=(J';\alpha',\xi')$ so defined is an element of
$\rmK(P)$. This can be done either by a direct computation or by the
following argument. Due to $\bfk'\cdot\bfm'=n$, $J'\in\rmK(n)$. Moreover,
$L'$ solves the system of equations \gref{GeqkD}, \gref{GeqmD}, \gref{Geqa}
with $\Delta$ being given by the Bratteli diagram \gref{Gspli0}.
Thus, $\SUJ$ is subconjugate to $\SUJp$ by some $D\in\rmSU(n)$ with this
inclusion matrix, and $\alpha'$ and $\xi'$ are the characteristic classes of
the extension $Q_L^D$ of $Q_L$ to structure group $\SUJp$. Hence,
$L'\in\rmK(P)$. 
We say that $L'$ arises from $L$ by a splitting of the $i_0$th member.
\vspace{0.2cm}

\noindent
{\it Merging:}
Choose $i_1<i_2$ such that $m_{i_1}=m_{i_2}$. Define $J' =
(\bfk',\bfm')$ and $\alpha'$ by 
\begin{eqnarray*}\label{Gmrgk}
\bfk'
& = &
(k_1,\dots, k_{i_1-1},k_{i_1}+k_{i_2},k_{i_1+1},\dots, \widehat{k_{i_2}}
,\dots, k_r)\,,
\\ \label{Gmrgm}
\bfm'
& = &
(m_1,\dots, m_{i_1-1},m_{i_1},m_{i_1+1},\dots, 
\widehat{m_{i_2}},\dots, m_r)\,,
\\ \label{Gmrga}
\alpha'
& = &
(\alpha_1,\dots,\alpha_{i_1-1},\alpha_{i_1}\smile\alpha_{i_2},
\alpha_{i_1+1},\dots,\widehat{\alpha_{i_2}},\dots,\alpha_r)\,,
\end{eqnarray*}
where $\widehat{\phantom{m_i}}$ indicates that the entry is omitted, as well
as $\xi' = \xi$.
To check that $L'=(J';\alpha',\xi')\in\rmK(P)$ we proceed analogously to the
case of splitting.
We say that $L'$ arises from $L$ by merging the $i_1$th and the $i_2$th
member.
\vspace{0.2cm}

We remark that it may happen that for certain elements of $\rmK(P)$ no
splittings or no mergings can be applied. Amongst these elements are, for
example, those with $m_1=\cdots=m_r=1$ (no splitting) and those having
pairwise distinct $m_i$ (no merging).

Next, we derive operations to create the direct predecessors of $[L]$. 
Direct predecessors are necessary to construct 
$\hat{\rmK}(P)$ from the unique maximal element (which is given by 
$P$ itself). Note that predecessors correspond to strata of higher 
symmetry. Similar to the situation above, in view of Theorem \rref{Tds}, we
have to go through all the diagrams \gref{Gdgrspl} and \gref{Gdgrmrg} and
find all $L'\in\rmK(P)$ that obey \gref{Geqx} and the system of equations
\gref{GeqkD}, \gref{GeqmD}, \gref{Geqa} -- where $L$ and $L'$ have to
be interchanged -- with $L$ being a representative of $[L]$ and $\Delta$ being 
given by the corresponding diagram. Again, we can
reduce this work by noting that it suffices to consider a fixed 
representative $L$ and by ignoring permutations, now of the upper vertices.
The remaining diagrams to be considered are 
\begin{eqnarray} \label{Ginvspli}
\unitlength1.5cm
\begin{array}{c}
\begin{picture}(5,2)
\put(0,1.5){
\plpen{0,0}{bc}{1}{tc}{1}{\scriptsize}
\put(0.5,0){\makebox(0,0)[cc]{$\cdots$}}
\put(0.5,-1){\makebox(0,0)[cc]{$\cdots$}}
\plpen{1,0}{bc}{i_1\!-\!1}{tc}{i_1\!-\!1}{\scriptsize}
\plpen{1.5,0}{bc}{i_1}{tc}{i_1}{\scriptsize}
\plpev{1.5,0}{}{}{tc}{i_2}{\scriptsize}
\plpen{2,0}{bc}{i_1\!+\!1}{tc}{i_1\!+\!1}{\scriptsize}
\put(2.5,0){\makebox(0,0)[cc]{$\cdots$}}
\put(2.5,-1){\makebox(0,0)[cc]{$\cdots$}}
\plpen{3,0}{bc}{i_2\!-\!1}{tc}{i_2\!-\!1}{\scriptsize}
\plpee{3.5,0}{bc}{i_2}{tc}{i_2\!+\!1}{\scriptsize}
\put(4,0){\makebox(0,0)[cc]{$\cdots$}}
\put(4.5,-1){\makebox(0,0)[cc]{$\cdots$}}
\plpee{4.5,0}{bc}{r}{tc}{r\!+\!1}{\scriptsize}
}
\end{picture}
\end{array}
\\ \label{Ginvmrg}
\unitlength1.5cm
\begin{array}{c}
\begin{picture}(3.5,2)
\put(0,1.5){
\plpen{0,0}{bc}{1}{tc}{1}{\scriptsize}
\put(0.5,0){\makebox(0,0)[cc]{$\cdots$}}
\put(0.5,-1){\makebox(0,0)[cc]{$\cdots$}}
\plpen{1,0}{bc}{i_0\!-\!1}{tc}{i_0\!-\!1}{\scriptsize}
\plpen{1.5,0}{bc}{i_0}{tc}{i_0}{\scriptsize}
\plpeme{2,0}{bc}{i_0\!+\!1~~}{}{}{\scriptsize}
\plpeme{2.5,0}{bc}{~~i_0\!+\!2}{tc}{i_0\!+\!1}{\scriptsize}
\put(3,0){\makebox(0,0)[cc]{$\cdots$}}
\put(2.5,-1){\makebox(0,0)[cc]{$\cdots$}}
\plpeme{3.5,0}{bc}{r}{tc}{r\!-\!1}{\scriptsize}
}
\end{picture}
\end{array}
\end{eqnarray}
with arbitrary $i_1<i_2$ and $i_0$, respectively. One can check that all 
necessary $L'$ are obtained by the following two kinds of operations, applied 
to $L$:
\vspace{0.2cm}

\noindent
{\it Inverse splitting:}
Choose $i_1<i_2$ such that $k_{i_1}=k_{i_2}$ and
$\alpha_{i_1}=\alpha_{i_2}$. Define $J'=(\bfk',\bfm')$ and $\alpha'$ by 
\begin{eqnarray*}\nonumber
\bfk'
& = &
(
k_1
,\dots,
k_{i_1-1},k_{i_1},k_{i_1+1}
,\dots,
\widehat{k_{i_2}}
,\dots,
k_r
)\,,
\\ 
\bfm'
& = &
(
m_1
,\dots,
m_{i_1-1},m_{i_1}+m_{i_2},m_{i_1+1}
,\dots,
\widehat{m_{i_2}}
,\dots,
m_r
)\,,
\\ \nonumber
\alpha'
& = &
(
\alpha_1
,\dots,
\alpha_{i_1-1},\alpha_{i_1},\alpha_{i_1+1}
,\dots,
\widehat{\alpha_{i_2}}
,\dots,
\alpha_r
)\,.
\end{eqnarray*}
Then $g$ divides the greatest common divisor $g'$ of $\bfm'$, so that 
$\varrho_g$ is well-defined. Choose
$\xi'\in H^1(M,\ZZ_{g'})$ such that
$\xi=\varrho_g(\xi')$ and 
$
\beta_{g'}(\xi') = E_{\twbfm'}^{(2)}(\alpha')
$.
By construction, $L'=(J';\alpha',\xi')$ is an element of $\rmK(P)$.
We say that it arises from $L$ by an inverse splitting of the $i_1$th
and the $i_2$th member.
\vspace{0.2cm}

\noindent
{\it Inverse Merging:}
Choose $i_0$ such that $k_{i_0}\neq 1$. Choose a decomposition
$k_{i_0}=k_{i_0,1}+k_{i_0,2}$ with strictly positive integers 
$k_{i_0,1}, k_{i_0,2}$. Choose cohomology elements
$\alpha_{i_0,1},\alpha_{i_0,2}\in H^\rmeven(M,\ZZ)$ such that
${\alpha}_{i_0,l}^{(2j)}=0$ for $j>k_{i_0,l}$, $l=1,2$, and
$\alpha_{i_0,1}\alpha_{i_0,2}=\alpha_{i_0}$.
Define $J'=(\bfk',\bfm')$ and $\alpha'$ by 
\begin{eqnarray*}\nonumber
\bfk'
& = &
(
k_1,\dots,k_{i_0-1},k_{i_0,1},k_{i_0,2},k_{i_0+1},\dots,k_r
)\,,
\\ \nonumber
\bfm'
& = &
(
m_1,\dots,m_{i_0-1},m_{i_0},m_{i_0},m_{i_0+1},\dots,m_r
)\,,
\\ \nonumber
\alpha'
& = &
(
\alpha_1
,\dots,
\alpha_{i_0-1},\alpha_{i_0,1},\alpha_{i_0,2},\alpha_{i_0+1}
,\dots,
\alpha_r
)\,,
\end{eqnarray*}
and $\xi'=\xi$. Again, by construction, $L' = (J';\alpha',\xi')\in\rmK(P)$.
We say that $L'$ arises from $L$ by an inverse merging of the $i_0$th
member.
\vspace{0.2cm}

\noindent
Let us summarize.
\begin{Theorem}\label{Tdsgen}
Let $[L]\in\hat{\rmK}(P)$ and let $L$ be a representative. The direct 
successors (predecessors) of $[L]$ are obtained by applying all possible 
splittings and mergings (inverse splittings and inverse mergings) to $L$ and 
passing to equivalence classes. 
\end{Theorem}
\subsection{Examples}
\label{examples}
In this subsection, let $P$ be a principal $\rmSU(4)$-bundle. 
\vspace{0.3cm}

\noindent
{\it Example} 1. Direct successors of $[L]$ for $J=(1,1|2,2)$. 
(Recall the notation from Subsection \rref{SSex}.) Note that $\alpha$ has 
components $\alpha_i=1+\alpha_i^{(2)}$, $i=1,2$. 
Let us start with splitting operations. For $i_0=1$, the only possible
splitting is given by the decomposition $m_1=2=1+1$. It yields
$L_a'=(J_a';\alpha_a',\xi_a')$, where 
$J_a'=(1,1,1|1,1,2)$, $\alpha_a'=(\alpha_1,\alpha_1,\alpha_2)$, and 
$\xi_a'=0$. The passage from $L$ to $L_a'$ can very easily performed 
on the level of a Bratteli diagram whose vertices are labelled by the 
respective 
quantities $k_i,m_i$ and $\alpha_i$ (rather than by the mere number $i$):
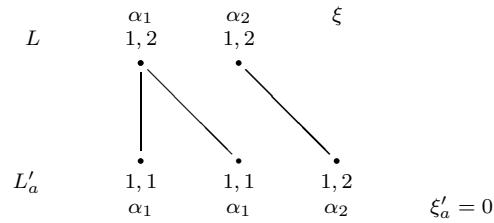
\begin{center}
\unitlength1.3cm
\begin{picture}(2.7,2.6)
\put(0,1.8){
\plpaen{0,0}{br}{L\hspace{1.1cm}}{}{tr}{L_a'\hspace{1.1cm}}{}{
   \scriptsize}
\plpaen{0,0}{bc}{1,2}{\alpha_1}{tc}{1,1}{\alpha_1}{\scriptsize}
\plpaez{0,0}{}{}{}{tc}{1,1}{\alpha_1}{\scriptsize}
\plpaez{1,0}{bc}{1,2}{\alpha_2}{tc}{1,2}{\alpha_2}{\scriptsize}
\plpaez{1,0}{bl}{\phantom{(1,2)}}{\hspace*{1cm}\xi}{
   tl}{\phantom{1,2}}{\hspace*{1cm}\xi_a'=0}{\scriptsize}
}
\end{picture}
\end{center}
For $i_0=2$, a similar splitting operation creates $L_b'$, given by
the labelled Bratteli diagram
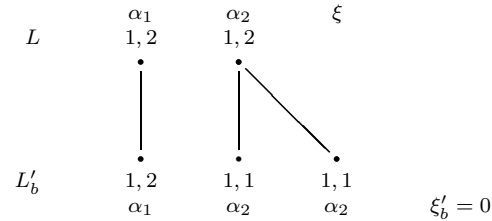
\begin{center}
\unitlength1.3cm
\begin{picture}(2.7,2.6)
\put(0,1.8){
\plpaen{0,0}{br}{L\hspace{1.1cm}}{}{tr}{L_b'\hspace{1.1cm}}{}{
   \scriptsize}
\plpaen{0,0}{bc}{1,2}{\alpha_1}{tc}{1,2}{\alpha_1}{\scriptsize}
\plpaen{1,0}{bc}{1,2}{\alpha_2}{tc}{1,1}{\alpha_2}{\scriptsize}
\plpaez{1,0}{}{}{}{tc}{1,1}{\alpha_2}{\scriptsize}
\plpaez{1,0}{bl}{\phantom{(1,2)}}{\hspace*{1cm}\xi}{
   tl}{\phantom{1,2}}{\hspace*{1cm}\xi_b'=0}{\scriptsize}
}
\end{picture}
\end{center}
As for merging operations, the only choice for $i_1$, $i_2$ is $i_1=1$,
$i_2=2$. This yields $L_c'$:
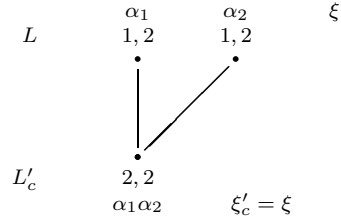
\begin{center}
\unitlength1.3cm
\begin{picture}(2.7,2.6)
\put(0,1.8){
\plpaen{0,0}{br}{L\hspace{1.1cm}}{}{tr}{L_c'\hspace{1.1cm}}{}{
   \scriptsize}
\plpaen{0,0}{bc}{1,2}{\alpha_1}{tc}{2,2}{\alpha_1\smile\alpha_2}{\scriptsize}
\plpaemz{1,0}{bc}{1,2}{\alpha_2}{}{}{}{\scriptsize}
\plpaemz{1,0}{bl}{\phantom{(1,2)}}{\hspace*{1cm}\xi}{
   tl}{\phantom{1,2}}{\hspace*{1cm}\xi_c'=\xi}{\scriptsize}
}
\end{picture}
\end{center}
Next, we have to pass to equivalence classes. Generically, $L_a'$, $L_b'$,
$L_c'$ generate their own classes. However, while $L_c'$ can never be
equivalent to $L_a'$ or $L_b'$, the latter are equivalent iff
$\alpha_1=\alpha_2$. In order to see for which bundle classes $P$ this can 
happen, consider equations \gref{GKMJ} and \gref{GKPJ}. The first one requires
$\alpha_1^{(2)}=\alpha_2^{(2)}$ to be a torsion element. Then, due to
$\alpha_1^{(4)}=\alpha_2^{(4)}=0$, the second one implies $c_2(P)=0$.
Thus, $L_a'$ and $L_b'$ can be (occasionally) equivalent only if
$P$ is trivial.
\vspace{0.3cm}

\noindent
{\it Example} 2. Direct predecessors of $[L]$ for $J=(1,1|2,2)$. 
Inverse splittings can be applied only if $\alpha_1=\alpha_2$. In this case, 
for any solution $\xi\in H^1(M,\ZZ_4)$ of the system of equations
\begin{eqnarray}\label{Gexinvmrg1}
\xi'\mod\, 2  & = & \xi\,,
\\ \label{Gexinvmrg2}
\beta_4(\xi') & = & \alpha_1^{(2)}\,,
\end{eqnarray}
we obtain an element $L'=(J';\alpha',\xi')$, where
$J'=(1|4)$ and $\alpha'=\alpha_1=\alpha_2$. The passage from
$L$ to $L'$ can be summarized in the labelled Bratteli diagram
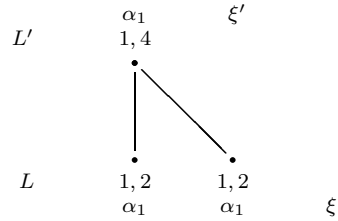
\begin{center}
\unitlength1.3cm
\begin{picture}(2.7,2.6)
\put(0,1.8){
\plpaen{0,0}{br}{L'\hspace{1.1cm}}{}{tr}{L\hspace{1.1cm}}{}{
   \scriptsize}
\plpaen{0,0}{bc}{1,4}{\alpha_1}{tc}{1,2}{\alpha_1}{\scriptsize}
\plpaez{0,0}{}{}{}{tc}{1,2}{\alpha_1}{\scriptsize}
\plpaez{0,0}{bl}{\phantom{(1,2)}}{\hspace*{1cm}\xi'}{
   tl}{\phantom{1,2}}{\hspace*{1cm}\xi}{\scriptsize}
}
\end{picture}
\end{center}
that has to be read upwards.
Each $L'$ generates its own equivalence class. Due to
$k_1=k_2=1$, inverse mergings cannot be applied to $L$. Thus, in the case
$\alpha_1=\alpha_2$ the direct predecessors of the equivalence class of $L$
are labelled by the solutions of equations \gref{Gexinvmrg1} and
\gref{Gexinvmrg2}, whereas in the case $\alpha_1\neq\alpha_2$ direct
predecessors do not exist. Recall that the
first case can only occur if $P$ is trivial.
\vspace{0.3cm}

\noindent
{\it Example} 3. Direct predecessors of $[L]$ for $J=(2|2)$. 
Here $\alpha=1+\alpha^{(2)}+\alpha^{(4)}$. 
Inverse mergings can be applied and yield elements $L'$ as follows:
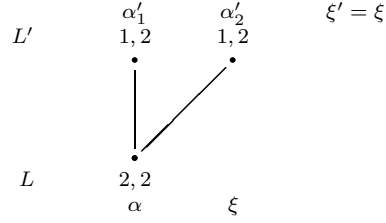
\begin{center}
\unitlength1.3cm
\begin{picture}(2.7,2.6)
\put(0,1.8){
\plpaen{0,0}{br}{L'\hspace{1.1cm}}{}{tr}{
   L\hspace{1.1cm}}{}{\scriptsize}
\plpaen{0,0}{bc}{1,2}{\alpha_1'}{tc}{2,2}{\alpha}{\scriptsize}
\plpaemz{1,0}{bc}{1,2}{\alpha_2'}{}{}{}{\scriptsize}
\plpaemz{1,0}{bl}{\phantom{(1,2)}}{\hspace*{1cm}\xi'=\xi}{
   tl}{\phantom{1,2}}{\hspace*{1cm}\xi}{\scriptsize}
}
\end{picture}
\end{center}
Here $\alpha_i' = 1 + {\alpha_i'}^{(2)}$, $i=1,2$, such 
that $\alpha_1'\alpha_2' = \alpha$. When passing to 
equivalence classes, elements $L'$ with 
$(\alpha_1',\alpha_2')$ 
and 
$(\alpha_2',\alpha_1')$ have to be identified.
Since $L$ does not allow for inverse splittings, there are no more direct 
predecessors.
%
%
%

\section{Application}
\subsection{The stratification for $\rmSU(2)$}
\label{Sex}


In Subsection \rref{SSex} we have discussed particular examples of orbit types.
In the present section we explain how to construct the Hasse diagram of the 
whole set of orbit types, starting from its maximal element. We restrict our 
attention to the simplest nontrivial case, the gauge group $\rmSU(2)$.
We start with simple examples of base manifolds, for which the orbit types are
known and proceed to more complicated ones, like lens spaces. 
This is intended to illustrate the technique. On the other hand, the means provided in 
Subsection \ref{generation} enable us to construct the Hasse diagram for any 
$\rmSU(n) \, .$ For $\rmSU(4) \, ,$ this was partially demonstrated in Subsection 
\ref{examples}. However, to present full Hasse diagrams for $\rmSU(4) \, ,$ or any other 
$\rmSU(n) \, ,$ in a transparent way needs some special graphical effort.

Let $L^\rmp$ denote the unique representative of the maximal element of 
$\hat\rmK(P)$. Since $Q_{L^\rmp} = P$, $L^\rmp$ is given by 
$J^\rmp=(2|1)$, $\alpha^\rmp = c(P)$, and $\xi^\rmp = 0$. Inverse
mergings yield elements $L$:

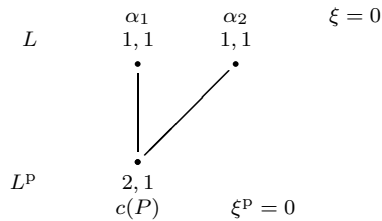
\begin{center}
\unitlength1.3cm
\begin{picture}(2.7,2.6)
\put(0,1.8){
\plpaen{0,0}{br}{L\hspace{1.1cm}}{}{tr}{
   L^\rmp\hspace{1.1cm}}{}{\scriptsize}
\plpaen{0,0}{bc}{1,1}{\alpha_1}{tc}{2,1}{c(P)}{\scriptsize}
\plpaemz{1,0}{bc}{1,1}{\alpha_2}{}{}{}{\scriptsize}
\plpaemz{1,0}{bl}{\phantom{(1,2)}}{\hspace*{1cm}\xi=0}{
   tl}{\phantom{1,2}}{\hspace*{1cm}\xi^\rmp=0}{\scriptsize}
}
\end{picture}
\end{center}
where $\alpha_i=1+\alpha_i^{(2)}$ such that
$\alpha_1\alpha_2 = c(P)$. Sorting by degree yields the 
equations 
$
\alpha_1^{(2)}
+
\alpha_2^{(2)}
=
0
$
and
$
\alpha_1^{(2)} \alpha_2^{(2)}
=
c_2(P)
$.
We obtain
$
\alpha_2^{(2)}=-\alpha_1^{(2)}
$
and 
\begin{equation}\label{Gex1}
-\left(\alpha_1^{(2)}\right)^2 = c_2(P)\,. 
\end{equation}
The solutions $\alpha_1^{(2)}$ and $-\alpha_1^{(2)}$ yield equivalent direct
predecessors. 
We note that the Howe subgroup labelled by $J=(1,1|1,1)$ is the toral 
subgroup $\rmU(1)$ of $\rmSU(2)$ and that $\alpha_1^{(2)}$ is 
just the first Chern class of the corresponding reduction of $P$. 
By virtue of 
this transliteration, equation \gref{Gex1} is consistent with the literature
\cite{Isham}.

Next, we determine the direct predecessors of the classes generated by
$L$. Inverse mergings cannot be applied. Inverse splittings can be applied 
provided $\alpha_1 = \alpha_2$, i.e., $2\alpha_1^{(2)}=0$. Then, for any 
solution $\xi'\in H^1(M,\ZZ_2)$ of the equation
\begin{equation}\label{Gex2}
\beta_2\left(\xi'\right)
=
\alpha_1^{(2)}\,,
\end{equation}
inverse merging yields an element $L'$ by
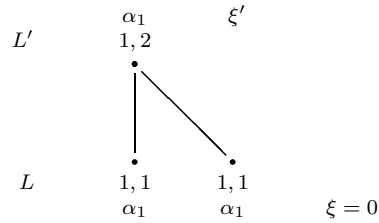
\begin{center}
\unitlength1.3cm
\begin{picture}(2.7,2.6)
\put(0,1.8){
\plpaen{0,0}{br}{L'\hspace{1.1cm}}{}{tr}{
   L\hspace{1.1cm}}{}{\scriptsize}
\plpaen{0,0}{bc}{1,2}{\alpha_1}{tc}{1,1}{\alpha_1}{\scriptsize}
\plpaez{0,0}{}{}{}{tc}{1,1}{\alpha_1}{\scriptsize}
\plpaez{0,0}{bl}{\phantom{1,2}}{\hspace*{1cm}\xi'}{
   tl}{\phantom{1,2}}{\hspace*{1cm}\xi=0}{\scriptsize}
}
\end{picture}
\end{center}
Each of these elements generates its own equivalence class. 
Recall that $J=(1|2)$ labels the center $\ZZ_2$ of $\rmSU(2)$ and that 
$\xi'$ is the natural characteristic class for principal 
$\ZZ_2$-bundles over $M$. 

Now let us draw Hasse diagrams of $\hat{\rmK}(P)$ for base manifolds
$M=\sphere{4}$, $\sphere{2}\times\sphere{2}$,
$\lens{2p}{3}
\times\sphere{1}$. In the following, vertices stand for the elements of
$\hat{\rmK}(P)$ and edges indicate the relation 'left vertex 
$\leq$ right vertex'. When viewing the elements of $\hat{\rmK}(P)$ 
as Howe subbundles, the vertex on the rhs.~represents the class corresponding 
to $P$ itself, the vertices in the middle and on the
lhs.~represent reductions of $P$ to the Howe subgroups $\rmU(1)$ and
$\ZZ_2$, respectively. When viewing the elements of $\hat{\rmK}(P)$ as orbit
types, or strata of the gauge orbit space, the vertex on the
rhs.~represents the generic stratum, whereas the vertices in the
middle and on the lhs.~represent $\rmU(1)$-strata and $\rmSU(2)$-strata
(the names refer to the isomorphy type of the corresponding 
stabilizer).
\\
\\
{\it Example} 1. $M=\sphere{4}$.~
If $c_2(P)=0$, 
equation \gref{Gex1} is
trivially satisfied by $\alpha_1^{(2)}=0$. Then equation \gref{Gex2} is
trivially satisfied by $\xi'=0$. Due to
$H^1(M,\ZZ_2)=0$ and $H^2(M,\ZZ)=0$, there are no more solutions for either
one. Thus, in the case where $P$ is trivial, the Hasse diagram of 
$\hat{\rmK}(P)$ is 
\begin{center}
\unitlength1.3cm
\begin{picture}(2,1.05)
\put(0,0.55){
\plene{0,0}{bc}{}
\plene{1,0}{bc}{}
\whole{2,0}{bc}{}
}
\end{picture}
\end{center}
If $P$ is nontrivial, $\hat{\rmK}(P)$ is trivial, i.e., it consists only of 
the class corresponding to $P$ itself. 

On the level of strata, the result means that in the sector of 
vanishing 
topological charge the gauge orbit space decomposes into the generic stratum, 
a $\rmU(1)$-stratum, and a $\rmSU(2)$-stratum. If, on the other hand, a 
topological charge is present, only the generic stratum survives.
\\
\\
{\it Example} 2. $M=\sphere{2}\times\sphere{2}$.~
Using the notation introduced in Example 3 (iii) of Subsection \rref{SSex},
equation \gref{Gex1} becomes 
$
-2ab~\gamma_{\sphere{2}}^{(2)}\!\times\!\gamma_{\sphere{2}}^{(2)}
=
c_2(P)\,.
$~
The discussion is similar to that of equation \gref{GS2S2ex}. Due
to $H^1(M,\ZZ_2)=0$, only the solution $a=b=0$ has a direct 
predecessor itself. Thus, in the case $c_2(P)=0$ the Hasse diagram of 
$\hat{\rmK}(P)$ is
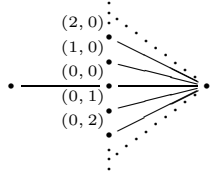
\begin{center}
\unitlength1.3cm
\begin{picture}(2,2.55)
\put(0,1.3){
\plene{0,0}{bc}{}
\pslvmde{1,0.75}
\vpunkte{1,0.75}
\plzmee{1,0.5}{}{}
\marke{1.01,0.46}{br}{(2,0)}
\plvmee{1,0.25}{}{}
\marke{1.01,0.21}{br}{(1,0)}
\plene{1,0}{}{}
\marke{1.01,-0.04}{br}{(0,0)}
\plvee{1,-0.25}{}{}
\marke{1.01,-0.29}{br}{(0,1)}
\plzee{1,-0.5}{}{}
\marke{1.01,-0.54}{br}{(0,2)}
\vpunkte{1,-0.75}
\pslvde{1,-0.75}
\whole{2,0}{bc}{}
}
\end{picture}
\end{center}
The vertices in the middle are labelled by the corresponding values of 
$(a,b)$. Note that passage to equivalence classes requires identification of 
solutions $(a,b)$ and $(-a,-b)$. In the case
$c_2(P)=2l~\gamma_{\sphere{2}}^{(2)}\!\times\!\gamma_{\sphere{2}}^{(2)}$, 
the Hasse diagram is
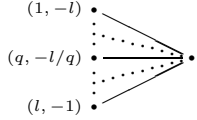
\begin{center}
\unitlength1.3cm
\begin{picture}(2,2.05)
\put(0,1.05){
\plzmee{1,0.5}{cr}{(1,-l)~}
\pslvmee{1,0.25}
\vpunkte{1,0.25}
\plene{1,0}{cr}{(q,-l/q)~}   
\vpunkte{1,-0.25}
\pslvee{1,-0.25}
\plzee{1,-0.5}{cr}{(l,-1)~}
\whole{2,0}{bc}{}
}
\end{picture}
\end{center}
where, due to the identification $(a,b)\sim(-a,-b)$, $q$ runs through the 
positive divisors of $l$ only. 
Finally, in the case 
$c_2(P)=(2l+1)\,\gamma_{\sphere{2}}^{(2)}\times\gamma_{\sphere{2}}^{(2)}$, 
$\hat{\rmK}(P)$ is trivial. 

The interpretation of the result in terms of strata of the gauge 
orbit space is similar to that for space time manifold $M=\sphere{4}$ above.
\\
\\
{\it Example} 3. $M=\lens{2p}{3}\times\sphere{1}$.~
Recall the notation from Subsection \rref{SSex}. We write
\begin{equation}\label{Gexdecompa}
\alpha_1^{(2)}=a\,\gamma_{\lens{2p}{3};\ZZ}^{(2)}\times 1_{\sphere{1}}\,.
\end{equation}
Due to $H^2(\lens{2p}{3},\ZZ)\cong\ZZ_{2p}$, $(\alpha_1^{(2)})^2 = 0$. 
Hence,
equation \gref{Gex1} is solvable iff $c_2(P)=0$, in which case the solutions are
given by $a\in\ZZ_{2p}$. Since when passing to equivalence classes we have
to identify solutions $a$ and $-a$, the direct predecessors are labelled by 
elements of $\ZZ_p$. 

Next, decomposing
~$
\xi'
=
\xi'_\rmL\,\gamma_{\rmL^3_{2p};\ZZ_2}^{(1)}\times 1_{\sphere{1}}
+
\xi'_\rmS\, 1_{\rmL^3_{2p};\ZZ_2}\times\gamma_{\sphere{1}}^{(1)}
$~
and using \gref{GbgLdpS}, equation \gref{Gex2} becomes ~$p\,\xi'_\rmL=a\,.$~
Thus, only the elements labelled by $a=0$ and $a=p$ have direct
predecessors. These are given by the values 
$\xi'_\rmL=0$, $\xi'_\rmS=0,1$ and
$\xi'_\rmL=1$, $\xi'_\rmS=0,1$, respectively. 
As a result, in the case $c_2(P)=0$, the Hasse diagram of $\hat{\rmK}(P)$ is 
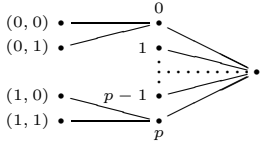
\begin{center}
\unitlength1.3cm
\begin{picture}(2,2.05)
\put(0,1.05){
\plene{0,0.5}{cr}{(0,0)~}
\plvee{0,0.25}{cr}{(0,1)~}
\plvmee{0,-0.25}{cr}{(1,0)~}
\plene{0,-0.5}{cr}{(1,1)~}
\plzmee{1,0.5}{bc}{0}  
\plvmee{1,0.25}{cr}{1~}
\vpunkte{1,0}
\pslene{1,0}
\plvee{1,-0.25}{cr}{p-1~}
\plzee{1,-0.5}{tc}{p}  
\whole{2,0}{bc}{}
}
\end{picture}
\end{center}
Here the vertices on the lhs.~are labelled by $(\xi'_\rmL,\xi'_\rmS)$, whereas
those in the middle are labelled by $a$. In the case $c_2(P)\neq 0$, 
$\hat{\rmK}(P)$ is trivial.  Again, the interpretation
in terms of strata of the gauge orbit space goes along the lines of Example
1 above.


\subsection
{Kinematical quantum nodes in Yang-Mills-Chern-Simons theory}
\label{SNodes}

%
%
Following \cite{AsoreyMitter:CS}, we consider gauge theory 
on the trivial bundle $\twP = (\Sigma\times\RR)\times\rmSU(n)$,
where $\Sigma$ is a Riemann surface,
in the Hamiltonian approach. The action functional consists of the
Yang-Mills and the Chern-Simons term,
$$
S(\twA)
~=~
\frac{1}{2} 
\int_{\Sigma\times\RR} \tr\left(\twF_{\twA}\wedge\ast\twF_{\twA}\right)
~+~
\frac{\lambda}{8\pi} 
\int_{\Sigma\times\RR}\tr\left(\twA\wedge\twF_{\twA} 
-
\frac{2}{3}\twA\wedge\twA\wedge\twA\right)
$$
where $\tw A\in\tw\con$, the space of $W^k$-connections in $\twP$, and
$\twF_{\tw A}$ denotes the curvature of $\twA$. The coupling 
$\lambda$ takes integer values.
By separating the time variable, we get the following Lagrangian
\begin{eqnarray*}
L(A,A_0,\dot A,\dot A_0)
& ~=~ &
\frac{1}{2}
\left(
\dot A\!\!-\!\!\nabla_AA_0
~,~
\dot A\!\!-\!\!\nabla_AA_0
\right)_0
-
\frac{1}{2}
\left(
F_A,F_A
\right)_0
\\&&
+
\frac{\lambda}{4\pi}
\left\{
2\left(A_0,\ast F_A\right)_0
+
\left(A,\ast \dot A\right)_0
\right\}\,.
\end{eqnarray*}
Here, $A_0 \in W^k(M,\rmsu(n)) \, ,$  $A$ is a $W^k$-connection form in the
trivial bundle $P = \Sigma\times\rmSU(n)$ and $(\cdot,\cdot)_0$ 
denotes the $L^2$-scalar product of $\rmsu(n)$-valued forms on $M$.
As usual, we denote the space of $W^k$-connections in $P$ by $\con \, .$ 
Constraint analysis yields the Gau{\ss} law
\begin{eqnarray*}
\nabla_A^\ast\Pi
-
\frac{\lambda}{4\pi}\ast\diff A
= 0\,,
\end{eqnarray*}
where $\Pi$ denotes the momentum conjugate to $A$. Performing canonical
quantization one finds that physical states are given by functions
$\psi:\con\rightarrow\CC$ that are contained in the kernel of the Gau{\ss} law 
operator
\begin{equation} \label{GGauss}
\nabla_{\hat A}^\ast\frac{\delta}{\delta A}
-
\frac{i\lambda}{4\pi}\ast\diff {\hat A}\,,
\end{equation}
where $\hat A$ means the multiplication operator, i.e., $(\hat A\psi)(A') =
A'\psi(A')$, $\forall A'\in\con$. On the physical states, the Hamiltonian 
is given by
\begin{eqnarray*}
H
=
-\frac{1}{2}
\left(
\frac{\delta}{\delta A} + i\frac{\lambda}{4\pi}\ast A,
\frac{\delta}{\delta A} + i\frac{\lambda}{4\pi}\ast A
\right)_0
+
\frac{1}{2}
\left(F_A,F_A\right)_0\,.
\end{eqnarray*}
Let us consider connections $A\in\con$ that can be reduced to some subbundle 
of $P$ with nontrivial first Chern class. That is, in physics language, $A$ 
carries a nontrivial magnetic charge. Thus, it may be viewed as monopole-like,
although it is not assumed to be a solution of the field equations.
In \cite{Asorey:Nodes} it was shown that if the Chern-Simons term is
present, i.e., $\lambda\neq 0$, then $\psi(A)=0$ for any such $A$ and any
physical state $\psi$. Therefore, such $A$ are called {\it kinematical}
quantum nodes. 
Note that by geometric reasons there also exist dynamical nodes which differ 
from state to state. Due to their monopole-like character, kinematical quantum 
nodes are expected to play a role in the confinement mechanism. 
In the following we shall show that being a node is a
property of strata. For that purpose, we reformulate the result of
\cite{Asorey:Nodes} in our language.
\begin{Theorem} \label{TNodes}
Let $A\in\con$ have orbit type $[(J;\alpha,\xi)]\in\hat\rmK(P)$. 
If $\alpha_i^{(2)}\neq 0$ for some $i$ then $A$
is a kinematical quantum node, i.e., $\psi(A)=0$ for all physical states 
$\psi$.
\end{Theorem}
We outline the proof, following \cite{Asorey:Nodes}. Let $L=(J;\alpha,\xi)$. 
Since $\Sigma$ is a compact
orientable $2$-manifold, $H^2(\Sigma,\ZZ)=\ZZ$. Let $\gamma^{(2)}$ be a
generator. Then $\alpha_i^{(2)}=c_i\gamma^{(2)}$ for certain $c_i\in\ZZ$.
Consider the following element of $\rmu(n)$:
$$
\twphi
:=
i\left[\left(\frac{c_1}{k_1}~\II_{k_1}\otimes\II_{m_1}\right)
\oplus\cdots\oplus
\left(\frac{c_r}{k_r}~\II_{k_r}\otimes\II_{m_r}\right)\right]\,.
$$
Due to $(\alpha,\xi)\in\rmK(P,J)$,
$
(m_1c_1+\cdots+m_rc_r)\gamma^{(2)}
=
E_{\bfm}^{(2)}(\alpha)
=
0\,.
$~
It follows $\tr(\twphi) = 0$, hence $\twphi\in\rmsu(n)$. By construction,
$\twphi$ is invariant under the adjoint action of the subgroup
$\SUJ\subseteq\rmSU(n)$. Thus, we can define an equivariant function
$\phi:P\rightarrow\rmsu(n)$ by assigning to any $q\in Q_L$ the constant value
$\twphi$ and extending equivariantly to $P$. By construction, 
$\nabla_A\phi = 0$. Consequently, for any state $\psi:\con\rightarrow\CC$,
$$
\left(
\phi,\left(\nabla_{\hat A}^\ast\frac{\delta}{\delta A}\psi\right)(A)
\right)_0
=
\left(
\phi,\nabla_A^\ast\left\{\left(\frac{\delta}{\delta A}\psi\right)(A)\right\}
\right)_0
=
0\,.
$$
For physical states, the Gauss law implies
\begin{equation} \label{GAso2}
\left(
\phi,\left(\ast\diff \hat A\psi\right)(A)
\right)_0
=
\left(
\phi,\ast\diff A
\right)_0
\psi(A)
=
0\,.
\end{equation}
Using $\nabla_A\phi=0$ and the structure equation
$F_A = \diff A+\frac{1}{2}[A,A]$, we obtain
\begin{equation} \label{GAso3}
\left(
\phi,\ast\diff A
\right)_0
\psi(A)
=
2\left(
\phi,\ast F_A
\right)_0
\psi(A)
\end{equation}
Since $A$ is reducible to $Q_L$ (recall that $Q_L$ contains a holonomy bundle
of $A$), $F_A$ has block
structure $\left((F_A)_1\otimes\II_{m_1}\right)
\oplus\cdots\oplus
\left((F_A)_r\otimes\II_{m_r}\right)$
with $(F_A)_j$ being $(k_j\times k_j)$-matrices. Thus, by construction of 
$\phi$,
\begin{equation} \label{GAso4}
\left(
\phi,\ast F_A
\right)_0
~=~
\int_\Sigma\Tr\left(\phi F_A\right)
~=~
i\sum_{j=1}^r ~\frac{m_j}{k_j}\, c_j~
\int_{\Sigma} \Tr ((F_A)_j)\,.
\end{equation}
Since $c_j$ are the first Chern classes of the Whitney factors of the
extension of $Q_L$ to structure group
$\UJ\cong\rmU(k_1)\times\cdots\times\rmU(k_r)$, the integrals on the 
r.h.s.~give $-2\pi i c_j$.
Thus, equations \gref{GAso2}--\gref{GAso4} imply
\begin{equation}\label{GAsoend}
\sum_{j=1}^r ~\frac{m_j}{k_j}\,c_j^2~\psi(A)
= 0\,.
\end{equation}
It follows that if one of the $c_j$ is nonzero then $\psi(A)=0$,
for all physical states $\psi$. 
\vspace{0.3cm}

\noindent
{\bf Remark:} Let us compare \gref{GAsoend} with formula $(6)$ in 
\cite{Asorey:Nodes}. Define $k'_i=k_im_i$ and $m'_i=1$.
Then $J'=(\bfk',\bfm')\in\rmK(n)$ and $\UJ\subseteq\UJp$. Let $Q_L'$ denote
the extension of $Q_L$ to structure group $\UJp$.
It is not hard to see that the Whitney factors of this subbundle have first
Chern classes $c'_i=m_ic_i$. Inserting
$k'_i$, $m'_i$, and $c'_i$ into \gref {GAsoend} one obtains formula
$(6)$ in \cite{Asorey:Nodes}. In fact, the authors of
\cite{Asorey:Nodes} use that $A$ is reducible to $Q_L'$, rather than that
it is even reducible to $Q_L$.
\vspace{0.3cm}

As a consequence of Theorem \rref{TNodes}, the property of being a
kinematical node is actually a property of strata. It can be read off
directly from the labels $L\in\rmK(P)$. 
As an example, we present the Hasse diagram of $\hat\rmK(P)$ for $\rmSU(2)$
(which can be derived analogously to the $4$-dimensional case explained in
Subsection \rref{Sex}), with the nodal strata marked by an additional circle:
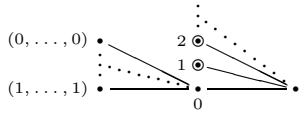
\begin{center}
\unitlength1.3cm
\begin{picture}(2,1.75)
\put(0,0.5){ 
\plzmee{0,0.5}{cr}{(0,\dots,0)~}
\pslvmee{0,0.25}
\vpunkte{0,0.25}
\plene{0,0}{cr}{(1,\dots,1)~}
%
\pslvmde{1,0.75}
\vpunkte{1,0.75}
\plzmee{1,0.5}{cr}{2~}
\plvmee{1,0.25}{cr}{1~}
\plene{1,0}{tc}{0}
\put(1,0.25){\circle{0.12}}
\put(1,0.5){\circle{0.12}}
\whole{2,0}{bc}{}
%
}
\end{picture} 
\end{center}  
The $\rmU(1)$-strata are labelled by the moduli of the first Chern classes of 
the corresponding $Q_L$. The $\ZZ_2$-strata are labelled by elements of 
$\ZZ_2^{2s}$, where $s$ is the genus of $\Sigma$. Thus, all but one
$\rmU 1$-strata are kinematical nodes. The non-nodal stratum is that with
zero topological charge. It is the only one which itself has singularities,
where the singularities are all non-nodal.
\section{Outlook}
In the present review we have given a survey on the stratified structure 
of the gauge orbit space. Based on the results presented, a lot of points 
deserve a detailed study, for example,

- the topology of strata, in particular w.r.t.~potential anomalies
\cite{Heil:Anom}, 

- the geometric properties of strata w.r.t.~the $L^2$-metric, in particular in
the vicinity of singularities,

- the study of other metrics, like the strong metrics $\gamma^k$ or $\eta^k$,
defined in Subsection \rref{Sbasic} or the (potentially degenerate)
information metric \cite{Friedrich:Info,GroisserMurray}.

From the viewpoint of physics, however, the most important question 
related to the stratified structure of the gauge orbit space is: what is
the physical relevance of the nongeneric strata, i.e., what physical 
effects do they produce? 
To study this question systematically, one needs a quantization in which
all strata are included on an equal footing and in which the stratification
is explicitly encoded. To achieve this, we propose to view the gauge theory
as an infinite dimensional Hamiltonian system with symmetry and to work out 
the following programme:

1. Try to carry over the procedure of singular Marsden-Weinstein
reduction, established in finite dimensions by Sjamaar and Lerman 
\cite{SjamaarLerman} to the infinite dimensional Hamiltonian system under
consideration (for an exposition of the method see 
\cite[Appendix B5]{CushmanBates} or \cite[\S IV.1.11]{Landsman:Book}).
Singular Marsden-Weinstein reduction equips the reduced phase space with the
structure of a stratified symplectic space ('singular Marsden-Weinstein
quotient'). A stratified symplectic space is a Poisson space $X$ together with 
a stratification $X=\cup_i X_i$ (of some given type) into symplectic manifolds
$X_i$ such that the embeddings $X_i\rightarrow X$ are Poisson space morphisms.

2. Develop a geometric quantization of the reduced phase space so obtained.
The generalization of methods of geometric quantization to stratified 
symplectic spaces is a field of active research. Besides the discussion of
specific examples, until now the following notions have been established in finite
dimensions: 

- prequantization of Poisson spaces \cite{Huebschmann90} (applies to $X$),

- prequantization of symplectic manifolds (standard, applies to the $X_i$),

- polarization of stratified symplectic spaces \cite{Huebschmann01}.

\noindent
Thus, to realize the concept of geometric quantization of a stratified symplectic 
space, the first problem to be solved consists in clarifying the relation between 
the prequantization of the Poisson space $X$ and the prequantizations of its 
symplectic strata $X_i \, .$  Next, using the above mentioned polarization concept, 
one can try to construct the full quantum theory. 
Then, it is still a big challenge to extend these methods to the infinite 
dimensional case.
\section{Acknowledgements}
The authors are indebted to M.~\v{C}adek, T.~Friedrich, J.~Hilgert, C.~Isham, H.-B.~Rademacher 
and L.M.~Woodward for useful suggestions and remarks.
I.~V.~is grateful to the Centre of Natural Sciences of the University
Leipzig for the warm hospitality extended to him during his stay in
Leipzig.

\newpage


\begin{appendix}

\section[~\\Some basic facts from bundle theory]{
Some basic facts from bundle theory}
\label{bunth}
\paragraph{Classifying spaces and classifying maps.}
Let $G$ be a Lie group. A principal $G$-bundle $E \rightarrow B$ is called 
universal for $G \,,$ iff $E$ is contractible. It can be shown that, 
for any Lie group $G \, ,$ there exists a universal bundle
$$
G\hookrightarrow\rmE G\stackrel{\pi_G}{\rightarrow}\rmB G
$$
with the following property: For any \CW-complex (hence, in particular, any
manifold) $X$ the assignment
\begin{equation} \label{Ghtpclf}
[X,\rmB G]\longrightarrow\Bun{X}{G}\,,~~f\mapsto f^\ast\rmE G\,,
\end{equation}
is a bijection. Here, $[\cdot,\cdot]$ denotes the set of homotopy classes of 
maps, $\Bun{X}{G}$ is the set of isomorphism classes of principal 
$G$-bundles over $X$ (where bundle morphisms are assumed to project 
to the identical mapping on $X$) and $f^\ast$ denotes the 
pull-back of bundles:
$
f^\ast\rmE G = \{ (x,\epsilon)\in X\times\rmE G~:~f(x) = \pi_G(\epsilon)\}\,.
$
$\rmB G$ is called the classifying space of $G$ and the homotopy class of maps 
$X\rightarrow\rmB G$ associated to $P\in\Bun{X}{G}$
by virtue of \gref{Ghtpclf} is called the classifying map of
$P$ . In this appendix, we will denote it by $\ka{P}$. Since the total space
of $\rmE G$ is contractible, the exact homotopy sequence of fibre
spaces implies
\begin{equation} \label{GGandBG}
\pi_i(G)\cong\pi_{i+1}(\rmB G)\,,~~i=0,1,2,\dots\,.
\end{equation}
\paragraph{Associated principal bundles defined by homomorphisms.}
Let $\varphi:G\rightarrow G'$ be a Lie group homomorphism and let
$P\in\Bun{X}{G}$. By virtue of the action
$$
G' \times G \rightarrow G' \,,~~(a',a) \mapsto \varphi(a^{-1})a'\,,
$$
$G'$ becomes a right $G$-space and we have an associated bundle
$P\times_G G'$. To indicate that this bundle is completely
given by $\vp$, we denote it by $\ab{P}{\vp}$. By setting
$[(p,a')]\cdot b' := [(p,a'b')]$, $\forall p\in P$, $a',b'\in G'$, a right
$G'$-action on $\ab{P}{\vp}$ is defined, thus making it a principal
$G'$-bundle over $X$.

In the main text of the review, two special cases of associated principal
bundles occur:

(i) $\vp$ is a Lie subgroup embedding. Here
$\ab{P}{\vp}$ represents the extension of $P$ to structure group $G'$.

(ii) $\vp$ is factorization by a normal Lie subgroup $N$. Here
$\ab{P}{\vp}$ represents the quotient bundle $P/N$.

Thus, the construction of associated principal bundle provides a
unifying viewpoint on the operations of extending the structure group
and factorizing by a normal subgroup. In particular, it allows to determine
the classifying map of both extensions and quotients. Namely, one has
\begin{equation} \label{Gclfmapassbun}
\ka{\ab{P}{\varphi}} = \rmB\varphi\circ\ka{P}\,,
\end{equation}
where $\rmB\varphi:\rmB G\rightarrow\rmB G'$ is the map of classifying spaces
associated to $\varphi$. It is defined as the classifying map of the
associated principal $G'$-bundle $\ab{(\rmE G)}{\varphi}$. Note the
following (covariant) functorial property: For $\varphi:G\rightarrow G'$ and
$\varphi':G'\rightarrow G''$ there holds
$$
\rmB(\psi\circ\varphi)=\rmB\psi\circ\rmB\varphi\,.
$$
\section[~\\Eilenberg-MacLane spaces and Postnikov tower]{
Eilenberg-MacLane spaces and Postnikov tower}
\label{AlgTop}
\paragraph{Eilenberg-MacLane Spaces.}
Let $\pi$ be a group and $n$ a positive integer. An arcwise connected
\CW-complex $X$ is called an Eilenberg-MacLane space of type
$K(\pi,n)$ iff $\pi_n(X)=\pi$ and $\pi_i(X)=0$ for $i\neq n$.
Eilenberg-MacLane spaces exist for any choice of $\pi$ and $n$, provided
$\pi$ is commutative for $n\geq 2$. They are unique up to homotopy
equivalence.

The simplest example of an Eilenberg-MacLane space is the $1$-sphere
$\sphere{1}$, which is of type $K(\ZZ,1)$. Two further examples, $K(\ZZ,2)$
and $K(\ZZ_g,1)$, are briefly discussed in \rref{Postnikov}.
Apart from very special examples, Eilenberg-MacLane
spaces are infinite dimensional.
Up to homotopy equivalence one has
$$
K(\pi_1\times\pi_2,n)=K(\pi_1,n)\times K(\pi_2,n)\,.
$$
Now assume $\pi$ to be commutative also in the case $n=1$.
Due to the Hurewicz and the universal coefficient theorems,
$H^n(K(\pi,n),\pi) = \Hom\left(H_n(K(\pi,n)),\pi\right)$.
Moreover, $H_n(K(\pi,n)) \cong \pi_n(K(\pi,n)) = \pi$. It follows that
$H^n(K(\pi,n),\pi)$ contains
elements which correspond to isomorphisms $H_n(K(\pi,n))\rightarrow\pi$.
Such elements are called {\it characteristic}.
If $\gamma\in H^n(K(\pi,n),\pi)$ is characteristic then for any \CW-complex
$X$, the map
\begin{equation} 
\label{G[]=H}
[X,K(\pi,n)]\rightarrow H^n(X,\pi)\,, ~~f\mapsto f^\ast\gamma\,,
\end{equation}
is a bijection \cite[\S VII.12]{Bredon:Top}.
In this sense, Eilenberg-MacLane spaces provide a link
between homotopy properties and cohomology.

Next, consider the path-loop fibration over $K(\pi,n)$,
$$
\Omega(K(\pi,n))\hookrightarrow P(K(\pi,n))\longrightarrow K(\pi,n)\,,
$$
where $\Omega(K(\pi,n))$ and $P(K(\pi,n))$ denote the loop space and the path
space of $K(\pi,n)$, respectively (both based at some point $x_0$). Since
$P(K(\pi,n))$ is
contractible, the exact homotopy sequence induced by the path-loop fibration
implies 
$
\pi_i\left(\Omega(K(\pi,n+1))\right) 
= 
\pi_{i+1}\left(K(\pi,n+1)\right)
$
Hence, $\Omega(K(\pi,n+1))=K(\pi,n)$, $\forall n$,
and the path-loop fibration
over $K(\pi,n+1)$ reads
\begin{equation} \label{Gpathloopfibr}
K(\pi,n)\hookrightarrow P(K(\pi,n+1))\longrightarrow K(\pi,n+1)\,.
\end{equation}
\paragraph{Postnikov Tower.}
\label{CHSBSprelimsSSPost}
A map $f:X\rightarrow X'$ of topological spaces is called an
$n$-equivalence iff the homomorphism induced on homotopy groups
$f_\ast:\pi_i(X)\rightarrow\pi_i(X')$ is an isomorphism for $i<n$ and
surjective for $i=n$. One also defines the notion of an $\infty$-equivalence,
which is often called weak homotopy equivalence.

Let $f:X\rightarrow X'$ be an $n$-equivalence and let $Y$ be a \CW-complex.
Then the map $[Y,X]\rightarrow[Y,X']$, $g\mapsto f\circ g$, is bijective for
$\dim Y<n$ and surjective for $\dim Y=n$
\cite[Ch.~VII,~Cor.~11.13]{Bredon:Top}.

A \CW-complex $Y$ is called $n$-simple iff it is arcwise connected and the
action of $\pi_1(Y)$ on $\pi_i(Y)$ is trivial for $1\leq i\leq n$. It is
called simple iff it is $n$-simple for all $n$.

The following theorem describes how a simple \CW-complex can be approximated 
by $n$-equivalent spaces constructed from Eilenberg-MacLane spaces.
\vspace{0.2cm}

\noindent
{\bf Theorem B.1.}~
{\it Let $Y$ be a simple \CW-complex. There exist
\\
{\rm (i)} a sequence of \CW-complexes $Y_n$ and principal fibrations
$$
K(\pi_n(Y),n)\hookrightarrow Y_{n+1}\stackrel{q_n}{\longrightarrow}Y_n\,,
~~n=1,2,3,\dots\,,
$$
given as the pull-back of the path-loop fibration \gref{Gpathloopfibr} over 
$K(\pi_n(Y),n+1)$ by some map 
$\theta_n:Y_n\rightarrow K\left(\pi_n(Y),n+1\right)$,
\\
{\rm (ii)} a sequence of $n$-equivalences $y_n:Y\rightarrow Y_n$,
$n=1,2,3,\dots$, 
\\
such that $Y_1=\onepoint$ (one point space) and $q_n\circ y_{n+1}=y_n$ for all 
$n$.}
\vspace{0.3cm}

The sequence of spaces and maps $\left(Y_n,y_n,q_n\right)$,
$n=1,2,3,\dots$, is called Postnikov tower (or Postnikov decomposition)
of $Y$.

We remark that the theorem follows from a more general theorem about simple 
maps \cite[Ch.~VII,~Thm.~13.7]{Bredon:Top} by noting that the assumption that
$Y$ be a simple \CW-complex implies that the constant map
$Y\rightarrow\onepoint$ is a simple map. See
\cite[Ch.~VII,~Def.~13.4]{Bredon:Top} for a definition of the latter.


\section[~\\Construction of $\BSUJ_5$]{Construction of $\BSUJ_5$}
\label{Postnikov}
In this appendix, let $J\in\rmK(n)$ and consider the classifying space
$\BSUJ$ of the Howe subgroup $\SUJ$. We are going to prove that $\BSUJ_5$,
i.e., the $5$th stage of the Postnikov tower of $\BSUJ$ is given by formula
(\ref{GBG5}).
\paragraph{Preparation.}
First, in order to be able to apply Theorem B.1, we have to check that
$\BSUJ$ is
a simple space. To see this, note that any inner automorphism of $\SUJ$ is
generated by an element of $\SUJ_0$, hence is homotopic to the identity
automorphism. Consequently, the natural action of $\pi_0(\SUJ)$ on
$\pi_{i-1}(\SUJ)$,
$i=1,2,3,\dots$ , induced by inner automorphisms, is trivial. Since the
natural isomorphisms $\pi_{i-1}(\SUJ)\cong\pi_i(\BSUJ)$ transform this 
action into that of $\pi_1(\BSUJ)$ on $\pi_i(\BSUJ)$,
the latter is trivial, too. Thus, $\BSUJ$ is a simple space, as asserted.

Second, we note the relevant homotopy groups of $\BSUJ$. 
According to \gref{Ghtpgr} and 
\gref{GGandBG}, these are
\begin{equation} \label{GhtpgrBSUJ}
\pi_1=\ZZ_g\,,~~~~
\pi_2=\ZZ^{\oplus (r-1)}\,,~~~~
\pi_3=0\,,~~~~
\pi_4=\ZZ^{\oplus r^\ast}\,,
\end{equation}
where $r^\ast$ denotes the number of $k_i>1$.

Third, we will need information about the integer-valued cohomology
groups of the Eilenberg-MacLane spaces $K(\ZZ_g,1)$ and $K(\ZZ,2)$. 

(i) $K(\ZZ,2)$: Consider the natural free action of $\rmU(1)$ on the sphere
$\sphere{\infty}$ (induced from $\rmU(1)$-action on
$\sphere{2n-1}\subseteq\CC^n$). The orbit space of this action is known as 
the infinite dimensional complex projective space $\CC\rmP^\infty$. 
Due to $\pi_i(\sphere{\infty}) = 0$, the exact homotopy
sequence of the principal bundle 
$\rmU(1)\hookrightarrow\sphere{\infty}\rightarrow\CC\rmP^\infty$ implies
$\pi_i(\CP^\infty) = \pi_{i-1}(\rmU(1)) = \ZZ \, ,$ for $i=2$ and $0$ otherwise.
Thus, $\CC\rmP^\infty$ is a model for $K(\ZZ,2)$.
It follows, see \cite[Ch.~VI, Prop.~10.2]{Bredon:Top},
\begin{equation}\label{GHiKZ2}
H^i(K(\ZZ,2),\ZZ) = \left\{\begin{array}{ccl}
\ZZ & | & i\mbox{ even,}\\
0 & | & i\mbox{ odd.}\,,
\end{array}\right.
\end{equation}

(ii) $K(\ZZ_g,1)$: Consider the restriction of the above action to the
subgroup $\ZZ_g\subseteq\rmU(1)$. The resulting orbit space is the 
infinite dimensional lens space $\rmL^\infty_g$. The exact homotopy
sequence of the corresponding principal bundle implies 
$\pi_i(\lens{g}{\infty}) = \pi_{i-1}(\ZZ_g) = \ZZ_g \, ,$ for $i=1$ and $0$
otherwise. Hence, $\lens{g}{\infty}$ is a model for $K(\ZZ_g,1)$. 
Consequently, see \cite[\S 24, p.~176]{FomenkoFuchs},
\begin{equation}\label{GHiKZg1}
H^i(K(\ZZ_g,1),\ZZ)
= \left\{\begin{array}{ccl}
\ZZ & | & i=0,\\
\ZZ_g & | & i\neq 0\mbox{, even,}\\
0 & | & i\neq 0\mbox{, odd.}\,,
\end{array}\right.
\end{equation}
(Note that the vanishing of all homotopy groups of $\sphere{\infty}$ 
also implies that $\CP^\infty$ and $\lens{g}{\infty}$ are models for the
classifying spaces $\rmB\rmU(1)$ and $\rmB\ZZ_g$, respectively.) 
\paragraph{Construction.}
We start with $\BSUJ_1=\onepoint$. 
Then $\BSUJ_2$ must coincide with the fibre which is $K(\ZZ_g,1)$, see
\gref{GhtpgrBSUJ}. Next, according to \gref{GhtpgrBSUJ}, $\BSUJ_3$ is the 
total space of a fibration
\begin{equation} \label{GBSUJ3}
K(\ZZ^{\oplus (r-1)},2) \hookrightarrow \BSUJ _3\stackrel{q_2} {\longrightarrow}
K(\ZZ_g,1)
\end{equation}
given by the pull-back of the path-loop fibration over 
$K(\ZZ^{\oplus (r-1)},3)$ by some map 
$
\theta_2:K(\ZZ_g,1)\rightarrow K(\ZZ^{\oplus (r-1)},3)
$. 
Since $K(\ZZ^{\oplus (r-1)},n)=\prod_{j=1}^{r-1}K(\ZZ,n)$, $\forall n$,
\gref{G[]=H} yields for the set of homotopy classes 
$$
[K(\ZZ_g,1),K(\ZZ^{\oplus (r-1)},3)]=\prod_{i=1}^{r-1} H^3(K(\ZZ_g,1),\ZZ)\,.
$$
Due to \gref{GHiKZg1}, the r.h.s.~is trivial. Hence, $\theta_2$ is 
homotopic to a constant map, so that the fibration \gref{GBSUJ3} is trivial.
Thus,
$$
\BSUJ_3=K(\ZZ_g,1)\times\prod_{j=1}^{r-1}K(\ZZ,2)\,.
$$
Then, in view of \gref{GhtpgrBSUJ}, $\BSUJ_4$ is given by a
fibration over $\BSUJ_3$ with fibre $K(0,3)=\onepoint$. Hence, it just
coincides with the base space. Finally, $\BSUJ_5$ is the total space of a 
fibration
\begin{equation}\label{GfibrBG5}
K\left(\ZZ^{\oplus r^\ast},4\right)\hookrightarrow\BSUJ_5
\stackrel{q_4}{\longrightarrow}
K\left(\ZZ_g,1\right)\times\prod_{j=1}^{r-1}K(\ZZ,2)\,,
\end{equation}
which is induced by a map $\theta_4$ from the base to $K\left(\ZZ^{\oplus
r^\ast},5\right)$. We have  
%
$$
\left[K(\ZZ_g,1)
\times
\prod_{j=1}^{r-1}K(\ZZ,2),K\left(\ZZ^{\oplus r^\ast},5\right)\right]
=
\prod_{i=1}^{r^\ast} H^5\left(
K(\ZZ_g,1)\times\prod_{j=1}^{r-1}K(\ZZ,2),\ZZ
\right)\,.
$$
Since $H^\ast(K(\ZZ,2),\ZZ)$ is torsion-free, see \gref{GHiKZ2}, we can apply 
the K\"unneth Theorem for cohomology to obtain
\begin{eqnarray*}
&&
\hspace{-1.5cm}H^5\left(K(\ZZ_g,1)\times\prod_{j=1}^{r-1}K(\ZZ,2),\ZZ\right)
\\
&&
\cong~
\bigoplus~
H^j(K(\ZZ_g,1),\ZZ)\otimes
H^{j_1}(K(\ZZ,2),\ZZ)\otimes\cdots\otimes H^{j_{r-1}}(K(\ZZ,2),
\ZZ)\,,
\end{eqnarray*}
with the direct sum running over all decompositions of $5$
into a sum of $r$ nonnegative integers $j,j_1,\dots,j_{r-1}$.
Each summand of the rhs. is trivial, because it contains tensor factors of 
odd degree, which are trivial due to \gref{GHiKZ2} and \gref{GHiKZg1}. 
Hence, $\theta_4$ is again homotopic a constant map and the fibration 
\gref{GfibrBG5} is trivial. This proves formula \gref{GBG5}, used in the
main text.
\end{appendix}

\newpage


\end{document}